\documentclass[a4paper,11pt]{article}

\usepackage{jinstpub}

\usepackage{hyperref}
\usepackage{graphicx}
\usepackage{amsmath}
\usepackage{wrapfig}
\usepackage{url}
\usepackage{xcolor}
\usepackage{upgreek}
\usepackage{makecell}

\usepackage{array}
\newcolumntype{L}[1]{>{\raggedright\let\newline\\\arraybackslash\hspace{0pt}}m{#1}}
\newcolumntype{C}[1]{>{\centering\let\newline\\\arraybackslash\hspace{0pt}}m{#1}}
\newcolumntype{R}[1]{>{\raggedleft\let\newline\\\arraybackslash\hspace{0pt}}m{#1}}

\title{Reciprocating probe measurements in the test divertor operation phase of Wendelstein 7-X}

\author[a]{C. Killer,}
\author[b]{P. Drews,}
\author[a,c]{O. Grulke,}
\author[b]{A. Knieps,}
\author[b]{D. Nicolai,}
\author[b]{G. Satheeswaran,}
\author[a,d]{W7-X Team}

\affiliation[a]{Max-Planck-Institut für Plasmaphysik, Greifswald, Germany}
\affiliation[b]{Forschungszentrum Jülich, IEK4-Plasmaphysik, Jülich, Germany}
\affiliation[c]{Department of Physics, Technical University of Denmark, Lyngby, Denmark}
\affiliation[d]{definition in T. Klinger et al. , Nucl. Fusion 59 (2019) 112004}

\emailAdd{carsten.killer@ipp.mpg.de}

\abstract{
Reciprocating probes are a classic and widespread tool for the investigation of the edge and Scrape-Off Layer of magnetic fusion plasmas. In the Wendelstein 7-X (W7-X) stellarator, the \textit{Multi-Purpose Manipulator} serves as a multi-user platform for probe measurements of various kinds. This paper presents a review on reciprocating probe operation during the first operation phase of W7-X with a test divertor (2017-2018). It gives an overview of the diverse zoo of probe heads and presents lessons learned about probe operation in complex magnetic geometries, operation safety, and probe head design. A few examples of probe measurements with a focus on unexpected observations are presented, e.g. indications for supra-thermal electrons and the propagation of plasma perturbations due to biased probes.
}


\begin{document}
\maketitle
\flushbottom

\section{Introduction}

Reciprocating probes are a classic and widespread tool for the investigation of the edge and Scrape-Off Layer (SOL) of magnetic fusion plasmas \cite{Davies1996,Watkins1997,Boedo1998,Pedrosa1999,Ezumi2003,Schubert2007,Smick2009,Boedo2009a,Zhang2010,Gunn2011,Tsui2012,deMarne2017,DeOliveira2021}. In contrast to many line-of-sight integrated diagnostics, they have the advantage of a highly localized measurement with typically good temporal resolution. Furthermore, in contrast to fixed probes mounted in wall or target elements, reciprocating probes offer a higher flexibility as they can access different radial positions in the plasma boundary region. 
\\
A significant drawback, however, is the often short measurement duration of reciprocating probes that results from the high heat loads onto probes inserted into the edge plasma. Being an invasive diagnostic, another disadvantage is the perturbation of the plasma caused by the insertion of the probe. Nevertheless, reciprocating probes have been and continue to be a major workhorse for plasma edge investigations in many fusion plasma devices.
\\
In the optimized stellarator Wendelstein 7-X (W7-X) \cite{Klinger2019_short}, the Multi-Purpose Manipulator (MPM) acts as a multi-user platform for reciprocating probe measurements. As W7-X employs an island divertor plasma exhaust concept and features a three-dimensional SOL, a good coverage of SOL diagnostics is even more important than in symmetric devices such as tokamaks. The aim of this paper is to provide a review of MPM operation in W7-X so far and to present a report on challenges and lessons learned in order to optimize reciprocating probe operation. The MPM and its probe heads are introduced in section \ref{sec:mpm}, including a discussion on materials and geometric design of probe heads. In section \ref{sec:operation}, the MPM operation is reviewed with respect to the role of the magnetic field geometry, motion schemes, heat loads, and operation safety. Finally, a few interesting measurement examples are presented in section \ref{sec:results}, with a focus on unpublished and unexpected results, including the perturbations from biased probes, electron emission by hot probes, and indications for supra-thermal electrons outside the plasma.

\begin{figure}[b]
  \centering
  \includegraphics[width=1\textwidth]{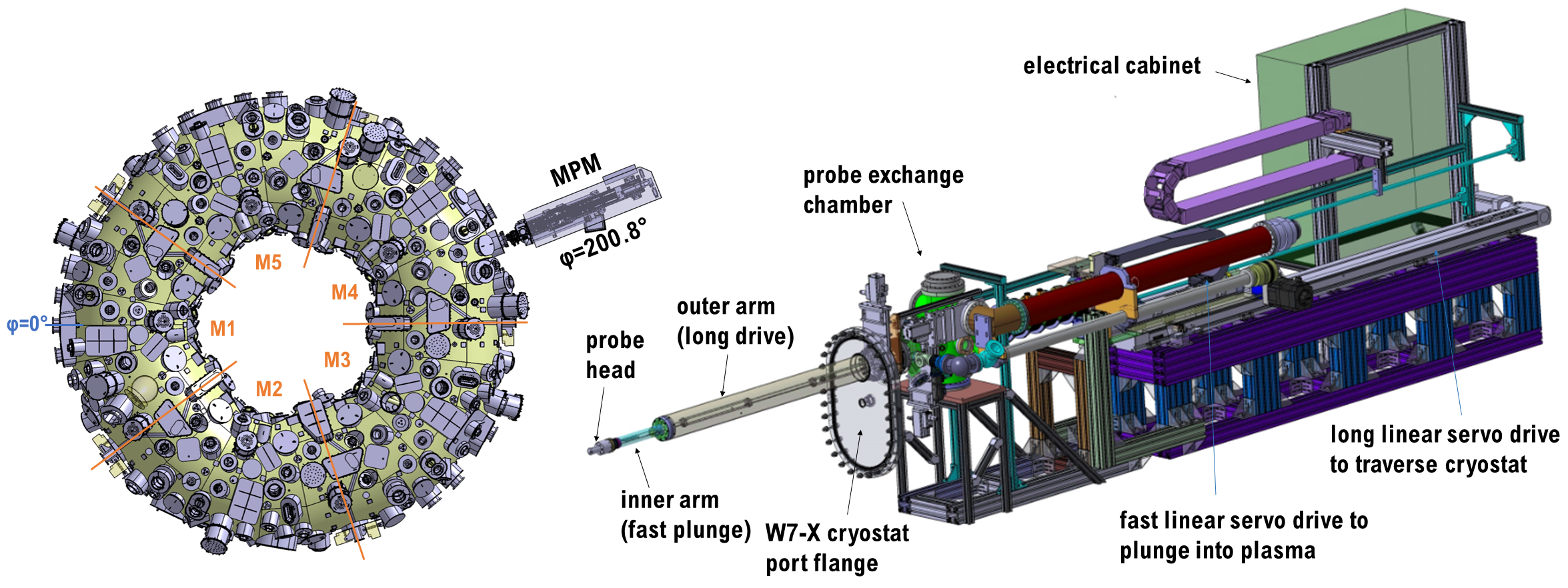}
  \caption[]{Schematic view of the MPM in a top view of W7-X (left) and major components (right)}
  \label{fig:mpm}
\end{figure}

\begin{table}[tb]
	\caption{List of MPM probe heads. Electric probe pins can be employed for different applications depending on electric operation mode, e.g. $T_e$, $n_e$ from swept Langmuir or triple probes. The column "measurements" corresponds to probe insertions into the plasma if not specified otherwise. The abbreviations for plasma-exposed materials in the final column are $C$ - Carbon (graphite), $BN$ - Boron Nitride, $W$ - Tungsten, $Mo$ - Molybdenum.}
	\centering
		\begin{tabular}{l  p{10mm}  p{25mm}  p{15mm} p{40mm} >{\raggedright\arraybackslash}p{19mm}}
			Probe head & cam-paign & measurements & reference & features & exposed materials \\ \hline
			FZJ-COMB1 & 1.1 & 24 & \cite{Drews2017,Liu2018,Liu2018a}  & 9 electric probe pins, magnetic pick-up probe & body: C pins: W \\
			FZJ-COMB2 & 1.2a/b & 378 & \cite{Drews2019,Liu2019,Drews2019a,Liu2020,Knieps2020} & 9 electric probe pins, magnetic pick-up probe, ion sensitive probe, material exposition, gas pipe & body: BN pins: W
			\\
			FZJ-GAS1 & 1.2a & 25 & \cite{Drews2021} & 4 electric probe pins, gas pipe & body: BN pins: W
			\\
			FZJ-GAS2 & 1.2b & 34 & \cite{Drews2021} & 4 electric probe pins, piezo valve for controlled gas injection & body: C pins: W
			\\
			FZJ-MACH1/2 & 1.2a/b & 75 & \cite{Cai2019}	& Polar (Gundestrup) + radial Mach probe array (28 electrodes) & body: BN/C pins: W
			\\
			FZJ-MAT1	& 1.2a &146 s exposition in 37 programs & & 8 samples for material exposition & body: steel
			\\
			FZJ-MAT2	& 1.2b & 158 s exposition in 37 programs & & 18 samples for material exposition & body: C
			\\
			FZJ-RFA1/2 & 1.2a/b & 128 & \cite{Henkel2018,Li2019,Drews2019a,Killer2019a,Henkel2020} & 6 retarding field analyzers, 2/4 electric probe pins & body: BN/C pins: W
			\\
			IPP-FLUC1  	& 1.2a/b & 502 & \cite{Killer2019,Killer2019a,Hammond2019,Killer2020,Perseo2020,Barbui2020,Ballinger2021,Killer2021} & 28 electric probe pins (poloidal array, parallel + poloidal Mach probe) & body: BN pins: Mo
			\\
			IPP-LBO1  & 1.2a/b & 650 injections in 298 programs& \cite{Wegner2018,Wegner2020a,Wegner2020} &	Holds four coated glass targets for laser ablation & 
			\\
			NIFS-FILD1  & 1.2b & 113 & \cite{Ogawa2019,Aekaeslompolo2019} &	8 Faraday films for fast ion loss detection & body: BN
			\\
			PPPL-PMPI1 	& 1.2b & 35\,s of injection in 10 programs & \cite{Nagy2019,Lunsford2021} & Horizontal powder flinger for boron impurity injection & body: C
			\\
			RFX-HRP1 & 1.2b & 105& \cite{Agostinetti2018,Spolaore2019,Lazerson2019} & 3 magnetic pick-up probes, 8 electric probe pins, 3 Mach probes & body: BN pins: C			
			\\
		\end{tabular}
	
	\label{tab:probes}
\end{table}

\section{The Multi-Purpose Manipulator}\label{sec:mpm}
The Multi-Purpose Manipulator (MPM) is a multi-user platform at W7-X that carries probes into the edge plasma of W7-X, allowing for a wide range of physics investigations \cite{Nicolai2017,Satheeswaran2017}. At its heart, it consists of two stacked linear drives that move the probe head from the exchange chamber outside of the W7-X cryostat into the boundary plasma (Fig. 1). A long ($\sim$2500\,mm) linear drive first transports the probe from the exchange chamber to a "parking position" that is approximately level with the plasma vessel wall \cite{Nicolai2017}. A second linear drive on top of the first one then performs fast plunges into the plasma vessel with a maximum plunge depth of 350\,mm and an acceleration of up to 30\,m/s$^2$, resulting in typical peak velocities of a few m/s. Both linear drives are operated with servo motors that allow arbitrary movement schemes and a fine positioning accuracy \cite{Satheeswaran2017}. The motors are located well outside the W7-X cryostat vessel where the residual magnetic field is small ($<1$\,mT).
\\
Due to its standardized plug-and-play probe interface \cite{Nicolai2017}, the MPM acts as a versatile multi-user platform that allows to operate a wide range of different probe types in frequent succession. After starting with one probe head in test operation in the first operation phase of W7-X in a limiter configuration (OP1.1, 2016)\cite{Drews2017}, already 14 different probe heads have been used in the following test-divertor operation phases 1.2a/b (2017-2018), see Table 1. The MPM was operated on 57 experiment days (out of 69 W7-X experiment days in total) with more than 1500 fast reciprocating measurements being conducted. With W7-X operation being still in its infancy, a wide range of physics questions (e.g. measurement of various plasma conditions, impurity injections, fast ion losses, material studies) had to be addressed in a short time period, necessitating in total 35 exchanges of the probe head in this period. These exchanges could only be performed when the (steady state, thanks to superconducting coils) magnetic field of W7-X was ramped down.
\\
The probe interface with 32 electric channels allows a flexible use of different probe head applications. Bias voltages for electric probes are provided by up to six ultra-capacitor modules (160\,V, 6\,F, for constant bias voltages in e.g. Mach probes) or conventional amplifiers using grid power to amplify signals defined by an arbitrary waveform generator (e.g. for operating swept Langmuir probes). Data acquisition (typically 2\,MHz), event control and synchronization to the W7-X central timing system \cite{Schacht2019} is controlled via MDSplus.

\subsection{MPM probe heads}
\begin{figure}[bt]
  \centering
  \includegraphics[width=0.9\textwidth]{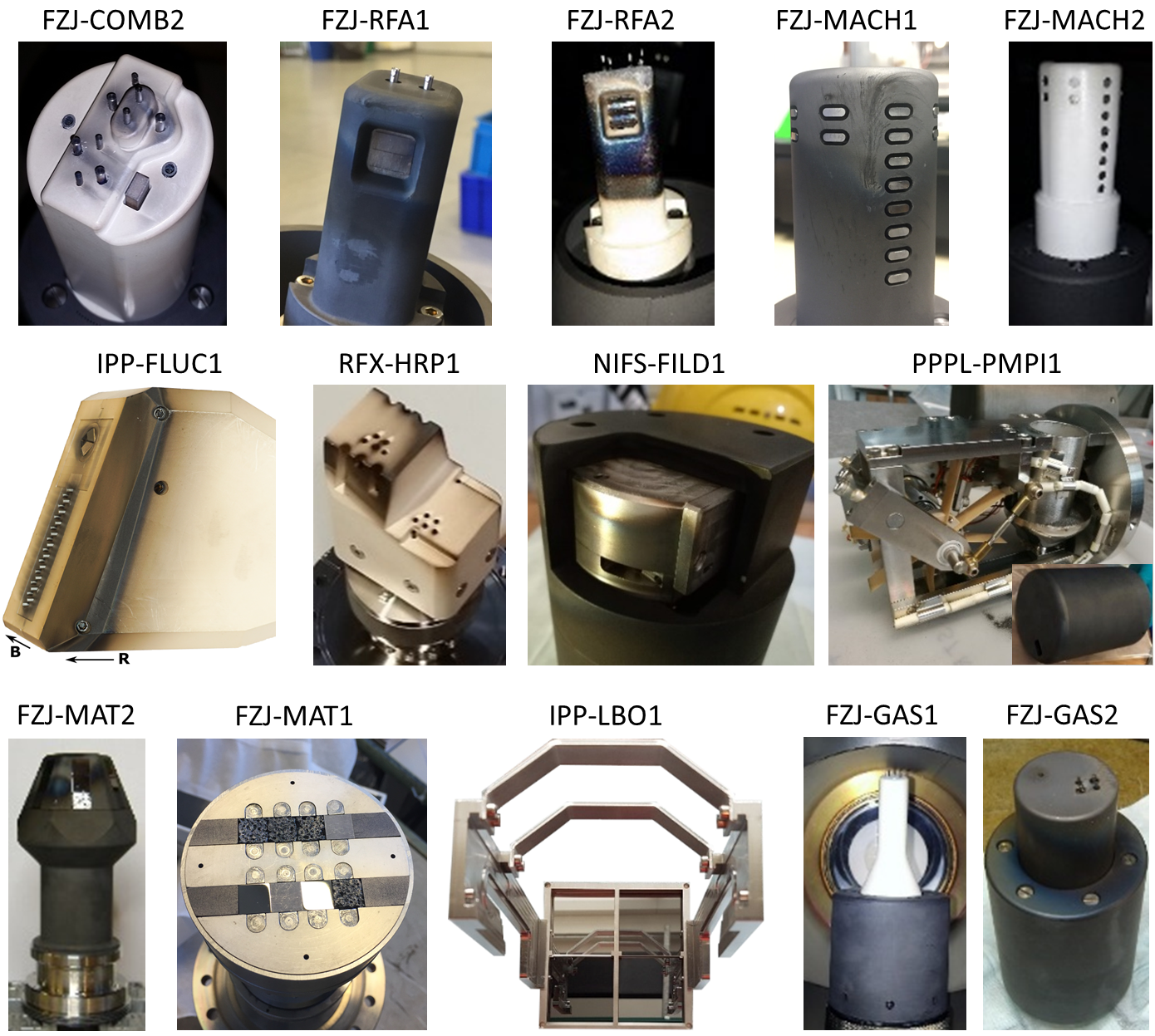}
  \caption[]{Photographs of probe heads listed in Table 1.}
  \label{fig:probes}
\end{figure}

An overview of probe heads used at the MPM so far is given in Table 1 and Fig. \ref{fig:probes}. The naming scheme for the probe heads is \textit{A-BN} where $A, B$ are abbreviations for the home institution and description of the probe head, respectively, and $N$ is an integer denominating different versions of a probe head.

\subsection{Materials}
Materials of reciprocating probes need to withstand high heat loads and mechanical forces associated with the probe movement. The plasma-facing parts of the probe bodies of MPM probe heads have been made of graphite or (hexagonal) boron nitride (BN), which both offer good thermal stability against the high loads experienced when inserted into the plasma. Both materials can easily be machined. 
\\
Graphite is advantageous in terms of impurity handling, as it can be obtained in and conditioned to a very clean state such that release of impurities due to heating of the material is small. The release of carbon in case of graphite erosion due to overheating is usually well tolerated by the plasma in W7-X \cite{Jakubowski2021}. However, a graphite body makes insulating the probe tips quite challenging, as arcs between the often electrically biased pins and the probe body must be avoided. In addition, a graphite body may perturb the plasma stronger than an dielectric body, as it offers cross-field electrical conductivity along the probe head, which can affect the electric fields in the plasma edge and therefore distort the measurement. Furthermore, currents along the probe body might exert $j\times B$ stress on the probe head \cite{Davies1996}.
\\
Boron nitride, in comparison, being a dielectric, greatly simplifies insulation of the probe pins and avoids currents in the probe body. However, even the highest purity BN products are not as pure as graphite, posing the risk of contaminating the plasma with impurities by outgassing upon heating. Therefore, thorough pre-conditioning of BN probes by heating and vacuum pumping in the lab was found to be crucial in order to reduce impurity release into the plasma. Even with these preparations, BN probe heads were observed to be more likely to affect the plasma by impurity release when becoming hot, see e.g. \cite{Killer2019a}.
\\
In summary, BN was favored in the majority of probe heads due to the electrical advantages over graphite. For experiments where probes are designed to receive extremely high heat loads, graphite might be more suited for its smaller impurity release. 
\\
The probe tips were made either of graphite or highly heat resistant metals. While graphite again offers certain advantages (only low $Z$ vapor ablation in case of overheating, and a larger work function reducing secondary electron emission), mostly tungsten (W) or molybdenum (Mo) probe tips were used, as metallic probes feature a better mechanical stability of fine structures. The choice between W and Mo was mostly due to machining considerations: W has a superior thermal performance ($T_{melt} \approx 3420$°), but is very brittle, so that delicate mechanical structures are difficult to machine. Mo, in contrast, is much easier to machine while still offering a good thermal stability ($T_{melt} \approx 2620$°).
\\
As per general policy of W7-X, only materials with low magnetic permeability are used for probe heads so that the magnetic field is not perturbed by the probe.

\subsection{Probe head geometry} \label{sec:design}

\begin{figure}[tb]
  \centering
  \includegraphics[width=0.8\textwidth]{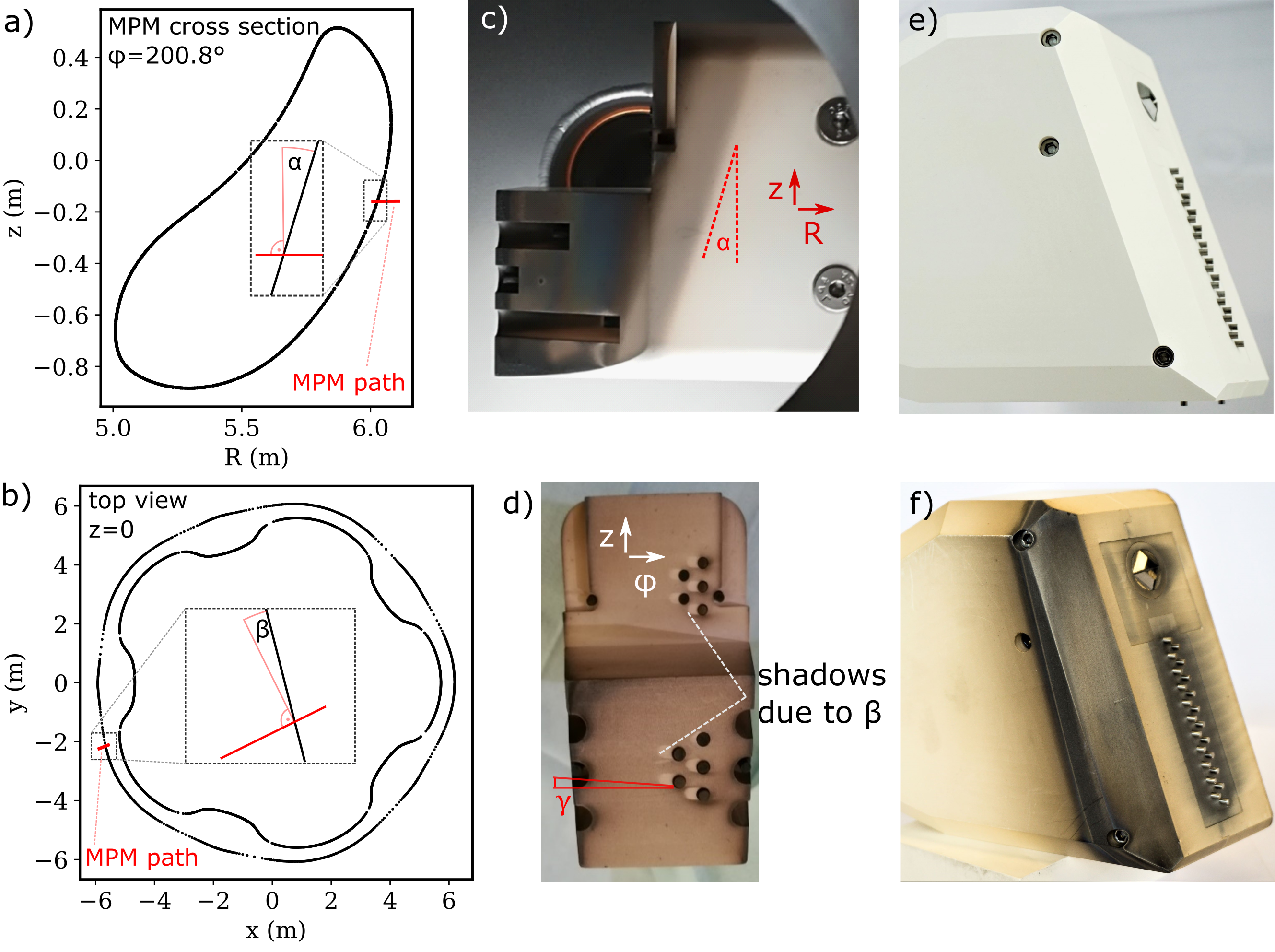}
  \caption[]{Angle between MPM vector and flux surface close to the LCFS in the magnetic standard configuration in a) $R$-$z$ plane and b) $x$-$y$ plane. The RFX-HRP1 probe depicted in c), d) is not shaped to align with the magnetic field structure and therefore stains from plasma operation reflect the angles $\alpha$ in c) and $\beta$ (implicitly) in d). The shape of the IPP-FLUC1 probe depicted in e) (before operation) and f) (after operation) mimics the magnetic field structure, which is confirmed as the discolorations from plasma operation agree with the probe geometry.}
  \label{fig:probe_geometry}
\end{figure}

While the vector of the MPM is approximately along the radial direction (i.e. pointing towards the torus center at $x$=$y$=0) at a constant vertical position, the shape of the plasma is three-dimensional and does not align with a cylindrical coordinate system. Therefore, the probe does not move along the normal vector of flux surfaces. 
\\
In the case of multi-pin probes, it is often desired to place multiple pins on the same flux surface, e.g. for Mach probes, Triple probes, or poloidal probe array for fluctuation studies. Fig. \ref{fig:probe_geometry} shows how the plane containing the multiple probes tips has to be oriented with respect to the MPM co-ordinates: 
\begin{itemize}
    \item vertical direction: angle $\alpha$ between the flux surface shape in the $R$-$z$ cross section and the $z$ axis, see Fig. \ref{fig:probe_geometry} a)
    \item toroidal direction: angle $\beta$ between flux surface shape in the $x$-$y$ cross section and the tangential vector of a hypothetical cylinder approximating W7-X, see Fig. \ref{fig:probe_geometry} b)
    \item "pitch" angle of field lines $\gamma$ against the $x$-$y$ plane (due to the non-planar axis of W-7X) can be important for interactions of probe pins with each other or direction-sensitive probes such as retarding field analyzers.
\end{itemize}

The relevance of considering $\alpha$ and $\beta$ for the probe head design is corroborated by the photographs in Fig. \ref{fig:probe_geometry}: The probe head depicted in c), d) has a front plane containing multiple pins that is orthogonal to the MPM vector, i.e. not aligned to the magnetic field structure. The dark stains from plasma operation on the (initially white) boron nitride body in Fig. \ref{fig:probe_geometry} c) nicely confirm $\alpha=18$° if the shape of the discoloration is assumed to be the mark of a flux surface. As a consequence, the individual probe tips on the front plane of the probe are on different flux surfaces at a given probe position. The role of $\beta$ is indicated by the bright spots next to the probe tips in Fig. \ref{fig:probe_geometry} d), which are due to the shadowing of magnetic field lines by the probe body to the left and probe tips to the right.
\\
In contrast, the probe head (also boron nitride) in Fig. \ref{fig:probe_geometry} e), f) attempted to mimic the flux surface shape. Comparing the pristine probe head with the probe head after plasma operation, it is found that $\alpha$, $\beta$ and $\gamma$ (the latter two not clearly visible in this photograph) were targeted reasonably based on the dark stains.
\\
The angles $\alpha$, $\beta$ $\gamma$ depend on the magnetic configuration and on the position of the probe. In Table \ref{tab:angles}, the angles at the LCFS are provided for four major magnetic configurations. The angles do not change significantly across the surface of the probe head, i.e. the curvature of the LCFS can be neglected when considering a patch the size of a probe head ($\approx$10\,cm).

\begin{table}[tb]
	\caption{Angles between LCFS and MPM path as defined in section \ref{sec:design} and Fig. \ref{fig:probe_geometry}.}
	\centering
		\begin{tabular}{l  c  c  c}
			configuration & $\alpha$ (°) & $\beta$ (°) & $\gamma$ (°)  \\ \hline
			standard & 18.2 & 12.5  & 4.0 \\
			high mirror & 18.1 & 12.1  & 4.4 \\
			high iota & 21.0 & 13.3  & 3.6 \\
			low iota & 15.0 & 11.7  & 4.4 \\
			\\
		\end{tabular}
	
	\label{tab:angles}
\end{table}

\section{Review of MPM operation}\label{sec:operation}
Operating a reciprocating probe in a fusion plasma is a delicate exercise: On the one hand, we seek to obtain long measurement durations and expand the radial coverage of the probe position as deep as possible into the plasma in order to maximize the physics gain. On the other hand, with typical electron temperatures of $T_e\approx 50$\,eV at plasma densities of $n_e \sim$ a few $10^{19}$m$^{-3}$ at the last closed flux surface (LCFS), the probe can experience convective heat loads of several MW/m$^2$ during insertion. Excess heat loads can result in the probe measurements becoming invalid (e.g. probe becoming emissive), damage to the probe, and impurity release into the plasma (potentially terminating the discharge). In this section, different aspects of probe operation will be addressed.

\begin{figure*}[tb]
  \centering
  \includegraphics[width=1\textwidth]{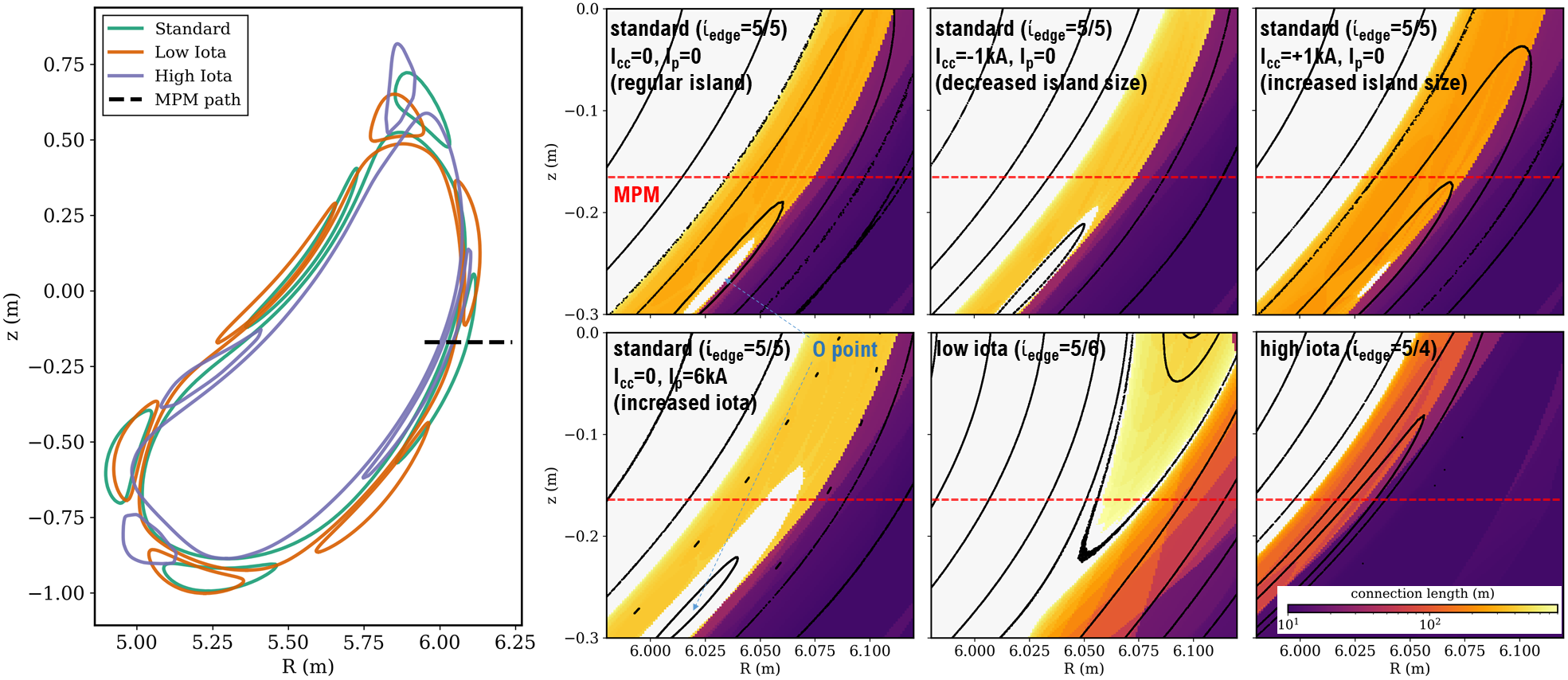}
  \caption[]{Poincare representation of MPM cross section in different major island divertor configurations (left) and combined Poincare plot and color-coded connection lengths in the MPM vicinity (right). The MPM path is indicated by the horizontal dashed line}
  \label{fig:Poincare}
\end{figure*}

\subsection{Magnetic field geometry}\label{sec:magnetic}

\begin{figure}[tb]
  \centering
  \includegraphics[width=0.5\textwidth]{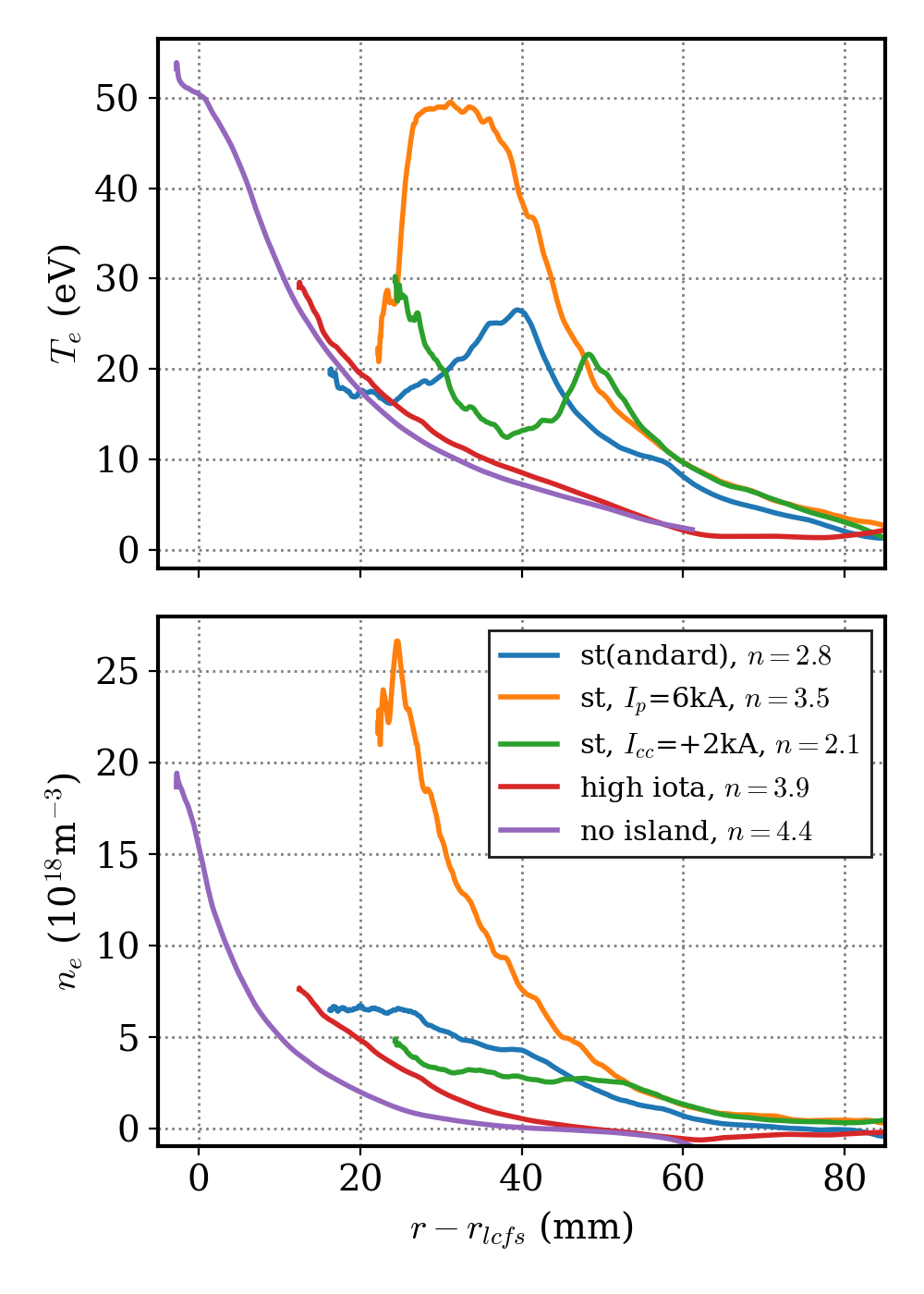}
  \caption[]{Electron temperature and density profile as a function of distance from the LCFS along the MPM path in different scenarios with ECR heating powers of 4-5\,MW. In the legend, $n$ refers to the volume averaged density and is given in units of $10^{19}$m$^{-3}$. The standard configuration is abbreviated as "st" and if not stated otherwise, the toroidal plasma current is $I_p \approx 0$ and the control coil current is $I_{cc}=0$.}
  \label{fig:profiles}
\end{figure}
The SOL of W7-X is dominated by the chain of magnetic islands to facilitate island divertor operation. Depending on the specific magnetic configuration, different island chains can be formed, most importantly 5/5 (standard), 5/4 (high iota), 5/6 (low iota). The edge magnetic topology for these configurations in the vicinity of the probe plunge are compiled in Fig. \ref{fig:Poincare}. For a given configuration, further fine adjustments are possible using the divertor control coils \cite{Renner2000,Barbui2020}, which barely affect the overall magnetic field but can significantly change the island size and position. In Fig. \ref{fig:Poincare}, modification of the island size is shown in two examples. For an increased island size, connection lengths decrease, the closed field line region (light gray color) around the O point (the center of the magnetic island flux surfaces, see marks by blue arrows in two subplots) becomes smaller, and the LCFS is pushed inwards (towards smaller R). For a reduced island size, connection lengths increase, the closed field line region around the O point becomes larger, and the LCFS moves outwards. A similar effect to reduced island sizes is observed when the rotational transform is slighty increased due to the toroidal plasma current \cite{Gao2019a}, here showcased for a situation with $I_p=$6\,kA. In this situation, the island moves inwards, away from the divertor (due to the low shear profile of W-7X \cite{Klinger2019_short}), resulting in longer connection lengths and a larger confined region around the O point.
\\
The radial profiles of $T_e$ and $n$ obtained via MPM probe measurements in Fig. \ref{fig:profiles} exemplarily show the effect of different magnetic configurations on the SOL plasma conditions. In a situation with no islands or the very narrow island in the high iota configuration, the profiles decay quickly from the LCFS into the SOL - similar to tokamaks - with typical pressure / heat flux decay lengths of 5-10\,mm \cite{Killer2019}. 
\\
In the standard configuration, where the probe crosses a magnetic island (see Fig. \ref{fig:Poincare}), the profiles are much broader. In the reference case ($I_p=0$, $I_{cc}=0$, blue in Fig. \ref{fig:profiles}), a local $T_e$ maximum is observed at the position of the transition from short connection lengths in the shadowed region of the island to long connection lengths in the main island SOL (here $\sim 40$\,mm) \cite{Killer2019,Killer2019a,Barbui2020,Killer2021}. While the position of this feature changes for slightly different iota / island position (orange) or an enlarged island (green), it is universal in the magnetic standard configuration and tied to the local magnetic field characteristics. The fine dependency of the LCFS position and island characteristics seen in Fig. \ref{fig:Poincare} is therefore reflected in the probe measurements and warrants special attention during probe operation. An example on high heat fluxes at decreased island sizes will be given in section \ref{sec:accident}. Figure \ref{fig:profiles} furthermore shows that it can be challenging to reach the LCFS with a reciprocating probe in the standard configuration, as the probe has to cross a wide SOL that even 4-5\,cm outside the LCFS has considerable heat loads of $\approx 1$\,MW/m$^2$ due to electron temperatures of some 10\,eV and densities of at least some $10^{18}$m$^{-3}$. 
\\
In addition to the plasma conditions, also the magnetic connection between different diagnostics depends on the magnetic field configuration. Mapping the experimental results of different diagnostics at different locations on the machine is particularly challenging in the 3D SOL of W7-X. In the standard configuration,  field lines starting from the MPM path cross the field of view of the 2D thermal He beam diagnostic that is located in the divertor region $\approx 7$\,m toroidally from the MPM \cite{Barbui2020}. In the standard, low iota, and other $\iota_{edge} \leq 1$ configurations, the MPM path passes very close to flux tubes from divertor target probes \cite{Killer2020}. Some first cross-validations of fundamental plasma parameters such as $T_e$, $n_e$ between these diagnostics have been started \cite{Barbui2020} but no systematic investigations have been made yet.

\subsection{Probe movement schemes}
\begin{figure}[tb]
  \centering
  \includegraphics[width=0.7\textwidth]{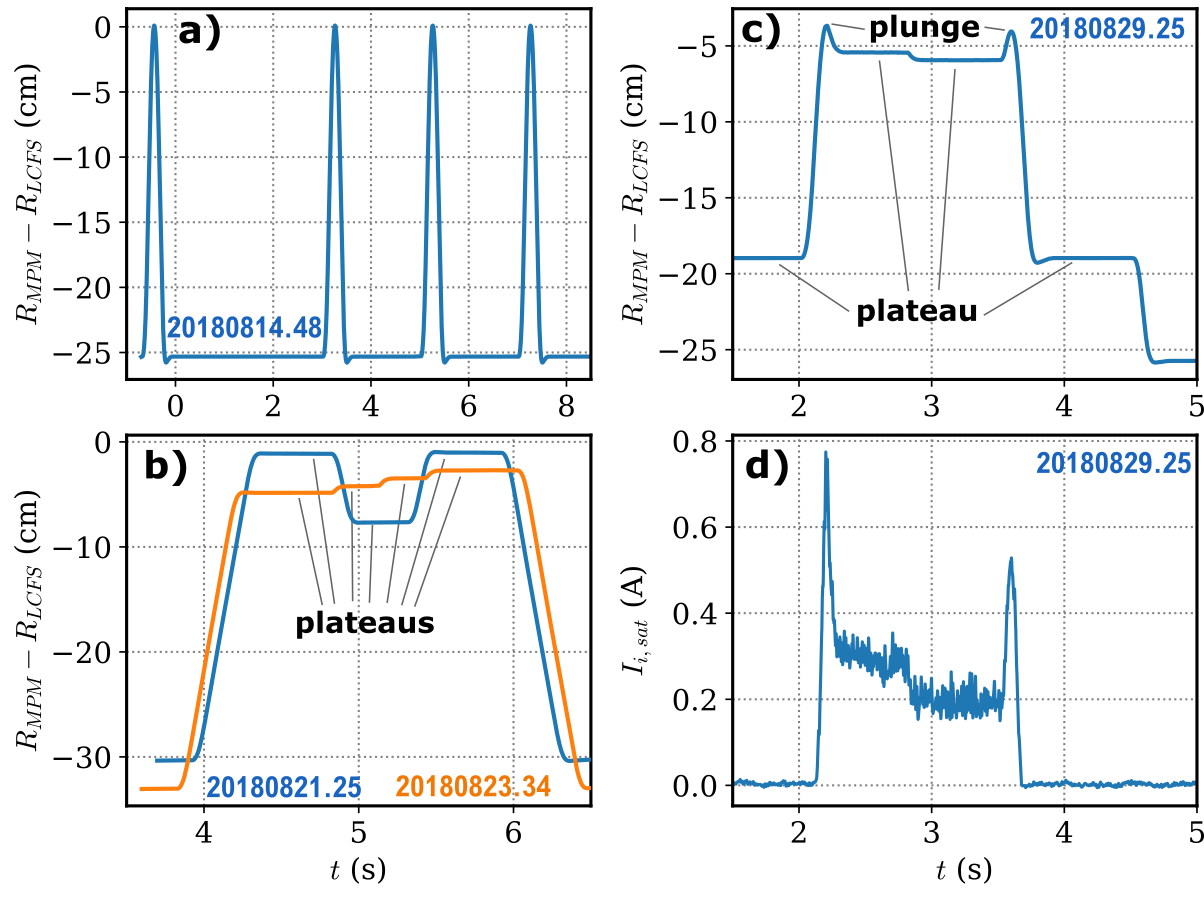}
  \caption[]{Example time traces of MPM probe position with respect to the LCFS. a) four fast plunges up to the LCFS with FZJ-COMB2, b) different constant position plateaus in two programs with NIFS-FILD1 scanning for fast ion losses, c) different constant position plateaus and two fast plunges with IPP-FLUC1, d) $I_{i,sat}$ trace corresponding to c).}
  \label{fig:plunges}
\end{figure}

The fine structure of the SOL shown in section \ref{sec:magnetic} warrants a versatile and accurate positioning system. The servo motor used at the MPM allows to arbitrarily position the probes with currently up to 18 motor actions being possible per plasma program, e.g. a fast plunge to a specified position or staying at a constant position ("plateau") for an extended duration.
\\
A few example probe position traces are presented in Fig. \ref{fig:plunges}: In a), four fast plunges up to the LCFS have been performed, where the first plunge was prior to the plasma start-up in order to calibrate the magnetic pick-up probe in the probe head FZJ-COMB2. In b), the probe head NIFS-FILD1 was positioned at different positions in the SOL for at least 100\,ms at each plateau in order to obtain better statistics of the fast ion loss detector signal. In c), a combined scheme of plateaus and plunges was performed with the IPP-FLUC1 probe head. While the two plunge peaks went up to the $E_r$ shear layer region in the low iota configuration \cite{Killer2020}, the two plateaus in between the plunges served to obtain good statistics of SOL turbulence measurements and were located around the mapped position of a target Langmuir probe \cite{Killer2020}. The plateau position 19\,cm from the LCFS was used to monitor the sporadic observation of fast electron losses (see section \ref{sec:electrons}). The $I_{i,sat}$ trace in Fig. \ref{fig:plunges} d) belongs to the position trace in c) and highlights the dependence of the SOL plasma conditions ($n \sim I_{i,sat}$) on the probe position, i.e. the strong gradients of the SOL.

\subsection{Heat loads}
\begin{figure}[tb]
  \centering
  \includegraphics[width=0.7\textwidth]{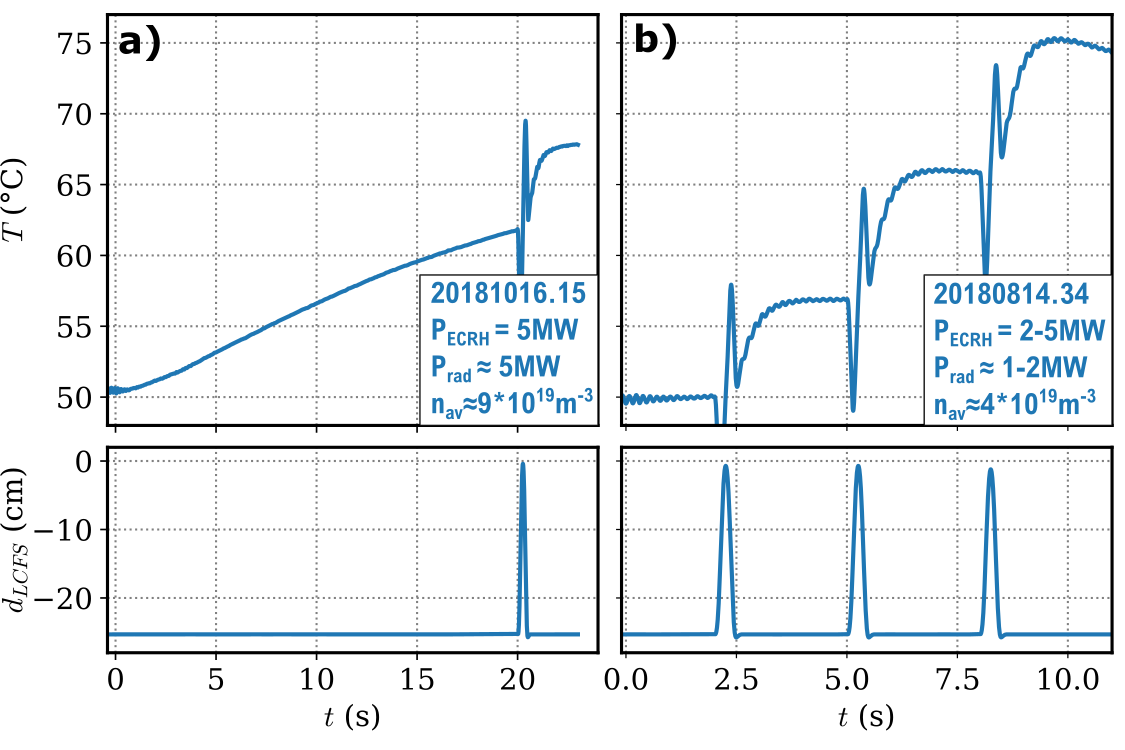}
  \caption[]{Temporal evolution of probe head temperature measured by thermocouples in the boron nitride body of FZJ-COMB2. a) detached plasma, duration 26\,s , b) attached plasma, duration 9\,s.  }
  \label{fig:probe_temperature}
\end{figure}

A reciprocating probe is subject to two major heat load contributions: convective loads due to the probe acting as a limiter for the field lines it is intersecting, and radiation from the plasma. The convective loads can easily reach a few MW/m$^2$, are highly directed along the magnetic field and only occur when the probe is inserted into the plasma, with a strong dependence on the local plasma conditions. The radiation from the plasma, in contrast, is typically two orders of magnitude smaller, but is applied to the probe in a rather isotropic geometry for the entire plasma duration (typically some 10\, and up to 100\,s so far in W-7X \cite{Pedersen2019}). 
\\
The effect of both contributions is presented in Fig. \ref{fig:probe_temperature} for two different scenarios, a) a detached plasma with high radiated power fraction\cite{Pedersen2019} and b) an attached plasma with low radiated power fractions. In a), stable detachment was achieved from $t=3$\,s on. During the first $\approx 20$\,s of the program, the probe (staying at the parking position at the plasma vessel wall) was subject to the high radiation levels associated with detachment, leading to a $\approx 12$\,K increase of the probe temperature. After 20\,s, a fast plunge to the LCFS was performed, adding another 5\,K. The data acquisition ended slightly ahead of the plasma termination at 26\,s. The sharp positive and negative peak feature in the temperature trace directly during the fast plunge is due to magnetic induction in the thermocouple acting as a pick-up probe while moving in the magnetic field gradient. In Fig. \ref{fig:probe_temperature} b), three fast plunges into the SOL were performed in attached plasma conditions, i.e. at low radiation levels. Here, each plunge adds 7-9\,K to the probe temperature although the probe did not move as close to the LCFS as in a), which is presumably due to the reduction of both SOL temperatures and densities during detachment \cite{Feng2021}. In between the plunges, the probe head temperature quickly equilibrates while no increase due to radiation is observed in this scenario. After the discharge ends at 9\,s, the probe immediately starts to slowly cool.
\\
While the temperatures shown here are not critical even for sensitive cabling inside the probe head, the extrapolation of the scenario in Fig. \ref{fig:probe_temperature} a) to discharge lengths of several minutes might pose an operational limit on the probe operation even when the probe remains at the parking position. It is further important to note that the temperatures presented here were taken in the body of the probe head, which depends on the thermal design of the probe head and the thermal conductivity to the water-cooled probe interface. The probe tips that are directly exposed to the highest heat loads during probe insertion can obtain much higher temperatures (e.g. melting of tungsten), which can degrade the quality of the probe measurement (e.g. by thermionic emission) and disturb the plasma due to impurity release.

\subsection{Operation safety} \label{sec:accident}

\begin{figure}[tb]
  \centering
  \includegraphics[width=0.7\textwidth]{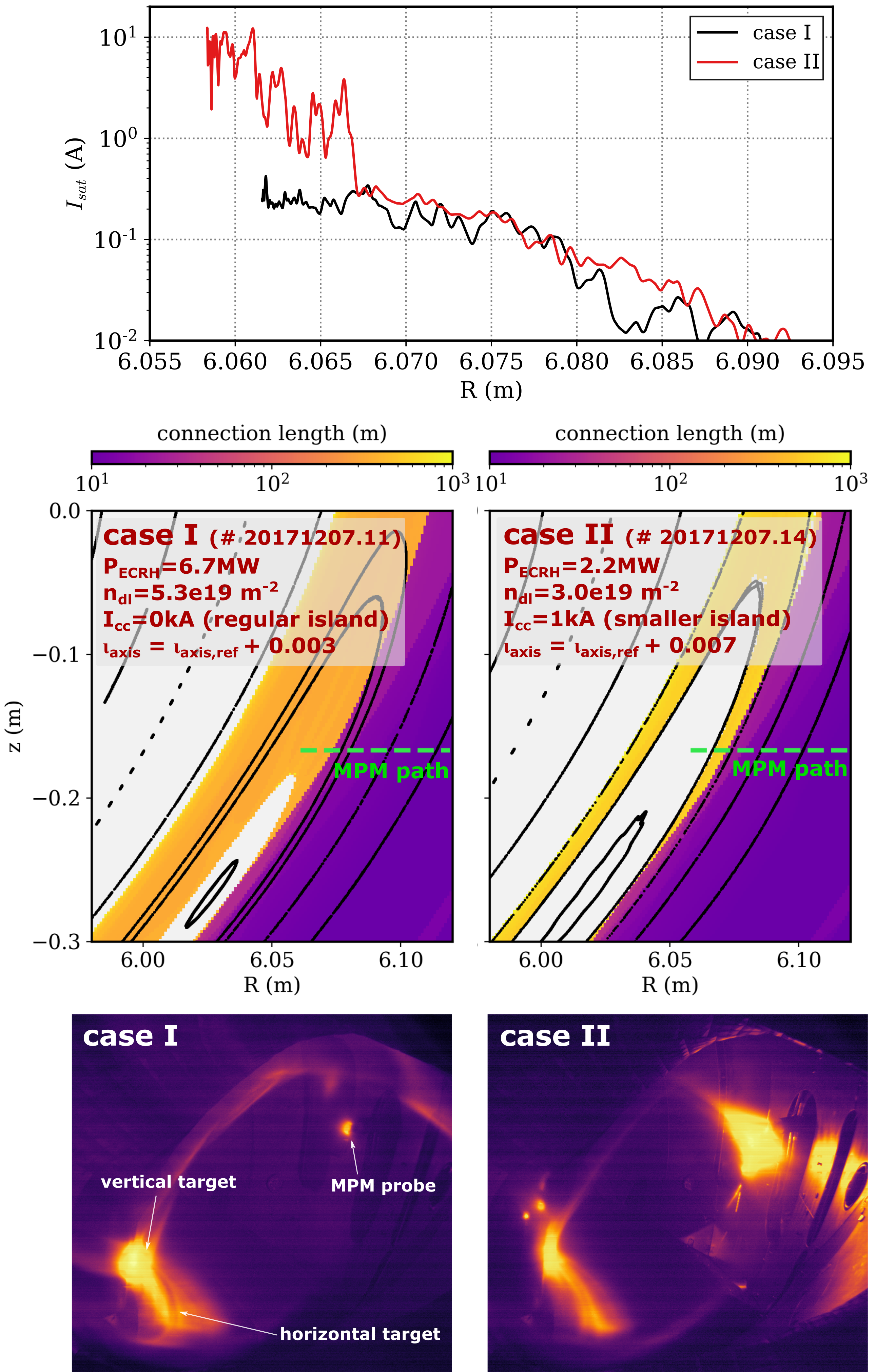}
  \caption[]{Top: Ion saturation current (note log scale) of a Mach probe pin along MPM path for two measurements in slightly different versions of the magnetic standard configuration. Middle: Connection length and Poincare plot in the vicinity of the MPM. Bottom: Video image of MPM at deepest insertion.}
  \label{fig:171207}
\end{figure}

Trying to maximize the physics gain of reciprocating probe measurements by inserting the probes as deep and long as possible increases the likelihood of probe overheating, possibly resulting in damage to the probe and contamination of the plasma. The task for the probe operator is to estimate the expected plasma conditions (e.g. $T_e$, $n_e$) for unknown operation scenarios in advance and adjust the probe motion accordingly. This is particularly challenging in early W7-X operation as the SOL is initially uncharted, and extensive 3D modeling is not yet validated by experiments and only available for a few of the many different plasma scenarios and magnetic configurations. As indicated by the blue and orange data in Fig. \ref{fig:profiles}, the SOL profiles can vary significantly even within one magnetic configuration due to small changes of the magnetic island position that are induced by the evolution of a toroidal plasma current: while measurement of the blue data in Fig.  \ref{fig:profiles} was safe, already a globally detectable contamination of the plasma occurred due to high heat loads onto the probe in the orange measurement, and in further experiments the impurity contamination became even larger up to plasma termination for higher plasma currents (i.e. stronger island displacement) as was presented in \cite{Killer2019a}.
\begin{figure}[tb]
  \centering
  \includegraphics[width=0.8\textwidth]{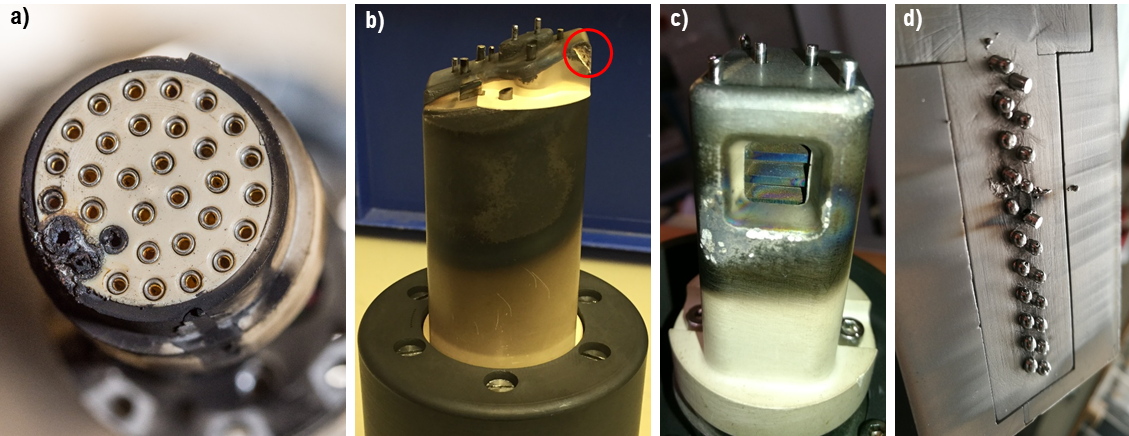}
  \caption[]{Damages after probe operation. a) arcs in probe interface due to insufficient high voltage insulation, losing in total 11 out of 32 channels for the 2018 campaign. b) FZJ-COMB2 after losing boron nitride flake (see circle top right), see Fig. \ref{fig:171207}. c) Damaged FZJ-RFA2 probe after incident at $I_p$=19\,kA described in \cite{Killer2019a}. d) Damaged IPP-FLUC1 probe \cite{Killer2019} with most of the pins molten after probe staying in the SOL too long.}
  \label{fig:damages}
\end{figure}
\\
Another example on the sensitivity of the SOL magnetic field structure (and therefore local plasma conditions) is presented in Fig. \ref{fig:171207}. Two reciprocating probe measurements with the same probe head (FZJ-COMB2) were performed in the magnetic standard configuration, where the rotational transform was slightly different due to different iota correction currents in the planar coils of W-7X \cite{Lazerson2019}. In addition, in case II the magnetic island size was reduced using the control coils. Due to the slightly higher iota and reduced island size in case II, the region of closed field lines in the island center is much larger and the connection lengths in the island SOL are generally longer. As a consequence, an order of magnitude larger ion saturation current is observed in case II as soon as the probe enters the long / closed connection length region, despite the larger heating power and larger density in case I, which would usually be expected to result in higher ion saturation currents in the SOL \cite{Killer2019}. The high heat loads associated with the strong $I_{sat}$ signal in case II further led to a sub-cm sized piece of boron nitride being sheared off from a recently machined leading edge of the probe. The debris piece was confined in the plasma by vertically oscillating along the outboard SOL at a constant toroidal position, presumably tracing out the magnetic island position for the remaining 250ms of the program, see video file in supplement. The impurity influx into the plasma led to significant visible light emission in the overview video camera (bottom row of Fig. \ref{fig:171207}), but caused only a small increase in total radiated power, so that the plasma continued to run until pre-scheduled ECRH shutdown 250ms after the probe accident. The main debris piece that was found on the divertor target after the operation phase had a mass of 25mg.
\\
In the total operation of W7-X so far, reciprocating probe operation resulted in about 15 events with significant impurity release, where impurities from the probe where detected in spectroscopic diagnostics and at least one global plasma parameter (e.g. line integrated density, diamagnetic energy, total plasma radiation) was affected. In seven of those cases, the impurity release led to plasma termination. The accidents were caused by too deep probe insertion or too long insertion durations. In most of these cases, this was due to unexpected changes in the fine structure of the magnetic field or of operation conditions, which led to unexpectedly high heat fluxes to the probe. In addition, a few accidents were due to technical or human errors on the manipulator control side. A few examples of damages on the probes or manipulator are presented in Fig. \ref{fig:damages}. There are no indications that probe damages are related to the design or materials of specific probes. In order to avoid such accidents in future operation, it is both necessary to study the expected plasma conditions in advance as well as to empirically test acceptable heat fluxes by gradually increase the probe insertion depth/duration for each operation scenario.

\section{MPM measurement examples}\label{sec:results}
In the following, a few measurements examples are briefly presented with (except for \ref{sec:pellets}) a focus on unexpected observations that are not yet covered in the regular research articles cited in Table \ref{tab:probes}.

\subsection{Exploitation of probe movement schemes}\label{sec:pellets}
\begin{figure}[tb]
  \centering
  \includegraphics[width=0.6\textwidth]{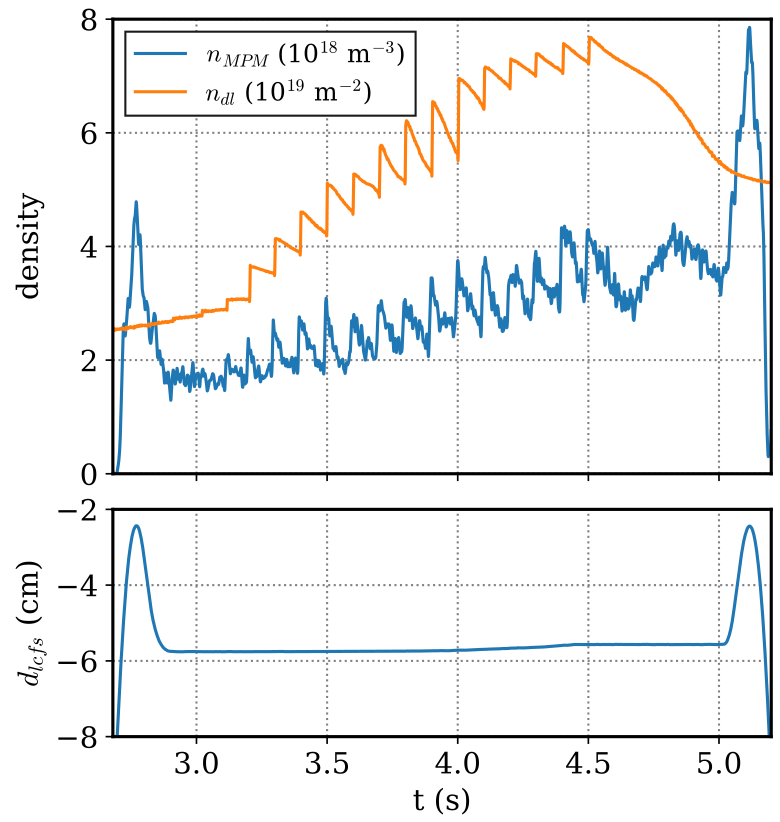}
  \caption[]{Time evolution of SOL plasma density from MPM and line-integrated plasma density $n_{dl}$. During the insertion of 17 hydrogen ice pellets \cite{Baldzuhn2019} from 3.0\,s to 4.5\,s the probe stayed at a constant position. Before and after the pellet phase, a fast plunge close to the LCFS is performed.}
  \label{fig:pellets}
\end{figure}
A physics application for the versatile probe movement capabilities of the MPM (see Fig. \ref{fig:plunges}) is presented in Fig. \ref{fig:pellets}. In this experiment, the plasma density was increased using hydrogen ice pellets \cite{Baldzuhn2019} from 3.0\,s to 4.5\,s, where each pellet causes a steep increase of the line-integrated density. During the pellet insertion phase, the MPM with probe head IPP-FLUC1 stayed at a constant position in the SOL to monitor the evolution of the SOL plasma conditions. During the pellet phase, the SOL density obtained from the probe clearly correlates to the overall plasma density. At around 4.7\,s, both density measurements deviate from each other which is probably related to changes in the core confinement quality due to the post-pellet reduced core turbulence regime \cite{Bozhenkov2020}. At 2.8\,s and 5.1\,s, the MPM performed a fast plunge closer to the LCFS, which provides density (and other) profiles before and after the pellet phase.

\subsection{High frequency fluctuations}\label{sec:spectra}

\begin{figure}[tb]
  \centering
  \includegraphics[width=0.6\textwidth]{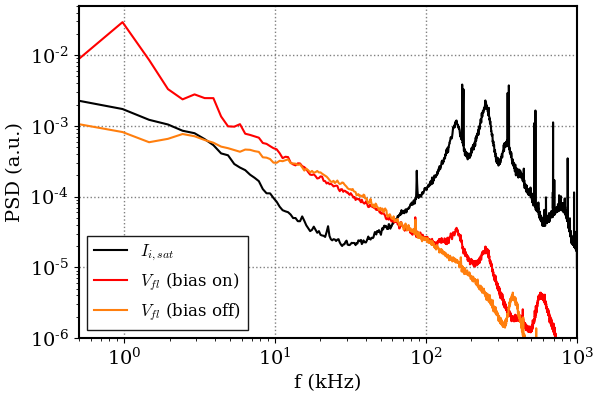}
  \caption[]{Power spectral density of probe signal fluctuations taken at a sampling frequency of 2\,MHz over a time series length of 500\,ms. The $V_{fl}$ signal is affected by switching bias voltages to $I_{i,sat}$ pins and the swept probe pin on and off. The sharp spikes at higher frequencies are an artifact of the measurement electronics.}
  \label{fig:spectra}
\end{figure}

The frequency spectrum of probe data fluctuations can provide fundamental insight into turbulence and mode activity. In a purely turbulent spectrum, a power law decay (in a double log plot such as Fig. \ref{fig:spectra}) towards higher frequencies is observed, whereas modes are typically identified as localized peaks in the spectrum. In the spectra in Fig. \ref{fig:spectra} multiple features are visible: In the ion saturation current spectrum, the PSD decreases only until $\approx$30\,kHz, and then has a wide region of higher fluctuation activity peaking at $\approx$250\,kHz. This behavior is very often seen in $I_{i,sat}$ measurement with the MPM, although the strength of this feature is mostly somewhat smaller and the number and position of peaks varies, with typically 2-3 peaks between 100\,kHz and 300\,kHz. Analyzing the spectra of multiple pins on a probe head, a strong coherence and only a very small phase delay is observed between the individual signals in this frequency range, possibly suggesting a mode with low mode number, i.e. the typical wavelength is larger than the extent of the probe array (which is not larger than 6\,cm for the probes heads existing so far).
\\
This unexpected fluctuation activity has been noted previously \cite{Liu2018a,Cai2019,Killer2021} but is not yet understood. While it had been conjectured that it is related to Alfvenic modes \cite{Killer2021,Rahbarnia2020}, it might also be due to a perturbation from the probe measurement itself, which is indicated by the two floating potential spectra in Fig. \ref{fig:spectra}. If the bias voltage of the neighboring current measurement probes is on, the $V_{fl}$ spectrum also reveals a few peaks $>100$\,kHz, although of a much smaller magnitude than in the $I_{i,sat}$ spectrum. If no bias voltages are applied to the current collecting pins, the peaks in the spectrum for $>100$\,kHz mostly vanish from the $V_{fl}$ pin data. Hence, the existence of these features is correlated to biased probes (here: negatively biased, collecting ions), but a solid explanation cannot be made here. This behavior is observed for different probe heads, although the specific shape of the $\sim 100$\,kHz peak structure slightly varies and might be influences by probe geometry and magnitude of bias voltage. However, due to limited operation time so far no statement about the role of probe geometry and bias voltage magnitude can be made yet. Unfortunately, the $I_{i,sat}$ spectrum cannot be obtained for the case without bias applied, as the electronics are set up such that the $I_{i,sat}$ measurement is not possible in this case. Indications for time-dependent modulation of the high frequency fluctuation amplitude due to electron collection of nearby probes are presented in section \ref{sec:perturbation}.
\\
Finally, the sharp spikes most prominently visible in the $I_{i,sat}$ spectrum in Fig. \ref{fig:spectra} for $>80$\,kHz are due to the probe electronics (switching frequency of DC-DC converters). These spikes are individual for each single signal and are not correlated to the broad peak of the PSD in the 100\,kHz range. In fact, here a rather strong example of these artifacts is shown in order to clearly show how they are distinguished from the mode-like broad peak. Further analysis reveals that the high coherence reported for the broad mode-like peak vanishes at the exact frequencies of the sharp peaks. For the upcoming operation phase(s) of W7-X, the electronics have been modified such that these features will disappear.

\subsection{Electron emission from hot probes}\label{sec:emission}

\begin{figure}[tb]
  \centering
  \includegraphics[width=1\textwidth]{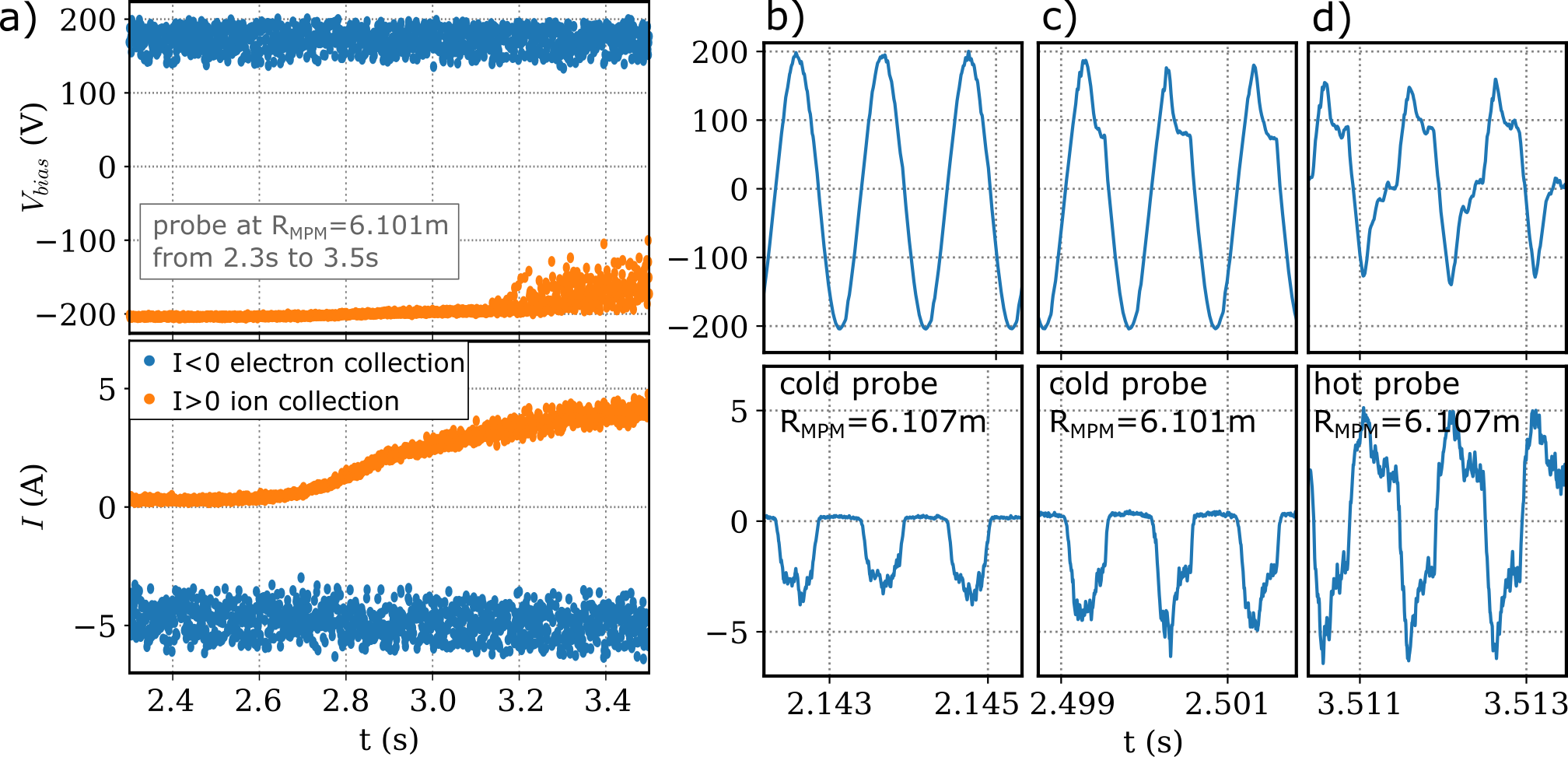}
  \caption[]{Voltage and current of a classic swept Langmuir that stayed at a constant position in the SOL ($R_{MPM}$=6.101m in the low iota configuration, $T_e \approx 25$\,eV, $n \approx 5\cdot 10^{18}$\,m$^{-3}$). In a), only the maxima and minima of each sweep of the $I$-$V$ curve are shown. In b)-d), the $V$ and $I$ traces for three sweeps at different stages of the experiment are shown. The current is defined such that ion saturation currents are positive.}
  \label{fig:emission}
\end{figure}

When the material temperature of a Langmuir probe in a plasma becomes sufficiently large, the probe can start to emit electrons, which is exploited in the concept of the emissive probe \cite{Sheehan2011}. In  W7-X experiments with the MPM, probe tips occasionally became self-emitting when they were exposed to the plasma for an extended period of time. Such an example is presented in Fig. \ref{fig:emission}, where the voltage and current of a classic swept Langmuir probe pin (cylindrical Molybdenum tip, $d$=$h$=2\,mm) are shown. 
\\
First, in Fig. \ref{fig:emission} b), at moderate plasma conditions in the outer SOL ($T_e \approx 15$\,eV, $n \approx 2\cdot 10^{18}$\,m$^{-3}$), the sinusoidal bias voltage ($f_{sweep}$=970\,Hz) applied to the probe tip results in a typical probe characteristic, where the amplitude of the electron saturation current is much larger than the ion saturation current (here: factor 20). Then, the probe was moved closer to the plasma and remained at constant position with $T_e \approx 25$\,eV, $n \approx 5\cdot 10^{18}$\,m$^{-3}$ for 1200m\,s. Initially, in Fig. \ref{fig:emission} c), the probe characteristic remains qualitatively similar except that the bias voltage cannot be kept up by the power supply during electron collection at positive bias voltages, as the supply device is nominally limited to 2\,A. 
\\
In the further course of the measurement, the probe becomes increasingly hot as indicated by continuously increasing light emission seen in a video camera. The probe characteristic towards the end of the measurement in Fig. \ref{fig:emission} d) now appears qualitatively different: The bias voltage does not reach the preset maxima at $\pm$200\,V and the current reveals similar amplitudes of up to 5\,A in electron and ion collection direction. At the same time, the plasma conditions remained unaffected as monitored by a nearby triple probe that was operating in parallel on the same probe head. The obvious interpretation is that during phases of negative bias voltage, the observed current is mostly due to electrons being emitted from the hot probe. 
\\
The probe tip did not appear damaged / molten after this experiment. This opens up the possibility to use self-emissive probes as a diagnostic (e.g. for plasma potential determination), which is however a technically challenging implementation \cite{Sheehan2011}. We finally emphasize that this kind of behavior was only seen in swept Langmuir probes that operated far into the electron collection branch, suggesting that the strong heating of the probe is due to the large electron collection currents.

\subsection{Perturbation by probes}\label{sec:perturbation}

\begin{figure}[tb]
  \centering
  \includegraphics[width=0.6\textwidth]{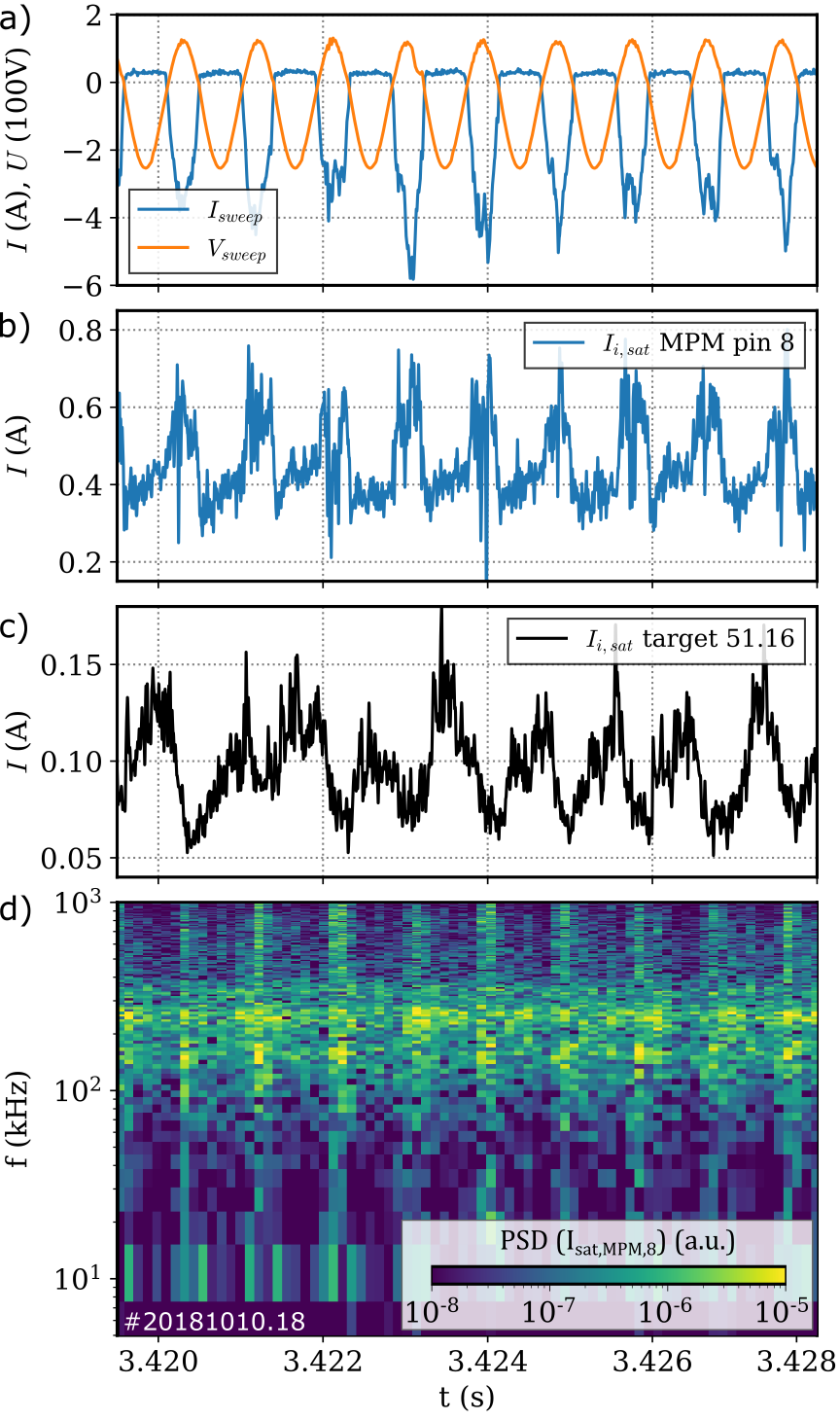}
  \caption[]{The perturbation from a swept Langmuir probe on probe head IPP-FLUC1 (a) is seen in ion saturation current signals of (b) a nearby tip on the same probe head and (c) at a magnetically connected target probe 10\,m along the field. (a-c) are low-pass filtered at 50\,kHz for visual clarity. Ion collection currents are defined to be positive. (d): power spectral density of the (unfiltered) trace in b). }
  \label{fig:sweep_perturbation}
\end{figure}

As an active diagnostic, electric probes have the disadvantage that their presence affects the plasma that we want to study with the probe. However, fusion plasmas are usually large and dense enough that typical (ion saturation) currents of up to $\sim$100\,mA drawn by the probe do not significantly affect the plasma. For DC and AC biasing experiments using probes as an active perturbation source \cite{Winslow1998,Thomsen2005}, typically larger currents are drawn.
\\
In W7-X, particularly strong perturbations by probes on the MPM have been observed in the case of swept Langmuir probes being in the electron collection mode, i.e. at positive bias voltages. Such an example is presented in Fig. \ref{fig:sweep_perturbation}, where voltage and current of 10 sweep cycles are shown in a). The current is defined such that ion currents are positive. In b) and c), ion saturation currents at another probe tip on the reciprocating probe head as well as from a target probe that is magnetically connected with a parallel distance of 10\,m are shown. Both signals are clearly modulated by the swept probe. Further data analysis that is beyond the scope of this paper shows that the perturbation from the swept probe is picked up by the other probes only when the swept probe is in electron collection mode. This might be explained by the much longer collection length of electron currents compared to ion currents.
\\
Furthermore, the fluctuation spectrum in \ref{fig:sweep_perturbation} d) of the signal shown in b) reveals that during each electron collection phase of the swept probe, the high frequency ($>$100\,kHz) fluctuations of the $I_{i,sat}$ pin (see section \ref{sec:spectra}) become stronger. 
\\
One possible explanation might be a beam-driven instability being triggered by the electron current collection where the electrons attracted by the biased probe pin act as beam. However, this is just a conjecture at this point as such instabilities are difficult to detect even in low temperature plasmas \cite{Rapson2014}. In a fusion plasma, the characteristic frequencies are in the GHz range and therefore require microwave diagnostics for further investigations.
\\
We finally emphasize that these rather strong perturbations (in contrast to the report on fluctuation spectra in section \ref{sec:spectra}) are purely observed when a probe is in electron collection mode. It is not directly depend on the bias voltage levels but only on the electron collection currents. Also, this strong perturbation is not observed during ion collection and not during electron emission (see section \ref{sec:emission}).

\subsection{Indications for supra-thermal electrons}\label{sec:electrons}

\begin{figure}[tb]
  \centering
  \includegraphics[width=0.65\textwidth]{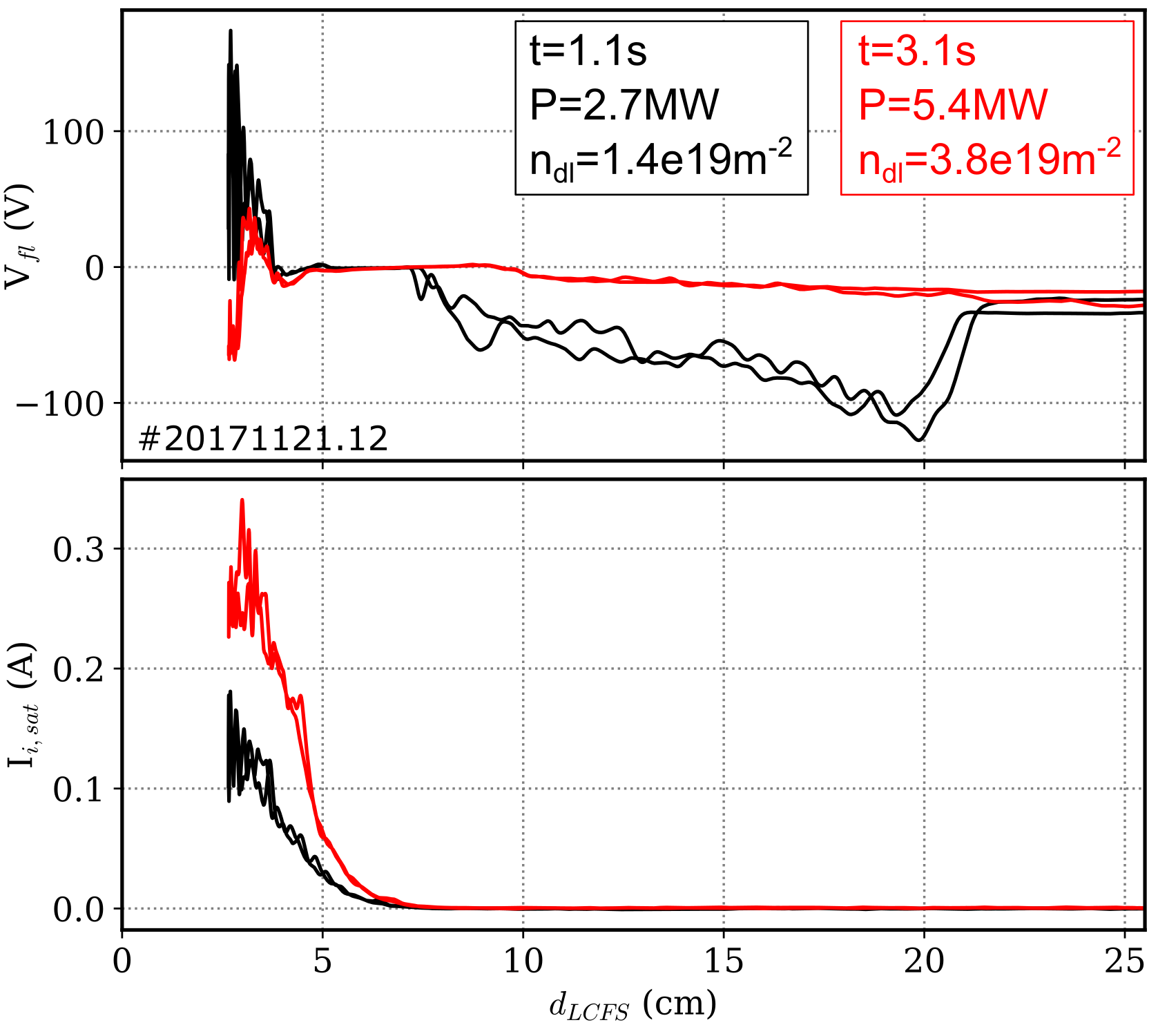}
  \caption[]{Radial profiles of a) $V_{fl}$ and b) $I_{i,sat}$ as a function of distance from the LCFS along MPM path taken in one plasma program in the magnetic standard configuration.}
  \label{fig:electrons}
\end{figure}

Fast (supra-thermal) electrons are a major concern for tokamaks, where they might be generated as runaway electrons after a disruption. Also in stellarators, supra-thermal electrons have been observed, where they have been seen to be confined on rational flux surfaces \cite{Medina2001}. The generation mechanism is mostly due to microwave heating, e.g. by directing the heating beam to trapped particles \cite{Hirsch2008} or by non-resonant heating \cite{Laqua2014}. 
\\
In W7-X, the MPM probe measurements often observe negative floating potential signals far outside the plasma in ECR-heated discharges where the general plasma density is rather low, typically $n_{dl}<2\cdot10^{19}$m$^{-2}$. An example is presented in Fig. \ref{fig:electrons}, where two measurements in the same plasma program at different operation conditions were taken. In the measurement at low density, the floating potential is negative from $d_{LCFS}$=8\,cm to 20\,cm, although there is no detectable plasma density ($I_{i,sat}\approx 0$) this far from the LCFS. Since a $V_{fl}$ probe is expected to measure 0\,V in vacuum without plasma, an additional electron population hitting the probe might be a possible explanation for the negative $V_{fl}$.
\\
Since the probe moves inwards and outwards, each probe plunge provides two radial profiles. The good agreement of both $V_{fl}$ profiles confirms the reproducible character of the wide $V_{fl}$ valley, also in other similar situations in ECR heated low density plasmas. This observation is surprising and cannot be completely explained yet, as the ECRH system at W7-X is not expected to create supra-thermal particles: the heating is entirely resonant and directed such that it does not heat particles trapped in the magnetic well. Although the origin of this phenomenon cannot be explained with certainty, the existence of an electron population well outside the actual plasma (possible trapped in the magnetic ripple) might cause the negative floating potentials and agrees in its dependence on plasma operation conditions (favored by low density and high ECRH power).

\section{Summary}
The Multi-Purpose Manipulator at W7-X was routinely operated in the 2017-2018 test divertor campaigns. The versatility and diversity of probe heads allowed to assess a wide range of physics questions. 
\\
From this experience, we have presented lessons learned on how to safely operate a reciprocating probe while at the same time optimizing the physics gain. It was shown that it is essential to adapt the probe operation to the fine details of the respective magnetic configuration as the SOL plasma conditions strongly depend on the particular details of the SOL magnetic field structure. This statement was supported by examples of SOL profiles in the configuration space of the W7-X magnetic field and case reports of probe accidents. In addition to convective heat loads by plunges into the SOL, radiative heat loads during longer plasma experiments can become significant, especially considering that W7-X aims towards discharge lengths of 30\,min. 
\\
Further, guidelines for optimized probe head shapes have been discussed. The value of reciprocating probe use can be further enhanced by conducting positioning schemes that are specifically tailored to the respective physics motivation of the measurement. 
\\
Finally, unexpected but robust observations by MPM probes have been presented: Fluctuation spectra typically show two separate features at higher frequencies ($>$100\,kHz), where the broad peak structure seems to be a true (although still unexplained) physics effect while sharp spikes are electronic artifacts. The former phenomenon is affected by the presence of bias voltages of nearby probe pins, where especially electron collection currents can temporally modulate the mode-like feature. The plasma perturbation by probes in electron collection is clearly seen in both neighboring probes on the MPM probe head but also in magnetically connected target Langmuir probes. Swept probes moving far into the electron collection branch of the Langmuir probe characteristic can furthermore heat up such that they become emissive. 
\\
In ECR heated plasmas with low density, negative floating potentials measured far outside the SOL (in supposed vacuum) indicate the presence of a (supra-thermal) electron population.

\section{Acknowledgments}

This work has been carried out within the framework of the EUROfusion Consortium and has received funding from the Euratom research and training programme 2014-2018 and 2019-2020 under grant agreement No 633053. The views and opinions expressed herein do not necessarily reflect those of the European Commission.

\begin{thebibliography}{63}%
\makeatletter
\providecommand \@ifxundefined [1]{%
 \@ifx{#1\undefined}
}%
\providecommand \@ifnum [1]{%
 \ifnum #1\expandafter \@firstoftwo
 \else \expandafter \@secondoftwo
 \fi
}%
\providecommand \@ifx [1]{%
 \ifx #1\expandafter \@firstoftwo
 \else \expandafter \@secondoftwo
 \fi
}%
\providecommand \natexlab [1]{#1}%
\providecommand \enquote  [1]{``#1''}%
\providecommand \bibnamefont  [1]{#1}%
\providecommand \bibfnamefont [1]{#1}%
\providecommand \citenamefont [1]{#1}%
\providecommand \href@noop [0]{\@secondoftwo}%
\providecommand \href [0]{\begingroup \@sanitize@url \@href}%
\providecommand \@href[1]{\@@startlink{#1}\@@href}%
\providecommand \@@href[1]{\endgroup#1\@@endlink}%
\providecommand \@sanitize@url [0]{\catcode `\\12\catcode `\$12\catcode
  `\&12\catcode `\#12\catcode `\^12\catcode `\_12\catcode `\%12\relax}%
\providecommand \@@startlink[1]{}%
\providecommand \@@endlink[0]{}%
\providecommand \url  [0]{\begingroup\@sanitize@url \@url }%
\providecommand \@url [1]{\endgroup\@href {#1}{\urlprefix }}%
\providecommand \urlprefix  [0]{URL }%
\providecommand \Eprint [0]{\href }%
\providecommand \doibase [0]{http://dx.doi.org/}%
\providecommand \selectlanguage [0]{\@gobble}%
\providecommand \bibinfo  [0]{\@secondoftwo}%
\providecommand \bibfield  [0]{\@secondoftwo}%
\providecommand \translation [1]{[#1]}%
\providecommand \BibitemOpen [0]{}%
\providecommand \bibitemStop [0]{}%
\providecommand \bibitemNoStop [0]{.\EOS\space}%
\providecommand \EOS [0]{\spacefactor3000\relax}%
\providecommand \BibitemShut  [1]{\csname bibitem#1\endcsname}%
\let\auto@bib@innerbib\@empty
\bibitem [{\citenamefont {Davies}\ \emph {et~al.}(1996)\citenamefont {Davies},
  \citenamefont {Erents}, \citenamefont {Loarte}, \citenamefont {Guo},
  \citenamefont {Matthews}, \citenamefont {Mccormick},\ and\ \citenamefont
  {Monk}}]{Davies1996}%
  \BibitemOpen
  \bibfield  {author} {\bibinfo {author} {\bibfnamefont {S.~J.}\ \bibnamefont
  {Davies}}, \bibinfo {author} {\bibfnamefont {S.~K.}\ \bibnamefont {Erents}},
  \bibinfo {author} {\bibfnamefont {A.}~\bibnamefont {Loarte}}, \bibinfo
  {author} {\bibfnamefont {H.~Y.}\ \bibnamefont {Guo}}, \bibinfo {author}
  {\bibfnamefont {G.~F.}\ \bibnamefont {Matthews}}, \bibinfo {author}
  {\bibfnamefont {K.}~\bibnamefont {Mccormick}}, \ and\ \bibinfo {author}
  {\bibfnamefont {R.~D.}\ \bibnamefont {Monk}},\ }\href {\doibase
  https://doi.org/10.1002/ctpp.19960360118} {\bibfield  {journal} {\bibinfo
  {journal} {Contributions to Plasma Physics}\ }\textbf {\bibinfo {volume}
  {36}},\ \bibinfo {pages} {117} (\bibinfo {year} {1996})}\BibitemShut
  {NoStop}%
\bibitem [{\citenamefont {Watkins}\ \emph {et~al.}(1997)\citenamefont
  {Watkins}, \citenamefont {Hunter}, \citenamefont {Tafoya}, \citenamefont
  {Ulrickson}, \citenamefont {Watson}, \citenamefont {Moyer}, \citenamefont
  {Cuthbertson}, \citenamefont {Gunner}, \citenamefont {Lehmer}, \citenamefont
  {Luong}, \citenamefont {Hill}, \citenamefont {Mascaro}, \citenamefont
  {Robinson}, \citenamefont {Snider},\ and\ \citenamefont
  {Stambaugh}}]{Watkins1997}%
  \BibitemOpen
  \bibfield  {author} {\bibinfo {author} {\bibfnamefont {J.~G.}\ \bibnamefont
  {Watkins}}, \bibinfo {author} {\bibfnamefont {J.}~\bibnamefont {Hunter}},
  \bibinfo {author} {\bibfnamefont {B.}~\bibnamefont {Tafoya}}, \bibinfo
  {author} {\bibfnamefont {M.}~\bibnamefont {Ulrickson}}, \bibinfo {author}
  {\bibfnamefont {R.~D.}\ \bibnamefont {Watson}}, \bibinfo {author}
  {\bibfnamefont {R.~A.}\ \bibnamefont {Moyer}}, \bibinfo {author}
  {\bibfnamefont {J.~W.}\ \bibnamefont {Cuthbertson}}, \bibinfo {author}
  {\bibfnamefont {G.}~\bibnamefont {Gunner}}, \bibinfo {author} {\bibfnamefont
  {R.}~\bibnamefont {Lehmer}}, \bibinfo {author} {\bibfnamefont
  {P.}~\bibnamefont {Luong}}, \bibinfo {author} {\bibfnamefont {D.~N.}\
  \bibnamefont {Hill}}, \bibinfo {author} {\bibfnamefont {M.}~\bibnamefont
  {Mascaro}}, \bibinfo {author} {\bibfnamefont {J.~I.}\ \bibnamefont
  {Robinson}}, \bibinfo {author} {\bibfnamefont {R.}~\bibnamefont {Snider}}, \
  and\ \bibinfo {author} {\bibfnamefont {R.}~\bibnamefont {Stambaugh}},\ }\href
  {\doibase 10.1063/1.1147833} {\bibfield  {journal} {\bibinfo  {journal}
  {Review of Scientific Instruments}\ }\textbf {\bibinfo {volume} {68}},\
  \bibinfo {pages} {373} (\bibinfo {year} {1997})},\ \Eprint
  {http://arxiv.org/abs/https://doi.org/10.1063/1.1147833}
  {https://doi.org/10.1063/1.1147833} \BibitemShut {NoStop}%
\bibitem [{\citenamefont {Boedo}\ \emph {et~al.}(1998)\citenamefont {Boedo},
  \citenamefont {Gray}, \citenamefont {Chousal}, \citenamefont {Conn},
  \citenamefont {Hiller},\ and\ \citenamefont {Finken}}]{Boedo1998}%
  \BibitemOpen
  \bibfield  {author} {\bibinfo {author} {\bibfnamefont {J.}~\bibnamefont
  {Boedo}}, \bibinfo {author} {\bibfnamefont {D.}~\bibnamefont {Gray}},
  \bibinfo {author} {\bibfnamefont {L.}~\bibnamefont {Chousal}}, \bibinfo
  {author} {\bibfnamefont {R.}~\bibnamefont {Conn}}, \bibinfo {author}
  {\bibfnamefont {B.}~\bibnamefont {Hiller}}, \ and\ \bibinfo {author}
  {\bibfnamefont {K.~H.}\ \bibnamefont {Finken}},\ }\href {\doibase
  10.1063/1.1148995} {\bibfield  {journal} {\bibinfo  {journal} {Review of
  Scientific Instruments}\ }\textbf {\bibinfo {volume} {69}},\ \bibinfo {pages}
  {2663} (\bibinfo {year} {1998})},\ \Eprint
  {http://arxiv.org/abs/https://doi.org/10.1063/1.1148995}
  {https://doi.org/10.1063/1.1148995} \BibitemShut {NoStop}%
\bibitem [{\citenamefont {Pedrosa}\ \emph {et~al.}(1999)\citenamefont
  {Pedrosa}, \citenamefont {López-Sánchez}, \citenamefont {Hidalgo},
  \citenamefont {Montoro}, \citenamefont {Gabriel}, \citenamefont {Encabo},
  \citenamefont {de~la Gama}, \citenamefont {Martinez}, \citenamefont
  {Sánchez}, \citenamefont {Pérez},\ and\ \citenamefont
  {Sierra}}]{Pedrosa1999}%
  \BibitemOpen
  \bibfield  {author} {\bibinfo {author} {\bibfnamefont {M.~A.}\ \bibnamefont
  {Pedrosa}}, \bibinfo {author} {\bibfnamefont {A.}~\bibnamefont
  {López-Sánchez}}, \bibinfo {author} {\bibfnamefont {C.}~\bibnamefont
  {Hidalgo}}, \bibinfo {author} {\bibfnamefont {A.}~\bibnamefont {Montoro}},
  \bibinfo {author} {\bibfnamefont {A.}~\bibnamefont {Gabriel}}, \bibinfo
  {author} {\bibfnamefont {J.}~\bibnamefont {Encabo}}, \bibinfo {author}
  {\bibfnamefont {J.}~\bibnamefont {de~la Gama}}, \bibinfo {author}
  {\bibfnamefont {L.~M.}\ \bibnamefont {Martinez}}, \bibinfo {author}
  {\bibfnamefont {E.}~\bibnamefont {Sánchez}}, \bibinfo {author}
  {\bibfnamefont {R.}~\bibnamefont {Pérez}}, \ and\ \bibinfo {author}
  {\bibfnamefont {C.}~\bibnamefont {Sierra}},\ }\href {\doibase
  10.1063/1.1149350} {\bibfield  {journal} {\bibinfo  {journal} {Review of
  Scientific Instruments}\ }\textbf {\bibinfo {volume} {70}},\ \bibinfo {pages}
  {415} (\bibinfo {year} {1999})},\ \Eprint
  {http://arxiv.org/abs/https://doi.org/10.1063/1.1149350}
  {https://doi.org/10.1063/1.1149350} \BibitemShut {NoStop}%
\bibitem [{\citenamefont {Ezumi}\ \emph {et~al.}(2003)\citenamefont {Ezumi},
  \citenamefont {Masuzaki}, \citenamefont {Ohno}, \citenamefont {Uesugi},
  \citenamefont {Takamura},\ and\ \citenamefont {{LHD Experimental
  Group}}}]{Ezumi2003}%
  \BibitemOpen
  \bibfield  {author} {\bibinfo {author} {\bibfnamefont {N.}~\bibnamefont
  {Ezumi}}, \bibinfo {author} {\bibfnamefont {S.}~\bibnamefont {Masuzaki}},
  \bibinfo {author} {\bibfnamefont {N.}~\bibnamefont {Ohno}}, \bibinfo {author}
  {\bibfnamefont {Y.}~\bibnamefont {Uesugi}}, \bibinfo {author} {\bibfnamefont
  {S.}~\bibnamefont {Takamura}}, \ and\ \bibinfo {author} {\bibnamefont {{LHD
  Experimental Group}}},\ }\href {\doibase
  https://doi.org/10.1016/S0022-3115(02)01364-8} {\bibfield  {journal}
  {\bibinfo  {journal} {Journal of Nuclear Materials}\ }\textbf {\bibinfo
  {volume} {313-316}},\ \bibinfo {pages} {696} (\bibinfo {year} {2003})},\
  \bibinfo {note} {plasma-Surface Interactions in Controlled Fusion Devices
  15}\BibitemShut {NoStop}%
\bibitem [{\citenamefont {Schubert}\ \emph {et~al.}(2007)\citenamefont
  {Schubert}, \citenamefont {Endler},\ and\ \citenamefont
  {Thomsen}}]{Schubert2007}%
  \BibitemOpen
  \bibfield  {author} {\bibinfo {author} {\bibfnamefont {M.}~\bibnamefont
  {Schubert}}, \bibinfo {author} {\bibfnamefont {M.}~\bibnamefont {Endler}}, \
  and\ \bibinfo {author} {\bibfnamefont {H.}~\bibnamefont {Thomsen}},\ }\href
  {\doibase 10.1063/1.2740785} {\bibfield  {journal} {\bibinfo  {journal}
  {Review of Scientific Instruments}\ }\textbf {\bibinfo {volume} {78}},\
  \bibinfo {pages} {053505} (\bibinfo {year} {2007})},\ \Eprint
  {http://arxiv.org/abs/https://doi.org/10.1063/1.2740785}
  {https://doi.org/10.1063/1.2740785} \BibitemShut {NoStop}%
\bibitem [{\citenamefont {Smick}\ and\ \citenamefont
  {LaBombard}(2009)}]{Smick2009}%
  \BibitemOpen
  \bibfield  {author} {\bibinfo {author} {\bibfnamefont {N.}~\bibnamefont
  {Smick}}\ and\ \bibinfo {author} {\bibfnamefont {B.}~\bibnamefont
  {LaBombard}},\ }\href {\doibase 10.1063/1.3069290} {\bibfield  {journal}
  {\bibinfo  {journal} {Review of Scientific Instruments}\ }\textbf {\bibinfo
  {volume} {80}},\ \bibinfo {pages} {023502} (\bibinfo {year}
  {2009})}\BibitemShut {NoStop}%
\bibitem [{\citenamefont {Boedo}\ \emph {et~al.}(2009)\citenamefont {Boedo},
  \citenamefont {Crocker}, \citenamefont {Chousal}, \citenamefont {Hernandez},
  \citenamefont {Chalfant}, \citenamefont {Kugel}, \citenamefont {Roney},\ and\
  \citenamefont {Wertenbaker}}]{Boedo2009a}%
  \BibitemOpen
  \bibfield  {author} {\bibinfo {author} {\bibfnamefont {J.~A.}\ \bibnamefont
  {Boedo}}, \bibinfo {author} {\bibfnamefont {N.}~\bibnamefont {Crocker}},
  \bibinfo {author} {\bibfnamefont {L.}~\bibnamefont {Chousal}}, \bibinfo
  {author} {\bibfnamefont {R.}~\bibnamefont {Hernandez}}, \bibinfo {author}
  {\bibfnamefont {J.}~\bibnamefont {Chalfant}}, \bibinfo {author}
  {\bibfnamefont {H.}~\bibnamefont {Kugel}}, \bibinfo {author} {\bibfnamefont
  {P.}~\bibnamefont {Roney}}, \ and\ \bibinfo {author} {\bibfnamefont
  {J.}~\bibnamefont {Wertenbaker}},\ }\href {\doibase 10.1063/1.3266065}
  {\bibfield  {journal} {\bibinfo  {journal} {Review of Scientific
  Instruments}\ }\textbf {\bibinfo {volume} {80}},\ \bibinfo {pages} {123506}
  (\bibinfo {year} {2009})},\ \Eprint
  {http://arxiv.org/abs/https://doi.org/10.1063/1.3266065}
  {https://doi.org/10.1063/1.3266065} \BibitemShut {NoStop}%
\bibitem [{\citenamefont {Zhang}\ \emph {et~al.}(2010)\citenamefont {Zhang},
  \citenamefont {Chang}, \citenamefont {Wan}, \citenamefont {Xu}, \citenamefont
  {Xiao}, \citenamefont {Li}, \citenamefont {Xu}, \citenamefont {Yan},
  \citenamefont {Wang}, \citenamefont {Liu}, \citenamefont {Jiang},\ and\
  \citenamefont {Liu}}]{Zhang2010}%
  \BibitemOpen
  \bibfield  {author} {\bibinfo {author} {\bibfnamefont {W.}~\bibnamefont
  {Zhang}}, \bibinfo {author} {\bibfnamefont {J.~F.}\ \bibnamefont {Chang}},
  \bibinfo {author} {\bibfnamefont {B.~N.}\ \bibnamefont {Wan}}, \bibinfo
  {author} {\bibfnamefont {G.~S.}\ \bibnamefont {Xu}}, \bibinfo {author}
  {\bibfnamefont {C.~J.}\ \bibnamefont {Xiao}}, \bibinfo {author}
  {\bibfnamefont {B.}~\bibnamefont {Li}}, \bibinfo {author} {\bibfnamefont
  {C.~S.}\ \bibnamefont {Xu}}, \bibinfo {author} {\bibfnamefont
  {N.}~\bibnamefont {Yan}}, \bibinfo {author} {\bibfnamefont {L.}~\bibnamefont
  {Wang}}, \bibinfo {author} {\bibfnamefont {S.~C.}\ \bibnamefont {Liu}},
  \bibinfo {author} {\bibfnamefont {M.}~\bibnamefont {Jiang}}, \ and\ \bibinfo
  {author} {\bibfnamefont {P.}~\bibnamefont {Liu}},\ }\href {\doibase
  10.1063/1.3499237} {\bibfield  {journal} {\bibinfo  {journal} {Review of
  Scientific Instruments}\ }\textbf {\bibinfo {volume} {81}},\ \bibinfo {pages}
  {113501} (\bibinfo {year} {2010})},\ \Eprint
  {http://arxiv.org/abs/https://doi.org/10.1063/1.3499237}
  {https://doi.org/10.1063/1.3499237} \BibitemShut {NoStop}%
\bibitem [{\citenamefont {Gunn}\ and\ \citenamefont {Pascal}(2011)}]{Gunn2011}%
  \BibitemOpen
  \bibfield  {author} {\bibinfo {author} {\bibfnamefont {J.~P.}\ \bibnamefont
  {Gunn}}\ and\ \bibinfo {author} {\bibfnamefont {J.-Y.}\ \bibnamefont
  {Pascal}},\ }\href {\doibase 10.1063/1.3661128} {\bibfield  {journal}
  {\bibinfo  {journal} {Review of Scientific Instruments}\ }\textbf {\bibinfo
  {volume} {82}},\ \bibinfo {pages} {123505} (\bibinfo {year} {2011})},\
  \Eprint {http://arxiv.org/abs/https://doi.org/10.1063/1.3661128}
  {https://doi.org/10.1063/1.3661128} \BibitemShut {NoStop}%
\bibitem [{\citenamefont {Tsui}\ \emph {et~al.}(2012)\citenamefont {Tsui},
  \citenamefont {Taussig}, \citenamefont {Watkins}, \citenamefont {Boivin},\
  and\ \citenamefont {Stangeby}}]{Tsui2012}%
  \BibitemOpen
  \bibfield  {author} {\bibinfo {author} {\bibfnamefont {C.~K.}\ \bibnamefont
  {Tsui}}, \bibinfo {author} {\bibfnamefont {D.~A.}\ \bibnamefont {Taussig}},
  \bibinfo {author} {\bibfnamefont {M.~G.}\ \bibnamefont {Watkins}}, \bibinfo
  {author} {\bibfnamefont {R.~L.}\ \bibnamefont {Boivin}}, \ and\ \bibinfo
  {author} {\bibfnamefont {P.~C.}\ \bibnamefont {Stangeby}},\ }\href {\doibase
  10.1063/1.4733571} {\bibfield  {journal} {\bibinfo  {journal} {Review of
  Scientific Instruments}\ }\textbf {\bibinfo {volume} {83}},\ \bibinfo {pages}
  {10D723} (\bibinfo {year} {2012})},\ \Eprint
  {http://arxiv.org/abs/https://doi.org/10.1063/1.4733571}
  {https://doi.org/10.1063/1.4733571} \BibitemShut {NoStop}%
\bibitem [{\citenamefont {{de Marné}}\ \emph {et~al.}(2017)\citenamefont {{de
  Marné}}, \citenamefont {Herrmann},\ and\ \citenamefont
  {Leitenstern}}]{deMarne2017}%
  \BibitemOpen
  \bibfield  {author} {\bibinfo {author} {\bibfnamefont {P.}~\bibnamefont {{de
  Marné}}}, \bibinfo {author} {\bibfnamefont {A.}~\bibnamefont {Herrmann}}, \
  and\ \bibinfo {author} {\bibfnamefont {P.}~\bibnamefont {Leitenstern}},\
  }\href {\doibase https://doi.org/10.1016/j.fusengdes.2017.05.056} {\bibfield
  {journal} {\bibinfo  {journal} {Fusion Engineering and Design}\ }\textbf
  {\bibinfo {volume} {123}},\ \bibinfo {pages} {754} (\bibinfo {year}
  {2017})},\ \bibinfo {note} {proceedings of the 29th Symposium on Fusion
  Technology (SOFT-29) Prague, Czech Republic, September 5-9, 2016}\BibitemShut
  {NoStop}%
\bibitem [{\citenamefont {De~Oliveira}\ \emph {et~al.}(2021)\citenamefont
  {De~Oliveira}, \citenamefont {Theiler},\ and\ \citenamefont
  {Elaian}}]{DeOliveira2021}%
  \BibitemOpen
  \bibfield  {author} {\bibinfo {author} {\bibfnamefont {H.}~\bibnamefont
  {De~Oliveira}}, \bibinfo {author} {\bibfnamefont {C.}~\bibnamefont
  {Theiler}}, \ and\ \bibinfo {author} {\bibfnamefont {H.}~\bibnamefont
  {Elaian}},\ }\href {\doibase 10.1063/5.0043523} {\bibfield  {journal}
  {\bibinfo  {journal} {Review of Scientific Instruments}\ }\textbf {\bibinfo
  {volume} {92}},\ \bibinfo {pages} {043547} (\bibinfo {year} {2021})},\
  \Eprint {http://arxiv.org/abs/https://doi.org/10.1063/5.0043523}
  {https://doi.org/10.1063/5.0043523} \BibitemShut {NoStop}%
\bibitem [{\citenamefont {Klinger}\ and\ \citenamefont {the
  W7-X~Team}(2019)}]{Klinger2019_short}%
  \BibitemOpen
  \bibfield  {author} {\bibinfo {author} {\bibfnamefont {T.}~\bibnamefont
  {Klinger}}\ and\ \bibinfo {author} {\bibnamefont {the W7-X~Team}},\ }\href
  {\doibase 10.1088/1741-4326/ab03a7} {\bibfield  {journal} {\bibinfo
  {journal} {Nuclear Fusion}\ }\textbf {\bibinfo {volume} {59}},\ \bibinfo
  {pages} {112004} (\bibinfo {year} {2019})}\BibitemShut {NoStop}%
\bibitem [{\citenamefont {Drews}\ \emph {et~al.}(2017)\citenamefont {Drews},
  \citenamefont {Liang}, \citenamefont {Liu}, \citenamefont {Krämer-Flecken},
  \citenamefont {Neubauer}, \citenamefont {Geiger}, \citenamefont {Rack},
  \citenamefont {Nicolai}, \citenamefont {Grulke}, \citenamefont {Killer},
  \citenamefont {Wang}, \citenamefont {Charl}, \citenamefont {Schweer},
  \citenamefont {Denner}, \citenamefont {Henkel}, \citenamefont {Gao},
  \citenamefont {Hollfeld}, \citenamefont {Satheeswaran}, \citenamefont
  {Sandri}, \citenamefont {Höschen},\ and\ \citenamefont {Team}}]{Drews2017}%
  \BibitemOpen
  \bibfield  {author} {\bibinfo {author} {\bibfnamefont {P.}~\bibnamefont
  {Drews}}, \bibinfo {author} {\bibfnamefont {Y.}~\bibnamefont {Liang}},
  \bibinfo {author} {\bibfnamefont {S.}~\bibnamefont {Liu}}, \bibinfo {author}
  {\bibfnamefont {A.}~\bibnamefont {Krämer-Flecken}}, \bibinfo {author}
  {\bibfnamefont {O.}~\bibnamefont {Neubauer}}, \bibinfo {author}
  {\bibfnamefont {J.}~\bibnamefont {Geiger}}, \bibinfo {author} {\bibfnamefont
  {M.}~\bibnamefont {Rack}}, \bibinfo {author} {\bibfnamefont {D.}~\bibnamefont
  {Nicolai}}, \bibinfo {author} {\bibfnamefont {O.}~\bibnamefont {Grulke}},
  \bibinfo {author} {\bibfnamefont {C.}~\bibnamefont {Killer}}, \bibinfo
  {author} {\bibfnamefont {N.}~\bibnamefont {Wang}}, \bibinfo {author}
  {\bibfnamefont {A.}~\bibnamefont {Charl}}, \bibinfo {author} {\bibfnamefont
  {B.}~\bibnamefont {Schweer}}, \bibinfo {author} {\bibfnamefont
  {P.}~\bibnamefont {Denner}}, \bibinfo {author} {\bibfnamefont
  {M.}~\bibnamefont {Henkel}}, \bibinfo {author} {\bibfnamefont
  {Y.}~\bibnamefont {Gao}}, \bibinfo {author} {\bibfnamefont {K.}~\bibnamefont
  {Hollfeld}}, \bibinfo {author} {\bibfnamefont {G.}~\bibnamefont
  {Satheeswaran}}, \bibinfo {author} {\bibfnamefont {N.}~\bibnamefont
  {Sandri}}, \bibinfo {author} {\bibfnamefont {D.}~\bibnamefont {Höschen}}, \
  and\ \bibinfo {author} {\bibfnamefont {T.~W.-X.}\ \bibnamefont {Team}},\
  }\href {http://stacks.iop.org/0029-5515/57/i=12/a=126020} {\bibfield
  {journal} {\bibinfo  {journal} {Nuclear Fusion}\ }\textbf {\bibinfo {volume}
  {57}},\ \bibinfo {pages} {126020} (\bibinfo {year} {2017})}\BibitemShut
  {NoStop}%
\bibitem [{\citenamefont {Liu}\ \emph {et~al.}(2018{\natexlab{a}})\citenamefont
  {Liu}, \citenamefont {Liang}, \citenamefont {Drews}, \citenamefont
  {Krämer-Flecken}, \citenamefont {Han}, \citenamefont {Nicolai},
  \citenamefont {Satheeswaran}, \citenamefont {Wang}, \citenamefont {Cai},
  \citenamefont {Charl}, \citenamefont {Cosfeld}, \citenamefont {Fuchert},
  \citenamefont {Gao}, \citenamefont {Geiger}, \citenamefont {Grulke},
  \citenamefont {Henkel}, \citenamefont {Hirsch}, \citenamefont {Hoefel},
  \citenamefont {Hollfeld}, \citenamefont {Höschen}, \citenamefont {Killer},
  \citenamefont {Knieps}, \citenamefont {König}, \citenamefont {Neubauer},
  \citenamefont {Pasch}, \citenamefont {Rahbarnia}, \citenamefont {Rack},
  \citenamefont {Sandri}, \citenamefont {Sereda}, \citenamefont {Schweer},
  \citenamefont {Wang}, \citenamefont {Wei}, \citenamefont {Weir},
  \citenamefont {Windisch},\ and\ \citenamefont {Team}}]{Liu2018}%
  \BibitemOpen
  \bibfield  {author} {\bibinfo {author} {\bibfnamefont {S.}~\bibnamefont
  {Liu}}, \bibinfo {author} {\bibfnamefont {Y.}~\bibnamefont {Liang}}, \bibinfo
  {author} {\bibfnamefont {P.}~\bibnamefont {Drews}}, \bibinfo {author}
  {\bibfnamefont {A.}~\bibnamefont {Krämer-Flecken}}, \bibinfo {author}
  {\bibfnamefont {X.}~\bibnamefont {Han}}, \bibinfo {author} {\bibfnamefont
  {D.}~\bibnamefont {Nicolai}}, \bibinfo {author} {\bibfnamefont
  {G.}~\bibnamefont {Satheeswaran}}, \bibinfo {author} {\bibfnamefont
  {N.}~\bibnamefont {Wang}}, \bibinfo {author} {\bibfnamefont {J.}~\bibnamefont
  {Cai}}, \bibinfo {author} {\bibfnamefont {A.}~\bibnamefont {Charl}}, \bibinfo
  {author} {\bibfnamefont {J.}~\bibnamefont {Cosfeld}}, \bibinfo {author}
  {\bibfnamefont {G.}~\bibnamefont {Fuchert}}, \bibinfo {author} {\bibfnamefont
  {Y.}~\bibnamefont {Gao}}, \bibinfo {author} {\bibfnamefont {J.}~\bibnamefont
  {Geiger}}, \bibinfo {author} {\bibfnamefont {O.}~\bibnamefont {Grulke}},
  \bibinfo {author} {\bibfnamefont {M.}~\bibnamefont {Henkel}}, \bibinfo
  {author} {\bibfnamefont {M.}~\bibnamefont {Hirsch}}, \bibinfo {author}
  {\bibfnamefont {U.}~\bibnamefont {Hoefel}}, \bibinfo {author} {\bibfnamefont
  {K.}~\bibnamefont {Hollfeld}}, \bibinfo {author} {\bibfnamefont
  {D.}~\bibnamefont {Höschen}}, \bibinfo {author} {\bibfnamefont
  {C.}~\bibnamefont {Killer}}, \bibinfo {author} {\bibfnamefont
  {A.}~\bibnamefont {Knieps}}, \bibinfo {author} {\bibfnamefont
  {R.}~\bibnamefont {König}}, \bibinfo {author} {\bibfnamefont
  {O.}~\bibnamefont {Neubauer}}, \bibinfo {author} {\bibfnamefont
  {E.}~\bibnamefont {Pasch}}, \bibinfo {author} {\bibfnamefont
  {K.}~\bibnamefont {Rahbarnia}}, \bibinfo {author} {\bibfnamefont
  {M.}~\bibnamefont {Rack}}, \bibinfo {author} {\bibfnamefont {N.}~\bibnamefont
  {Sandri}}, \bibinfo {author} {\bibfnamefont {S.}~\bibnamefont {Sereda}},
  \bibinfo {author} {\bibfnamefont {B.}~\bibnamefont {Schweer}}, \bibinfo
  {author} {\bibfnamefont {E.}~\bibnamefont {Wang}}, \bibinfo {author}
  {\bibfnamefont {Y.}~\bibnamefont {Wei}}, \bibinfo {author} {\bibfnamefont
  {G.}~\bibnamefont {Weir}}, \bibinfo {author} {\bibfnamefont {T.}~\bibnamefont
  {Windisch}}, \ and\ \bibinfo {author} {\bibfnamefont {W.-X.}\ \bibnamefont
  {Team}},\ }\href {http://stacks.iop.org/0029-5515/58/i=4/a=046002} {\bibfield
   {journal} {\bibinfo  {journal} {Nuclear Fusion}\ }\textbf {\bibinfo {volume}
  {58}},\ \bibinfo {pages} {046002} (\bibinfo {year}
  {2018}{\natexlab{a}})}\BibitemShut {NoStop}%
\bibitem [{\citenamefont {Liu}\ \emph {et~al.}(2018{\natexlab{b}})\citenamefont
  {Liu}, \citenamefont {Liang}, \citenamefont {Drews}, \citenamefont
  {Krämer-Flecken}, \citenamefont {Han}, \citenamefont {Nicolai},
  \citenamefont {Satheeswaran}, \citenamefont {Wang}, \citenamefont {Cai},
  \citenamefont {Charl}, \citenamefont {Cosfeld}, \citenamefont {Gao},
  \citenamefont {Grulke}, \citenamefont {Henkel}, \citenamefont {Hollfeld},
  \citenamefont {Killer}, \citenamefont {Knieps}, \citenamefont {König},
  \citenamefont {Neubauer}, \citenamefont {Rack}, \citenamefont {Sandri},
  \citenamefont {Sereda}, \citenamefont {Schweer}, \citenamefont {Wang},\ and\
  \citenamefont {Wei}}]{Liu2018a}%
  \BibitemOpen
  \bibfield  {author} {\bibinfo {author} {\bibfnamefont {S.~C.}\ \bibnamefont
  {Liu}}, \bibinfo {author} {\bibfnamefont {Y.}~\bibnamefont {Liang}}, \bibinfo
  {author} {\bibfnamefont {P.}~\bibnamefont {Drews}}, \bibinfo {author}
  {\bibfnamefont {A.}~\bibnamefont {Krämer-Flecken}}, \bibinfo {author}
  {\bibfnamefont {X.}~\bibnamefont {Han}}, \bibinfo {author} {\bibfnamefont
  {D.}~\bibnamefont {Nicolai}}, \bibinfo {author} {\bibfnamefont
  {G.}~\bibnamefont {Satheeswaran}}, \bibinfo {author} {\bibfnamefont {N.~C.}\
  \bibnamefont {Wang}}, \bibinfo {author} {\bibfnamefont {J.~Q.}\ \bibnamefont
  {Cai}}, \bibinfo {author} {\bibfnamefont {A.}~\bibnamefont {Charl}}, \bibinfo
  {author} {\bibfnamefont {J.}~\bibnamefont {Cosfeld}}, \bibinfo {author}
  {\bibfnamefont {Y.}~\bibnamefont {Gao}}, \bibinfo {author} {\bibfnamefont
  {O.}~\bibnamefont {Grulke}}, \bibinfo {author} {\bibfnamefont
  {M.}~\bibnamefont {Henkel}}, \bibinfo {author} {\bibfnamefont {K.~P.}\
  \bibnamefont {Hollfeld}}, \bibinfo {author} {\bibfnamefont {C.}~\bibnamefont
  {Killer}}, \bibinfo {author} {\bibfnamefont {A.}~\bibnamefont {Knieps}},
  \bibinfo {author} {\bibfnamefont {R.}~\bibnamefont {König}}, \bibinfo
  {author} {\bibfnamefont {O.}~\bibnamefont {Neubauer}}, \bibinfo {author}
  {\bibfnamefont {M.}~\bibnamefont {Rack}}, \bibinfo {author} {\bibfnamefont
  {N.}~\bibnamefont {Sandri}}, \bibinfo {author} {\bibfnamefont
  {S.}~\bibnamefont {Sereda}}, \bibinfo {author} {\bibfnamefont
  {B.}~\bibnamefont {Schweer}}, \bibinfo {author} {\bibfnamefont {E.~H.}\
  \bibnamefont {Wang}}, \ and\ \bibinfo {author} {\bibfnamefont {Y.~L.}\
  \bibnamefont {Wei}},\ }\href {\doibase 10.1063/1.5033353} {\bibfield
  {journal} {\bibinfo  {journal} {Physics of Plasmas}\ }\textbf {\bibinfo
  {volume} {25}},\ \bibinfo {pages} {072502} (\bibinfo {year}
  {2018}{\natexlab{b}})}\BibitemShut {NoStop}%
\bibitem [{\citenamefont {Drews}\ \emph
  {et~al.}(2019{\natexlab{a}})\citenamefont {Drews}, \citenamefont {Killer},
  \citenamefont {Cosfeld}, \citenamefont {Knieps}, \citenamefont {Brezinsek},
  \citenamefont {Jakubowski}, \citenamefont {Brandt}, \citenamefont
  {Bozhenkov}, \citenamefont {Dinklage}, \citenamefont {Cai}, \citenamefont
  {Endler}, \citenamefont {Hammond}, \citenamefont {Henkel}, \citenamefont
  {Gao}, \citenamefont {Geiger}, \citenamefont {Grulke}, \citenamefont
  {Höschen}, \citenamefont {König}, \citenamefont {Krämer-Flecken},
  \citenamefont {Liang}, \citenamefont {Li}, \citenamefont {Liu}, \citenamefont
  {Niemann}, \citenamefont {Nicolai}, \citenamefont {Neubauer}, \citenamefont
  {Neuner}, \citenamefont {Rack}, \citenamefont {Rahbarnia}, \citenamefont
  {Rudischhauser}, \citenamefont {Sandri}, \citenamefont {Satheeswaran},
  \citenamefont {Schilling}, \citenamefont {Thomsen}, \citenamefont
  {Windisch},\ and\ \citenamefont {Sereda}}]{Drews2019}%
  \BibitemOpen
  \bibfield  {author} {\bibinfo {author} {\bibfnamefont {P.}~\bibnamefont
  {Drews}}, \bibinfo {author} {\bibfnamefont {C.}~\bibnamefont {Killer}},
  \bibinfo {author} {\bibfnamefont {J.}~\bibnamefont {Cosfeld}}, \bibinfo
  {author} {\bibfnamefont {A.}~\bibnamefont {Knieps}}, \bibinfo {author}
  {\bibfnamefont {S.}~\bibnamefont {Brezinsek}}, \bibinfo {author}
  {\bibfnamefont {M.}~\bibnamefont {Jakubowski}}, \bibinfo {author}
  {\bibfnamefont {C.}~\bibnamefont {Brandt}}, \bibinfo {author} {\bibfnamefont
  {S.}~\bibnamefont {Bozhenkov}}, \bibinfo {author} {\bibfnamefont
  {A.}~\bibnamefont {Dinklage}}, \bibinfo {author} {\bibfnamefont
  {J.}~\bibnamefont {Cai}}, \bibinfo {author} {\bibfnamefont {M.}~\bibnamefont
  {Endler}}, \bibinfo {author} {\bibfnamefont {K.}~\bibnamefont {Hammond}},
  \bibinfo {author} {\bibfnamefont {M.}~\bibnamefont {Henkel}}, \bibinfo
  {author} {\bibfnamefont {Y.}~\bibnamefont {Gao}}, \bibinfo {author}
  {\bibfnamefont {J.}~\bibnamefont {Geiger}}, \bibinfo {author} {\bibfnamefont
  {O.}~\bibnamefont {Grulke}}, \bibinfo {author} {\bibfnamefont
  {D.}~\bibnamefont {Höschen}}, \bibinfo {author} {\bibfnamefont
  {R.}~\bibnamefont {König}}, \bibinfo {author} {\bibfnamefont
  {A.}~\bibnamefont {Krämer-Flecken}}, \bibinfo {author} {\bibfnamefont
  {Y.}~\bibnamefont {Liang}}, \bibinfo {author} {\bibfnamefont
  {Y.}~\bibnamefont {Li}}, \bibinfo {author} {\bibfnamefont {S.}~\bibnamefont
  {Liu}}, \bibinfo {author} {\bibfnamefont {H.}~\bibnamefont {Niemann}},
  \bibinfo {author} {\bibfnamefont {D.}~\bibnamefont {Nicolai}}, \bibinfo
  {author} {\bibfnamefont {O.}~\bibnamefont {Neubauer}}, \bibinfo {author}
  {\bibfnamefont {U.}~\bibnamefont {Neuner}}, \bibinfo {author} {\bibfnamefont
  {M.}~\bibnamefont {Rack}}, \bibinfo {author} {\bibfnamefont {K.}~\bibnamefont
  {Rahbarnia}}, \bibinfo {author} {\bibfnamefont {L.}~\bibnamefont
  {Rudischhauser}}, \bibinfo {author} {\bibfnamefont {N.}~\bibnamefont
  {Sandri}}, \bibinfo {author} {\bibfnamefont {G.}~\bibnamefont
  {Satheeswaran}}, \bibinfo {author} {\bibfnamefont {S.}~\bibnamefont
  {Schilling}}, \bibinfo {author} {\bibfnamefont {H.}~\bibnamefont {Thomsen}},
  \bibinfo {author} {\bibfnamefont {T.}~\bibnamefont {Windisch}}, \ and\
  \bibinfo {author} {\bibfnamefont {S.}~\bibnamefont {Sereda}},\ }\href
  {\doibase https://doi.org/10.1016/j.nme.2019.02.012} {\bibfield  {journal}
  {\bibinfo  {journal} {Nuclear Materials and Energy}\ }\textbf {\bibinfo
  {volume} {19}},\ \bibinfo {pages} {179 } (\bibinfo {year}
  {2019}{\natexlab{a}})}\BibitemShut {NoStop}%
\bibitem [{\citenamefont {Liu}\ \emph {et~al.}(2019)\citenamefont {Liu},
  \citenamefont {Liang}, \citenamefont {Drews}, \citenamefont {Killer},
  \citenamefont {Knieps}, \citenamefont {Xu}, \citenamefont {Wang},
  \citenamefont {Yan}, \citenamefont {Han}, \citenamefont {Höschen},
  \citenamefont {Kraemer-Flecken}, \citenamefont {Nicolai}, \citenamefont
  {Satheeswaran}, \citenamefont {Hammond}, \citenamefont {Cai}, \citenamefont
  {Charl}, \citenamefont {Cosfeld}, \citenamefont {Fuchert}, \citenamefont
  {Gao}, \citenamefont {Geiger}, \citenamefont {Grulke}, \citenamefont
  {Henkel}, \citenamefont {Hirsch}, \citenamefont {Höfel}, \citenamefont
  {König}, \citenamefont {Li}, \citenamefont {Neubauer}, \citenamefont
  {Pasch}, \citenamefont {Rahbarnia}, \citenamefont {Rack}, \citenamefont
  {Sandri}, \citenamefont {Sereda}, \citenamefont {Schweer}, \citenamefont
  {Wang}, \citenamefont {Xu},\ and\ \citenamefont {Gao}}]{Liu2019}%
  \BibitemOpen
  \bibfield  {author} {\bibinfo {author} {\bibfnamefont {S.}~\bibnamefont
  {Liu}}, \bibinfo {author} {\bibfnamefont {Y.}~\bibnamefont {Liang}}, \bibinfo
  {author} {\bibfnamefont {P.}~\bibnamefont {Drews}}, \bibinfo {author}
  {\bibfnamefont {C.}~\bibnamefont {Killer}}, \bibinfo {author} {\bibfnamefont
  {A.}~\bibnamefont {Knieps}}, \bibinfo {author} {\bibfnamefont
  {G.}~\bibnamefont {Xu}}, \bibinfo {author} {\bibfnamefont {H.}~\bibnamefont
  {Wang}}, \bibinfo {author} {\bibfnamefont {N.}~\bibnamefont {Yan}}, \bibinfo
  {author} {\bibfnamefont {X.}~\bibnamefont {Han}}, \bibinfo {author}
  {\bibfnamefont {D.}~\bibnamefont {Höschen}}, \bibinfo {author}
  {\bibfnamefont {A.}~\bibnamefont {Kraemer-Flecken}}, \bibinfo {author}
  {\bibfnamefont {D.}~\bibnamefont {Nicolai}}, \bibinfo {author} {\bibfnamefont
  {G.}~\bibnamefont {Satheeswaran}}, \bibinfo {author} {\bibfnamefont
  {K.}~\bibnamefont {Hammond}}, \bibinfo {author} {\bibfnamefont
  {J.}~\bibnamefont {Cai}}, \bibinfo {author} {\bibfnamefont {A.}~\bibnamefont
  {Charl}}, \bibinfo {author} {\bibfnamefont {J.}~\bibnamefont {Cosfeld}},
  \bibinfo {author} {\bibfnamefont {G.}~\bibnamefont {Fuchert}}, \bibinfo
  {author} {\bibfnamefont {Y.}~\bibnamefont {Gao}}, \bibinfo {author}
  {\bibfnamefont {J.}~\bibnamefont {Geiger}}, \bibinfo {author} {\bibfnamefont
  {O.}~\bibnamefont {Grulke}}, \bibinfo {author} {\bibfnamefont
  {M.}~\bibnamefont {Henkel}}, \bibinfo {author} {\bibfnamefont
  {M.}~\bibnamefont {Hirsch}}, \bibinfo {author} {\bibfnamefont
  {U.}~\bibnamefont {Höfel}}, \bibinfo {author} {\bibfnamefont
  {R.}~\bibnamefont {König}}, \bibinfo {author} {\bibfnamefont
  {Y.}~\bibnamefont {Li}}, \bibinfo {author} {\bibfnamefont {O.}~\bibnamefont
  {Neubauer}}, \bibinfo {author} {\bibfnamefont {E.}~\bibnamefont {Pasch}},
  \bibinfo {author} {\bibfnamefont {K.}~\bibnamefont {Rahbarnia}}, \bibinfo
  {author} {\bibfnamefont {M.}~\bibnamefont {Rack}}, \bibinfo {author}
  {\bibfnamefont {N.}~\bibnamefont {Sandri}}, \bibinfo {author} {\bibfnamefont
  {S.}~\bibnamefont {Sereda}}, \bibinfo {author} {\bibfnamefont
  {B.}~\bibnamefont {Schweer}}, \bibinfo {author} {\bibfnamefont
  {E.}~\bibnamefont {Wang}}, \bibinfo {author} {\bibfnamefont {S.}~\bibnamefont
  {Xu}}, \ and\ \bibinfo {author} {\bibfnamefont {X.}~\bibnamefont {Gao}},\
  }\href {http://iopscience.iop.org/10.1088/1741-4326/ab0d29} {\bibfield
  {journal} {\bibinfo  {journal} {Nuclear Fusion}\ } (\bibinfo {year}
  {2019})}\BibitemShut {NoStop}%
\bibitem [{\citenamefont {Drews}\ \emph
  {et~al.}(2019{\natexlab{b}})\citenamefont {Drews}, \citenamefont {Killer},
  \citenamefont {Knieps}, \citenamefont {Henkel}, \citenamefont {Brezinsek},
  \citenamefont {Dittmar}, \citenamefont {Cai}, \citenamefont {Card},
  \citenamefont {Grulke}, \citenamefont {Höschen}, \citenamefont {Jakubowski},
  \citenamefont {König}, \citenamefont {Krämer-Flecken}, \citenamefont
  {Liang}, \citenamefont {Li}, \citenamefont {Linsmeier}, \citenamefont {Liu},
  \citenamefont {Nicolai}, \citenamefont {Neubauer}, \citenamefont
  {Satheeswaran}, \citenamefont {Sandri}, \citenamefont {Schweer},\ and\
  \citenamefont {Schröder}}]{Drews2019a}%
  \BibitemOpen
  \bibfield  {author} {\bibinfo {author} {\bibfnamefont {P.}~\bibnamefont
  {Drews}}, \bibinfo {author} {\bibfnamefont {C.}~\bibnamefont {Killer}},
  \bibinfo {author} {\bibfnamefont {A.}~\bibnamefont {Knieps}}, \bibinfo
  {author} {\bibfnamefont {M.}~\bibnamefont {Henkel}}, \bibinfo {author}
  {\bibfnamefont {S.}~\bibnamefont {Brezinsek}}, \bibinfo {author}
  {\bibfnamefont {T.}~\bibnamefont {Dittmar}}, \bibinfo {author} {\bibfnamefont
  {J.}~\bibnamefont {Cai}}, \bibinfo {author} {\bibfnamefont {A.}~\bibnamefont
  {Card}}, \bibinfo {author} {\bibfnamefont {O.}~\bibnamefont {Grulke}},
  \bibinfo {author} {\bibfnamefont {D.}~\bibnamefont {Höschen}}, \bibinfo
  {author} {\bibfnamefont {M.}~\bibnamefont {Jakubowski}}, \bibinfo {author}
  {\bibfnamefont {R.}~\bibnamefont {König}}, \bibinfo {author} {\bibfnamefont
  {A.}~\bibnamefont {Krämer-Flecken}}, \bibinfo {author} {\bibfnamefont
  {Y.}~\bibnamefont {Liang}}, \bibinfo {author} {\bibfnamefont
  {Y.}~\bibnamefont {Li}}, \bibinfo {author} {\bibfnamefont {C.}~\bibnamefont
  {Linsmeier}}, \bibinfo {author} {\bibfnamefont {S.}~\bibnamefont {Liu}},
  \bibinfo {author} {\bibfnamefont {D.}~\bibnamefont {Nicolai}}, \bibinfo
  {author} {\bibfnamefont {O.}~\bibnamefont {Neubauer}}, \bibinfo {author}
  {\bibfnamefont {G.}~\bibnamefont {Satheeswaran}}, \bibinfo {author}
  {\bibfnamefont {N.}~\bibnamefont {Sandri}}, \bibinfo {author} {\bibfnamefont
  {B.}~\bibnamefont {Schweer}}, \ and\ \bibinfo {author} {\bibfnamefont
  {T.}~\bibnamefont {Schröder}},\ }\href {\doibase
  https://doi.org/10.1016/j.fusengdes.2019.03.188} {\bibfield  {journal}
  {\bibinfo  {journal} {Fusion Engineering and Design}\ }\textbf {\bibinfo
  {volume} {146}},\ \bibinfo {pages} {2353} (\bibinfo {year}
  {2019}{\natexlab{b}})},\ \bibinfo {note} {sI:SOFT-30}\BibitemShut {NoStop}%
\bibitem [{\citenamefont {Liu}\ \emph {et~al.}(2020)\citenamefont {Liu},
  \citenamefont {Liang}, \citenamefont {Wang}, \citenamefont {Killer},
  \citenamefont {Drews}, \citenamefont {Knieps}, \citenamefont {Han},
  \citenamefont {Grulke}, \citenamefont {Krämer-Flecken}, \citenamefont {Xu},
  \citenamefont {Yan}, \citenamefont {Höschen}, \citenamefont {Nicolai},
  \citenamefont {Satheeswaran}, \citenamefont {Geiger}, \citenamefont {Henkel},
  \citenamefont {Huang}, \citenamefont {König}, \citenamefont {Li},
  \citenamefont {Neubauer}, \citenamefont {Rahbarnia}, \citenamefont {Sandri},
  \citenamefont {Schweer}, \citenamefont {Wang}, \citenamefont {Wang},
  \citenamefont {Xu},\ and\ \citenamefont {Gao}}]{Liu2020}%
  \BibitemOpen
  \bibfield  {author} {\bibinfo {author} {\bibfnamefont {S.~C.}\ \bibnamefont
  {Liu}}, \bibinfo {author} {\bibfnamefont {Y.}~\bibnamefont {Liang}}, \bibinfo
  {author} {\bibfnamefont {H.~Q.}\ \bibnamefont {Wang}}, \bibinfo {author}
  {\bibfnamefont {C.}~\bibnamefont {Killer}}, \bibinfo {author} {\bibfnamefont
  {P.}~\bibnamefont {Drews}}, \bibinfo {author} {\bibfnamefont
  {A.}~\bibnamefont {Knieps}}, \bibinfo {author} {\bibfnamefont
  {X.}~\bibnamefont {Han}}, \bibinfo {author} {\bibfnamefont {O.}~\bibnamefont
  {Grulke}}, \bibinfo {author} {\bibfnamefont {A.}~\bibnamefont
  {Krämer-Flecken}}, \bibinfo {author} {\bibfnamefont {G.~S.}\ \bibnamefont
  {Xu}}, \bibinfo {author} {\bibfnamefont {N.}~\bibnamefont {Yan}}, \bibinfo
  {author} {\bibfnamefont {D.}~\bibnamefont {Höschen}}, \bibinfo {author}
  {\bibfnamefont {D.}~\bibnamefont {Nicolai}}, \bibinfo {author} {\bibfnamefont
  {G.}~\bibnamefont {Satheeswaran}}, \bibinfo {author} {\bibfnamefont
  {J.}~\bibnamefont {Geiger}}, \bibinfo {author} {\bibfnamefont
  {M.}~\bibnamefont {Henkel}}, \bibinfo {author} {\bibfnamefont
  {Z.}~\bibnamefont {Huang}}, \bibinfo {author} {\bibfnamefont
  {R.}~\bibnamefont {König}}, \bibinfo {author} {\bibfnamefont
  {Y.}~\bibnamefont {Li}}, \bibinfo {author} {\bibfnamefont {O.}~\bibnamefont
  {Neubauer}}, \bibinfo {author} {\bibfnamefont {K.}~\bibnamefont {Rahbarnia}},
  \bibinfo {author} {\bibfnamefont {N.}~\bibnamefont {Sandri}}, \bibinfo
  {author} {\bibfnamefont {B.}~\bibnamefont {Schweer}}, \bibinfo {author}
  {\bibfnamefont {E.~H.}\ \bibnamefont {Wang}}, \bibinfo {author}
  {\bibfnamefont {Y.~M.}\ \bibnamefont {Wang}}, \bibinfo {author}
  {\bibfnamefont {S.}~\bibnamefont {Xu}}, \ and\ \bibinfo {author}
  {\bibfnamefont {X.}~\bibnamefont {Gao}},\ }\href {\doibase 10.1063/5.0016151}
  {\bibfield  {journal} {\bibinfo  {journal} {Physics of Plasmas}\ }\textbf
  {\bibinfo {volume} {27}},\ \bibinfo {pages} {122504} (\bibinfo {year}
  {2020})},\ \Eprint {http://arxiv.org/abs/https://doi.org/10.1063/5.0016151}
  {https://doi.org/10.1063/5.0016151} \BibitemShut {NoStop}%
\bibitem [{\citenamefont {Knieps}\ \emph {et~al.}(2020)\citenamefont {Knieps},
  \citenamefont {Liang}, \citenamefont {Drews}, \citenamefont {Endler},
  \citenamefont {Grulke}, \citenamefont {Huang}, \citenamefont {Killer},
  \citenamefont {Liu}, \citenamefont {Nicolai}, \citenamefont {Rahbarnia},
  \citenamefont {Sandri},\ and\ \citenamefont {Satheeswaran}}]{Knieps2020}%
  \BibitemOpen
  \bibfield  {author} {\bibinfo {author} {\bibfnamefont {A.}~\bibnamefont
  {Knieps}}, \bibinfo {author} {\bibfnamefont {Y.}~\bibnamefont {Liang}},
  \bibinfo {author} {\bibfnamefont {P.}~\bibnamefont {Drews}}, \bibinfo
  {author} {\bibfnamefont {M.}~\bibnamefont {Endler}}, \bibinfo {author}
  {\bibfnamefont {O.}~\bibnamefont {Grulke}}, \bibinfo {author} {\bibfnamefont
  {Z.}~\bibnamefont {Huang}}, \bibinfo {author} {\bibfnamefont
  {C.}~\bibnamefont {Killer}}, \bibinfo {author} {\bibfnamefont
  {S.}~\bibnamefont {Liu}}, \bibinfo {author} {\bibfnamefont {D.}~\bibnamefont
  {Nicolai}}, \bibinfo {author} {\bibfnamefont {K.}~\bibnamefont {Rahbarnia}},
  \bibinfo {author} {\bibfnamefont {N.}~\bibnamefont {Sandri}}, \ and\ \bibinfo
  {author} {\bibfnamefont {G.}~\bibnamefont {Satheeswaran}},\ }\href {\doibase
  10.1063/5.0002193} {\bibfield  {journal} {\bibinfo  {journal} {Review of
  Scientific Instruments}\ }\textbf {\bibinfo {volume} {91}},\ \bibinfo {pages}
  {073506} (\bibinfo {year} {2020})},\ \Eprint
  {http://arxiv.org/abs/https://doi.org/10.1063/5.0002193}
  {https://doi.org/10.1063/5.0002193} \BibitemShut {NoStop}%
\bibitem [{\citenamefont {Drews}\ \emph {et~al.}(2021)\citenamefont {Drews},
  \citenamefont {Dittmar}, \citenamefont {Killer}, \citenamefont {Winters},
  \citenamefont {Kirschner}, \citenamefont {Brezinsek}, \citenamefont {Xu},
  \citenamefont {Wang}, \citenamefont {Jakubowski}, \citenamefont {Brunner},
  \citenamefont {Knauer}, \citenamefont {Grulke}, \citenamefont {Höschen},
  \citenamefont {Knieps}, \citenamefont {Nicolai}, \citenamefont {Neubauer},
  \citenamefont {Satheeswaran}, \citenamefont {Hirsch}, \citenamefont
  {Höfel},\ and\ \citenamefont {Liang}}]{Drews2021}%
  \BibitemOpen
  \bibfield  {author} {\bibinfo {author} {\bibfnamefont {P.}~\bibnamefont
  {Drews}}, \bibinfo {author} {\bibfnamefont {T.}~\bibnamefont {Dittmar}},
  \bibinfo {author} {\bibfnamefont {C.}~\bibnamefont {Killer}}, \bibinfo
  {author} {\bibfnamefont {V.}~\bibnamefont {Winters}}, \bibinfo {author}
  {\bibfnamefont {A.}~\bibnamefont {Kirschner}}, \bibinfo {author}
  {\bibfnamefont {S.}~\bibnamefont {Brezinsek}}, \bibinfo {author}
  {\bibfnamefont {S.}~\bibnamefont {Xu}}, \bibinfo {author} {\bibfnamefont
  {E.}~\bibnamefont {Wang}}, \bibinfo {author} {\bibfnamefont {M.}~\bibnamefont
  {Jakubowski}}, \bibinfo {author} {\bibfnamefont {K.}~\bibnamefont {Brunner}},
  \bibinfo {author} {\bibfnamefont {J.}~\bibnamefont {Knauer}}, \bibinfo
  {author} {\bibfnamefont {O.}~\bibnamefont {Grulke}}, \bibinfo {author}
  {\bibfnamefont {D.}~\bibnamefont {Höschen}}, \bibinfo {author}
  {\bibfnamefont {A.}~\bibnamefont {Knieps}}, \bibinfo {author} {\bibfnamefont
  {D.}~\bibnamefont {Nicolai}}, \bibinfo {author} {\bibfnamefont
  {O.}~\bibnamefont {Neubauer}}, \bibinfo {author} {\bibfnamefont
  {G.}~\bibnamefont {Satheeswaran}}, \bibinfo {author} {\bibfnamefont
  {M.}~\bibnamefont {Hirsch}}, \bibinfo {author} {\bibfnamefont
  {U.}~\bibnamefont {Höfel}}, \ and\ \bibinfo {author} {\bibfnamefont
  {Y.}~\bibnamefont {Liang}},\ }\href {\doibase
  https://doi.org/10.1016/j.fusengdes.2021.112786} {\bibfield  {journal}
  {\bibinfo  {journal} {Fusion Engineering and Design}\ }\textbf {\bibinfo
  {volume} {173}},\ \bibinfo {pages} {112786} (\bibinfo {year}
  {2021})}\BibitemShut {NoStop}%
\bibitem [{\citenamefont {Cai}\ \emph {et~al.}(2019)\citenamefont {Cai},
  \citenamefont {Liang}, \citenamefont {Killer}, \citenamefont {Liu},
  \citenamefont {Hiller}, \citenamefont {Knieps}, \citenamefont {Schweer},
  \citenamefont {Höschen}, \citenamefont {Nicolai}, \citenamefont
  {Offermanns}, \citenamefont {Satheeswaran}, \citenamefont {Henkel},
  \citenamefont {Hollfeld}, \citenamefont {Grulke}, \citenamefont {Drews},
  \citenamefont {Krings},\ and\ \citenamefont {Li}}]{Cai2019}%
  \BibitemOpen
  \bibfield  {author} {\bibinfo {author} {\bibfnamefont {J.}~\bibnamefont
  {Cai}}, \bibinfo {author} {\bibfnamefont {Y.}~\bibnamefont {Liang}}, \bibinfo
  {author} {\bibfnamefont {C.}~\bibnamefont {Killer}}, \bibinfo {author}
  {\bibfnamefont {S.}~\bibnamefont {Liu}}, \bibinfo {author} {\bibfnamefont
  {A.}~\bibnamefont {Hiller}}, \bibinfo {author} {\bibfnamefont
  {A.}~\bibnamefont {Knieps}}, \bibinfo {author} {\bibfnamefont
  {B.}~\bibnamefont {Schweer}}, \bibinfo {author} {\bibfnamefont
  {D.}~\bibnamefont {Höschen}}, \bibinfo {author} {\bibfnamefont
  {D.}~\bibnamefont {Nicolai}}, \bibinfo {author} {\bibfnamefont
  {G.}~\bibnamefont {Offermanns}}, \bibinfo {author} {\bibfnamefont
  {G.}~\bibnamefont {Satheeswaran}}, \bibinfo {author} {\bibfnamefont
  {M.}~\bibnamefont {Henkel}}, \bibinfo {author} {\bibfnamefont
  {K.}~\bibnamefont {Hollfeld}}, \bibinfo {author} {\bibfnamefont
  {O.}~\bibnamefont {Grulke}}, \bibinfo {author} {\bibfnamefont
  {P.}~\bibnamefont {Drews}}, \bibinfo {author} {\bibfnamefont
  {T.}~\bibnamefont {Krings}}, \ and\ \bibinfo {author} {\bibfnamefont
  {Y.}~\bibnamefont {Li}},\ }\href {\doibase 10.1063/1.5054279} {\bibfield
  {journal} {\bibinfo  {journal} {Review of Scientific Instruments}\ }\textbf
  {\bibinfo {volume} {90}},\ \bibinfo {pages} {033502} (\bibinfo {year}
  {2019})}\BibitemShut {NoStop}%
\bibitem [{\citenamefont {Henkel}\ \emph {et~al.}(2018)\citenamefont {Henkel},
  \citenamefont {Höschen}, \citenamefont {Liang}, \citenamefont {Li},
  \citenamefont {Liu}, \citenamefont {Nicolai}, \citenamefont {Sandri},
  \citenamefont {Satheeswaran}, \citenamefont {Yan}, \citenamefont {Zhang},\
  and\ \citenamefont {the EAST~team}}]{Henkel2018}%
  \BibitemOpen
  \bibfield  {author} {\bibinfo {author} {\bibfnamefont {M.}~\bibnamefont
  {Henkel}}, \bibinfo {author} {\bibfnamefont {D.}~\bibnamefont {Höschen}},
  \bibinfo {author} {\bibfnamefont {Y.}~\bibnamefont {Liang}}, \bibinfo
  {author} {\bibfnamefont {Y.}~\bibnamefont {Li}}, \bibinfo {author}
  {\bibfnamefont {S.}~\bibnamefont {Liu}}, \bibinfo {author} {\bibfnamefont
  {D.}~\bibnamefont {Nicolai}}, \bibinfo {author} {\bibfnamefont
  {N.}~\bibnamefont {Sandri}}, \bibinfo {author} {\bibfnamefont
  {G.}~\bibnamefont {Satheeswaran}}, \bibinfo {author} {\bibfnamefont
  {N.}~\bibnamefont {Yan}}, \bibinfo {author} {\bibfnamefont {H.~X.}\
  \bibnamefont {Zhang}}, \ and\ \bibinfo {author} {\bibnamefont {the
  EAST~team}},\ }\href {http://stacks.iop.org/1009-0630/20/i=5/a=054001}
  {\bibfield  {journal} {\bibinfo  {journal} {Plasma Science and Technology}\
  }\textbf {\bibinfo {volume} {20}},\ \bibinfo {pages} {054001} (\bibinfo
  {year} {2018})}\BibitemShut {NoStop}%
\bibitem [{\citenamefont {Li}\ \emph {et~al.}(2019)\citenamefont {Li},
  \citenamefont {Henkel}, \citenamefont {Liang}, \citenamefont {Knieps},
  \citenamefont {Drews}, \citenamefont {Killer}, \citenamefont {Nicolai},
  \citenamefont {Cosfeld}, \citenamefont {Geiger}, \citenamefont {Feng},
  \citenamefont {Effenberg}, \citenamefont {Zhang}, \citenamefont {Hacker},
  \citenamefont {Höschen}, \citenamefont {Satheeswaran}, \citenamefont {Liu},
  \citenamefont {Grulke}, \citenamefont {Jakubowski}, \citenamefont
  {Brezinsek}, \citenamefont {Otte}, \citenamefont {Neubauer}, \citenamefont
  {Schweer}, \citenamefont {Xu}, \citenamefont {Cai},\ and\ \citenamefont
  {and}}]{Li2019}%
  \BibitemOpen
  \bibfield  {author} {\bibinfo {author} {\bibfnamefont {Y.}~\bibnamefont
  {Li}}, \bibinfo {author} {\bibfnamefont {M.}~\bibnamefont {Henkel}}, \bibinfo
  {author} {\bibfnamefont {Y.}~\bibnamefont {Liang}}, \bibinfo {author}
  {\bibfnamefont {A.}~\bibnamefont {Knieps}}, \bibinfo {author} {\bibfnamefont
  {P.}~\bibnamefont {Drews}}, \bibinfo {author} {\bibfnamefont
  {C.}~\bibnamefont {Killer}}, \bibinfo {author} {\bibfnamefont
  {D.}~\bibnamefont {Nicolai}}, \bibinfo {author} {\bibfnamefont
  {J.}~\bibnamefont {Cosfeld}}, \bibinfo {author} {\bibfnamefont
  {J.}~\bibnamefont {Geiger}}, \bibinfo {author} {\bibfnamefont
  {Y.}~\bibnamefont {Feng}}, \bibinfo {author} {\bibfnamefont {F.}~\bibnamefont
  {Effenberg}}, \bibinfo {author} {\bibfnamefont {D.}~\bibnamefont {Zhang}},
  \bibinfo {author} {\bibfnamefont {P.}~\bibnamefont {Hacker}}, \bibinfo
  {author} {\bibfnamefont {D.}~\bibnamefont {Höschen}}, \bibinfo {author}
  {\bibfnamefont {G.}~\bibnamefont {Satheeswaran}}, \bibinfo {author}
  {\bibfnamefont {S.}~\bibnamefont {Liu}}, \bibinfo {author} {\bibfnamefont
  {O.}~\bibnamefont {Grulke}}, \bibinfo {author} {\bibfnamefont
  {M.}~\bibnamefont {Jakubowski}}, \bibinfo {author} {\bibfnamefont
  {S.}~\bibnamefont {Brezinsek}}, \bibinfo {author} {\bibfnamefont
  {M.}~\bibnamefont {Otte}}, \bibinfo {author} {\bibfnamefont {O.}~\bibnamefont
  {Neubauer}}, \bibinfo {author} {\bibfnamefont {B.}~\bibnamefont {Schweer}},
  \bibinfo {author} {\bibfnamefont {G.}~\bibnamefont {Xu}}, \bibinfo {author}
  {\bibfnamefont {J.}~\bibnamefont {Cai}}, \ and\ \bibinfo {author}
  {\bibfnamefont {Z.~H.}\ \bibnamefont {and}},\ }\href {\doibase
  10.1088/1741-4326/ab3a79} {\bibfield  {journal} {\bibinfo  {journal} {Nuclear
  Fusion}\ }\textbf {\bibinfo {volume} {59}},\ \bibinfo {pages} {126002}
  (\bibinfo {year} {2019})}\BibitemShut {NoStop}%
\bibitem [{\citenamefont {Killer}\ \emph
  {et~al.}(2019{\natexlab{a}})\citenamefont {Killer}, \citenamefont {Gao},
  \citenamefont {Perseo}, \citenamefont {Rudischhauser}, \citenamefont
  {Hammond}, \citenamefont {Buttenschön}, \citenamefont {Barbui},
  \citenamefont {Blackwell}, \citenamefont {Brunner}, \citenamefont {Drews},
  \citenamefont {Endler}, \citenamefont {Geiger}, \citenamefont {Grulke},
  \citenamefont {Jakubowski}, \citenamefont {Klose}, \citenamefont {Knauer},
  \citenamefont {Knieps}, \citenamefont {König}, \citenamefont {Li},
  \citenamefont {Neuner}, \citenamefont {Niemann}, \citenamefont {Otte},
  \citenamefont {Schilling}, \citenamefont {Sitjes}, \citenamefont {Rahbarnia},
  \citenamefont {Stange},\ and\ \citenamefont {Team}}]{Killer2019a}%
  \BibitemOpen
  \bibfield  {author} {\bibinfo {author} {\bibfnamefont {C.}~\bibnamefont
  {Killer}}, \bibinfo {author} {\bibfnamefont {Y.}~\bibnamefont {Gao}},
  \bibinfo {author} {\bibfnamefont {V.}~\bibnamefont {Perseo}}, \bibinfo
  {author} {\bibfnamefont {L.}~\bibnamefont {Rudischhauser}}, \bibinfo {author}
  {\bibfnamefont {K.}~\bibnamefont {Hammond}}, \bibinfo {author} {\bibfnamefont
  {B.}~\bibnamefont {Buttenschön}}, \bibinfo {author} {\bibfnamefont
  {T.}~\bibnamefont {Barbui}}, \bibinfo {author} {\bibfnamefont {B.~D.}\
  \bibnamefont {Blackwell}}, \bibinfo {author} {\bibfnamefont {K.-J.}\
  \bibnamefont {Brunner}}, \bibinfo {author} {\bibfnamefont {P.}~\bibnamefont
  {Drews}}, \bibinfo {author} {\bibfnamefont {M.}~\bibnamefont {Endler}},
  \bibinfo {author} {\bibfnamefont {J.}~\bibnamefont {Geiger}}, \bibinfo
  {author} {\bibfnamefont {O.}~\bibnamefont {Grulke}}, \bibinfo {author}
  {\bibfnamefont {M.}~\bibnamefont {Jakubowski}}, \bibinfo {author}
  {\bibfnamefont {S.}~\bibnamefont {Klose}}, \bibinfo {author} {\bibfnamefont
  {J.}~\bibnamefont {Knauer}}, \bibinfo {author} {\bibfnamefont
  {A.}~\bibnamefont {Knieps}}, \bibinfo {author} {\bibfnamefont
  {R.}~\bibnamefont {König}}, \bibinfo {author} {\bibfnamefont
  {Y.}~\bibnamefont {Li}}, \bibinfo {author} {\bibfnamefont {U.}~\bibnamefont
  {Neuner}}, \bibinfo {author} {\bibfnamefont {H.}~\bibnamefont {Niemann}},
  \bibinfo {author} {\bibfnamefont {M.}~\bibnamefont {Otte}}, \bibinfo {author}
  {\bibfnamefont {J.}~\bibnamefont {Schilling}}, \bibinfo {author}
  {\bibfnamefont {A.~P.}\ \bibnamefont {Sitjes}}, \bibinfo {author}
  {\bibfnamefont {K.}~\bibnamefont {Rahbarnia}}, \bibinfo {author}
  {\bibfnamefont {T.}~\bibnamefont {Stange}}, \ and\ \bibinfo {author}
  {\bibfnamefont {W.-X.}\ \bibnamefont {Team}},\ }\href {\doibase
  10.1088/1361-6587/ab4f2d} {\bibfield  {journal} {\bibinfo  {journal} {Plasma
  Physics and Controlled Fusion}\ }\textbf {\bibinfo {volume} {61}},\ \bibinfo
  {pages} {125014} (\bibinfo {year} {2019}{\natexlab{a}})}\BibitemShut
  {NoStop}%
\bibitem [{\citenamefont {Henkel}\ \emph {et~al.}(2020)\citenamefont {Henkel},
  \citenamefont {Li}, \citenamefont {Liang}, \citenamefont {Drews},
  \citenamefont {Knieps}, \citenamefont {Killer}, \citenamefont {Nicolai},
  \citenamefont {Höschen}, \citenamefont {Geiger}, \citenamefont {Xiao},
  \citenamefont {Sandri}, \citenamefont {Satheeswaran}, \citenamefont {Liu},
  \citenamefont {Grulke}, \citenamefont {Jakubowski}, \citenamefont
  {Brezinsek}, \citenamefont {Otte}, \citenamefont {Neubauer}, \citenamefont
  {Schweer}, \citenamefont {Xu},\ and\ \citenamefont {Cai}}]{Henkel2020}%
  \BibitemOpen
  \bibfield  {author} {\bibinfo {author} {\bibfnamefont {M.}~\bibnamefont
  {Henkel}}, \bibinfo {author} {\bibfnamefont {Y.}~\bibnamefont {Li}}, \bibinfo
  {author} {\bibfnamefont {Y.}~\bibnamefont {Liang}}, \bibinfo {author}
  {\bibfnamefont {P.}~\bibnamefont {Drews}}, \bibinfo {author} {\bibfnamefont
  {A.}~\bibnamefont {Knieps}}, \bibinfo {author} {\bibfnamefont
  {C.}~\bibnamefont {Killer}}, \bibinfo {author} {\bibfnamefont
  {D.}~\bibnamefont {Nicolai}}, \bibinfo {author} {\bibfnamefont
  {D.}~\bibnamefont {Höschen}}, \bibinfo {author} {\bibfnamefont
  {J.}~\bibnamefont {Geiger}}, \bibinfo {author} {\bibfnamefont
  {C.}~\bibnamefont {Xiao}}, \bibinfo {author} {\bibfnamefont {N.}~\bibnamefont
  {Sandri}}, \bibinfo {author} {\bibfnamefont {G.}~\bibnamefont
  {Satheeswaran}}, \bibinfo {author} {\bibfnamefont {S.}~\bibnamefont {Liu}},
  \bibinfo {author} {\bibfnamefont {O.}~\bibnamefont {Grulke}}, \bibinfo
  {author} {\bibfnamefont {M.}~\bibnamefont {Jakubowski}}, \bibinfo {author}
  {\bibfnamefont {S.}~\bibnamefont {Brezinsek}}, \bibinfo {author}
  {\bibfnamefont {M.}~\bibnamefont {Otte}}, \bibinfo {author} {\bibfnamefont
  {O.}~\bibnamefont {Neubauer}}, \bibinfo {author} {\bibfnamefont
  {B.}~\bibnamefont {Schweer}}, \bibinfo {author} {\bibfnamefont
  {G.}~\bibnamefont {Xu}}, \ and\ \bibinfo {author} {\bibfnamefont
  {J.}~\bibnamefont {Cai}},\ }\href {\doibase
  https://doi.org/10.1016/j.fusengdes.2020.111623} {\bibfield  {journal}
  {\bibinfo  {journal} {Fusion Engineering and Design}\ }\textbf {\bibinfo
  {volume} {157}},\ \bibinfo {pages} {111623} (\bibinfo {year}
  {2020})}\BibitemShut {NoStop}%
\bibitem [{\citenamefont {Killer}\ \emph
  {et~al.}(2019{\natexlab{b}})\citenamefont {Killer}, \citenamefont {Grulke},
  \citenamefont {Drews}, \citenamefont {Gao}, \citenamefont {Jakubowski},
  \citenamefont {Knieps}, \citenamefont {Nicolai}, \citenamefont {Niemann},
  \citenamefont {Sitjes}, \citenamefont {Satheeswaran},\ and\ \citenamefont
  {Team}}]{Killer2019}%
  \BibitemOpen
  \bibfield  {author} {\bibinfo {author} {\bibfnamefont {C.}~\bibnamefont
  {Killer}}, \bibinfo {author} {\bibfnamefont {O.}~\bibnamefont {Grulke}},
  \bibinfo {author} {\bibfnamefont {P.}~\bibnamefont {Drews}}, \bibinfo
  {author} {\bibfnamefont {Y.}~\bibnamefont {Gao}}, \bibinfo {author}
  {\bibfnamefont {M.}~\bibnamefont {Jakubowski}}, \bibinfo {author}
  {\bibfnamefont {A.}~\bibnamefont {Knieps}}, \bibinfo {author} {\bibfnamefont
  {D.}~\bibnamefont {Nicolai}}, \bibinfo {author} {\bibfnamefont
  {H.}~\bibnamefont {Niemann}}, \bibinfo {author} {\bibfnamefont {A.~P.}\
  \bibnamefont {Sitjes}}, \bibinfo {author} {\bibfnamefont {G.}~\bibnamefont
  {Satheeswaran}}, \ and\ \bibinfo {author} {\bibfnamefont {W.-X.}\
  \bibnamefont {Team}},\ }\href {\doibase 10.1088/1741-4326/ab2272} {\bibfield
  {journal} {\bibinfo  {journal} {Nuclear Fusion}\ }\textbf {\bibinfo {volume}
  {59}},\ \bibinfo {pages} {086013} (\bibinfo {year}
  {2019}{\natexlab{b}})}\BibitemShut {NoStop}%
\bibitem [{\citenamefont {Hammond}\ \emph {et~al.}(2019)\citenamefont
  {Hammond}, \citenamefont {Gao}, \citenamefont {Jakubowski}, \citenamefont
  {Killer}, \citenamefont {Niemann}, \citenamefont {Rudischhauser},
  \citenamefont {Ali}, \citenamefont {Andreeva}, \citenamefont {Blackwell},
  \citenamefont {Brunner}, \citenamefont {Cannas}, \citenamefont {Drewelow},
  \citenamefont {Drews}, \citenamefont {Endler}, \citenamefont {Feng},
  \citenamefont {Geiger}, \citenamefont {Grulke}, \citenamefont {Knauer},
  \citenamefont {Klose}, \citenamefont {Lazerson}, \citenamefont {Otte},
  \citenamefont {Pisano}, \citenamefont {Neuner}, \citenamefont {Sitjes},
  \citenamefont {Rahbarnia}, \citenamefont {Schilling}, \citenamefont
  {Thomsen},\ and\ \citenamefont {and}}]{Hammond2019}%
  \BibitemOpen
  \bibfield  {author} {\bibinfo {author} {\bibfnamefont {K.~C.}\ \bibnamefont
  {Hammond}}, \bibinfo {author} {\bibfnamefont {Y.}~\bibnamefont {Gao}},
  \bibinfo {author} {\bibfnamefont {M.}~\bibnamefont {Jakubowski}}, \bibinfo
  {author} {\bibfnamefont {C.}~\bibnamefont {Killer}}, \bibinfo {author}
  {\bibfnamefont {H.}~\bibnamefont {Niemann}}, \bibinfo {author} {\bibfnamefont
  {L.}~\bibnamefont {Rudischhauser}}, \bibinfo {author} {\bibfnamefont
  {A.}~\bibnamefont {Ali}}, \bibinfo {author} {\bibfnamefont {T.}~\bibnamefont
  {Andreeva}}, \bibinfo {author} {\bibfnamefont {B.~D.}\ \bibnamefont
  {Blackwell}}, \bibinfo {author} {\bibfnamefont {K.~J.}\ \bibnamefont
  {Brunner}}, \bibinfo {author} {\bibfnamefont {B.}~\bibnamefont {Cannas}},
  \bibinfo {author} {\bibfnamefont {P.}~\bibnamefont {Drewelow}}, \bibinfo
  {author} {\bibfnamefont {P.}~\bibnamefont {Drews}}, \bibinfo {author}
  {\bibfnamefont {M.}~\bibnamefont {Endler}}, \bibinfo {author} {\bibfnamefont
  {Y.}~\bibnamefont {Feng}}, \bibinfo {author} {\bibfnamefont {J.}~\bibnamefont
  {Geiger}}, \bibinfo {author} {\bibfnamefont {O.}~\bibnamefont {Grulke}},
  \bibinfo {author} {\bibfnamefont {J.}~\bibnamefont {Knauer}}, \bibinfo
  {author} {\bibfnamefont {S.}~\bibnamefont {Klose}}, \bibinfo {author}
  {\bibfnamefont {S.}~\bibnamefont {Lazerson}}, \bibinfo {author}
  {\bibfnamefont {M.}~\bibnamefont {Otte}}, \bibinfo {author} {\bibfnamefont
  {F.}~\bibnamefont {Pisano}}, \bibinfo {author} {\bibfnamefont
  {U.}~\bibnamefont {Neuner}}, \bibinfo {author} {\bibfnamefont {A.~P.}\
  \bibnamefont {Sitjes}}, \bibinfo {author} {\bibfnamefont {K.}~\bibnamefont
  {Rahbarnia}}, \bibinfo {author} {\bibfnamefont {J.}~\bibnamefont
  {Schilling}}, \bibinfo {author} {\bibfnamefont {H.}~\bibnamefont {Thomsen}},
  \ and\ \bibinfo {author} {\bibfnamefont {G.~A.~W.}\ \bibnamefont {and}},\
  }\href {\doibase 10.1088/1361-6587/ab4825} {\bibfield  {journal} {\bibinfo
  {journal} {Plasma Physics and Controlled Fusion}\ }\textbf {\bibinfo {volume}
  {61}},\ \bibinfo {pages} {125001} (\bibinfo {year} {2019})}\BibitemShut
  {NoStop}%
\bibitem [{\citenamefont {Killer}\ \emph {et~al.}(2020)\citenamefont {Killer},
  \citenamefont {Shanahan}, \citenamefont {Grulke}, \citenamefont {Endler},
  \citenamefont {Hammond}, \citenamefont {Rudischhauser},\ and\ \citenamefont
  {Team}}]{Killer2020}%
  \BibitemOpen
  \bibfield  {author} {\bibinfo {author} {\bibfnamefont {C.}~\bibnamefont
  {Killer}}, \bibinfo {author} {\bibfnamefont {B.}~\bibnamefont {Shanahan}},
  \bibinfo {author} {\bibfnamefont {O.}~\bibnamefont {Grulke}}, \bibinfo
  {author} {\bibfnamefont {M.}~\bibnamefont {Endler}}, \bibinfo {author}
  {\bibfnamefont {K.}~\bibnamefont {Hammond}}, \bibinfo {author} {\bibfnamefont
  {L.}~\bibnamefont {Rudischhauser}}, \ and\ \bibinfo {author} {\bibfnamefont
  {W.-X.}\ \bibnamefont {Team}},\ }\href {\doibase 10.1088/1361-6587/ab9313}
  {\bibfield  {journal} {\bibinfo  {journal} {Plasma Physics and Controlled
  Fusion}\ }\textbf {\bibinfo {volume} {62}},\ \bibinfo {pages} {085003}
  (\bibinfo {year} {2020})}\BibitemShut {NoStop}%
\bibitem [{\citenamefont {Perseo}\ \emph {et~al.}(2020)\citenamefont {Perseo},
  \citenamefont {Gradic}, \citenamefont {König}, \citenamefont {Ford},
  \citenamefont {Killer}, \citenamefont {Grulke},\ and\ \citenamefont
  {Ennis}}]{Perseo2020}%
  \BibitemOpen
  \bibfield  {author} {\bibinfo {author} {\bibfnamefont {V.}~\bibnamefont
  {Perseo}}, \bibinfo {author} {\bibfnamefont {D.}~\bibnamefont {Gradic}},
  \bibinfo {author} {\bibfnamefont {R.}~\bibnamefont {König}}, \bibinfo
  {author} {\bibfnamefont {O.~P.}\ \bibnamefont {Ford}}, \bibinfo {author}
  {\bibfnamefont {C.}~\bibnamefont {Killer}}, \bibinfo {author} {\bibfnamefont
  {O.}~\bibnamefont {Grulke}}, \ and\ \bibinfo {author} {\bibfnamefont {D.~A.}\
  \bibnamefont {Ennis}},\ }\href {\doibase 10.1063/1.5126098} {\bibfield
  {journal} {\bibinfo  {journal} {Review of Scientific Instruments}\ }\textbf
  {\bibinfo {volume} {91}},\ \bibinfo {pages} {013501} (\bibinfo {year}
  {2020})},\ \Eprint {http://arxiv.org/abs/https://doi.org/10.1063/1.5126098}
  {https://doi.org/10.1063/1.5126098} \BibitemShut {NoStop}%
\bibitem [{\citenamefont {Barbui}\ \emph {et~al.}(2020)\citenamefont {Barbui},
  \citenamefont {Krychowiak}, \citenamefont {Schmitz}, \citenamefont
  {Bozhenkov}, \citenamefont {Flom}, \citenamefont {Fuchert}, \citenamefont
  {Killer}, \citenamefont {König}, \citenamefont {Jakubowski}, \citenamefont
  {Burgos}, \citenamefont {Pasch}, \citenamefont {Scott},\ and\ \citenamefont
  {Team}}]{Barbui2020}%
  \BibitemOpen
  \bibfield  {author} {\bibinfo {author} {\bibfnamefont {T.}~\bibnamefont
  {Barbui}}, \bibinfo {author} {\bibfnamefont {M.}~\bibnamefont {Krychowiak}},
  \bibinfo {author} {\bibfnamefont {O.}~\bibnamefont {Schmitz}}, \bibinfo
  {author} {\bibfnamefont {S.}~\bibnamefont {Bozhenkov}}, \bibinfo {author}
  {\bibfnamefont {E.}~\bibnamefont {Flom}}, \bibinfo {author} {\bibfnamefont
  {G.}~\bibnamefont {Fuchert}}, \bibinfo {author} {\bibfnamefont
  {C.}~\bibnamefont {Killer}}, \bibinfo {author} {\bibfnamefont
  {R.}~\bibnamefont {König}}, \bibinfo {author} {\bibfnamefont
  {M.}~\bibnamefont {Jakubowski}}, \bibinfo {author} {\bibfnamefont {J.~M.}\
  \bibnamefont {Burgos}}, \bibinfo {author} {\bibfnamefont {E.}~\bibnamefont
  {Pasch}}, \bibinfo {author} {\bibfnamefont {E.}~\bibnamefont {Scott}}, \ and\
  \bibinfo {author} {\bibfnamefont {W.-X.}\ \bibnamefont {Team}},\ }\href
  {\doibase 10.1088/1741-4326/aba9eb} {\bibfield  {journal} {\bibinfo
  {journal} {Nuclear Fusion}\ }\textbf {\bibinfo {volume} {60}},\ \bibinfo
  {pages} {106014} (\bibinfo {year} {2020})}\BibitemShut {NoStop}%
\bibitem [{\citenamefont {Ballinger}\ \emph {et~al.}(2021)\citenamefont
  {Ballinger}, \citenamefont {Terry}, \citenamefont {Baek}, \citenamefont
  {Beurskens}, \citenamefont {Brunner}, \citenamefont {Fuchert}, \citenamefont
  {Knauer}, \citenamefont {Killer}, \citenamefont {Pasch}, \citenamefont
  {Rahbarnia}, \citenamefont {Schilling}, \citenamefont {Scott}, \citenamefont
  {{von Stechow}}, \citenamefont {Thomsen}, \citenamefont {Grulke},
  \citenamefont {Wurden}, \citenamefont {Kocsis}, \citenamefont {Szepesi},\
  and\ \citenamefont {Zsuga}}]{Ballinger2021}%
  \BibitemOpen
  \bibfield  {author} {\bibinfo {author} {\bibfnamefont {S.}~\bibnamefont
  {Ballinger}}, \bibinfo {author} {\bibfnamefont {J.}~\bibnamefont {Terry}},
  \bibinfo {author} {\bibfnamefont {S.}~\bibnamefont {Baek}}, \bibinfo {author}
  {\bibfnamefont {M.}~\bibnamefont {Beurskens}}, \bibinfo {author}
  {\bibfnamefont {K.}~\bibnamefont {Brunner}}, \bibinfo {author} {\bibfnamefont
  {G.}~\bibnamefont {Fuchert}}, \bibinfo {author} {\bibfnamefont
  {J.}~\bibnamefont {Knauer}}, \bibinfo {author} {\bibfnamefont
  {C.}~\bibnamefont {Killer}}, \bibinfo {author} {\bibfnamefont
  {E.}~\bibnamefont {Pasch}}, \bibinfo {author} {\bibfnamefont
  {K.}~\bibnamefont {Rahbarnia}}, \bibinfo {author} {\bibfnamefont
  {J.}~\bibnamefont {Schilling}}, \bibinfo {author} {\bibfnamefont
  {E.}~\bibnamefont {Scott}}, \bibinfo {author} {\bibfnamefont
  {A.}~\bibnamefont {{von Stechow}}}, \bibinfo {author} {\bibfnamefont
  {H.}~\bibnamefont {Thomsen}}, \bibinfo {author} {\bibfnamefont
  {O.}~\bibnamefont {Grulke}}, \bibinfo {author} {\bibfnamefont
  {G.}~\bibnamefont {Wurden}}, \bibinfo {author} {\bibfnamefont
  {G.}~\bibnamefont {Kocsis}}, \bibinfo {author} {\bibfnamefont
  {T.}~\bibnamefont {Szepesi}}, \ and\ \bibinfo {author} {\bibfnamefont
  {L.}~\bibnamefont {Zsuga}},\ }\href {\doibase
  https://doi.org/10.1016/j.nme.2021.100967} {\bibfield  {journal} {\bibinfo
  {journal} {Nuclear Materials and Energy}\ }\textbf {\bibinfo {volume} {27}},\
  \bibinfo {pages} {100967} (\bibinfo {year} {2021})}\BibitemShut {NoStop}%
\bibitem [{\citenamefont {Killer}\ \emph {et~al.}(2021)\citenamefont {Killer},
  \citenamefont {Narbutt}, \citenamefont {Grulke},\ and\ \citenamefont {the
  W7-X~Team}}]{Killer2021}%
  \BibitemOpen
  \bibfield  {author} {\bibinfo {author} {\bibfnamefont {C.}~\bibnamefont
  {Killer}}, \bibinfo {author} {\bibfnamefont {Y.}~\bibnamefont {Narbutt}},
  \bibinfo {author} {\bibfnamefont {O.}~\bibnamefont {Grulke}}, \ and\ \bibinfo
  {author} {\bibnamefont {the W7-X~Team}},\ }\href {\doibase
  10.1088/1741-4326/ac1ae3} {\bibfield  {journal} {\bibinfo  {journal} {Nuclear
  Fusion}\ }\textbf {\bibinfo {volume} {61}},\ \bibinfo {pages} {096038}
  (\bibinfo {year} {2021})}\BibitemShut {NoStop}%
\bibitem [{\citenamefont {Wegner}\ \emph {et~al.}(2018)\citenamefont {Wegner},
  \citenamefont {Geiger}, \citenamefont {Kunkel}, \citenamefont {Burhenn},
  \citenamefont {Schröder}, \citenamefont {Biedermann}, \citenamefont
  {Buttenschön}, \citenamefont {Cseh}, \citenamefont {Drews}, \citenamefont
  {Grulke}, \citenamefont {Hollfeld}, \citenamefont {Killer}, \citenamefont
  {Kocsis}, \citenamefont {Krings}, \citenamefont {Langenberg}, \citenamefont
  {Marchuk}, \citenamefont {Neuner}, \citenamefont {Nicolai}, \citenamefont
  {Offermanns}, \citenamefont {Pablant}, \citenamefont {Rahbarnia},
  \citenamefont {Satheeswaran}, \citenamefont {Schilling}, \citenamefont
  {Schweer}, \citenamefont {Szepesi},\ and\ \citenamefont
  {Thomsen}}]{Wegner2018}%
  \BibitemOpen
  \bibfield  {author} {\bibinfo {author} {\bibfnamefont {T.}~\bibnamefont
  {Wegner}}, \bibinfo {author} {\bibfnamefont {B.}~\bibnamefont {Geiger}},
  \bibinfo {author} {\bibfnamefont {F.}~\bibnamefont {Kunkel}}, \bibinfo
  {author} {\bibfnamefont {R.}~\bibnamefont {Burhenn}}, \bibinfo {author}
  {\bibfnamefont {T.}~\bibnamefont {Schröder}}, \bibinfo {author}
  {\bibfnamefont {C.}~\bibnamefont {Biedermann}}, \bibinfo {author}
  {\bibfnamefont {B.}~\bibnamefont {Buttenschön}}, \bibinfo {author}
  {\bibfnamefont {G.}~\bibnamefont {Cseh}}, \bibinfo {author} {\bibfnamefont
  {P.}~\bibnamefont {Drews}}, \bibinfo {author} {\bibfnamefont
  {O.}~\bibnamefont {Grulke}}, \bibinfo {author} {\bibfnamefont
  {K.}~\bibnamefont {Hollfeld}}, \bibinfo {author} {\bibfnamefont
  {C.}~\bibnamefont {Killer}}, \bibinfo {author} {\bibfnamefont
  {G.}~\bibnamefont {Kocsis}}, \bibinfo {author} {\bibfnamefont
  {T.}~\bibnamefont {Krings}}, \bibinfo {author} {\bibfnamefont
  {A.}~\bibnamefont {Langenberg}}, \bibinfo {author} {\bibfnamefont
  {O.}~\bibnamefont {Marchuk}}, \bibinfo {author} {\bibfnamefont
  {U.}~\bibnamefont {Neuner}}, \bibinfo {author} {\bibfnamefont
  {D.}~\bibnamefont {Nicolai}}, \bibinfo {author} {\bibfnamefont
  {G.}~\bibnamefont {Offermanns}}, \bibinfo {author} {\bibfnamefont {N.~A.}\
  \bibnamefont {Pablant}}, \bibinfo {author} {\bibfnamefont {K.}~\bibnamefont
  {Rahbarnia}}, \bibinfo {author} {\bibfnamefont {G.}~\bibnamefont
  {Satheeswaran}}, \bibinfo {author} {\bibfnamefont {J.}~\bibnamefont
  {Schilling}}, \bibinfo {author} {\bibfnamefont {B.}~\bibnamefont {Schweer}},
  \bibinfo {author} {\bibfnamefont {T.}~\bibnamefont {Szepesi}}, \ and\
  \bibinfo {author} {\bibfnamefont {H.}~\bibnamefont {Thomsen}},\ }\href
  {\doibase 10.1063/1.5037543} {\bibfield  {journal} {\bibinfo  {journal}
  {Review of Scientific Instruments}\ }\textbf {\bibinfo {volume} {89}},\
  \bibinfo {pages} {073505} (\bibinfo {year} {2018})}\BibitemShut {NoStop}%
\bibitem [{\citenamefont {{Th}omas Wegner}\ \emph {et~al.}(2020)\citenamefont
  {{Th}omas Wegner}, \citenamefont {Alcuson}, \citenamefont {Geiger},
  \citenamefont {von Stechow}, \citenamefont {Xanthopoulos}, \citenamefont
  {Angioni}, \citenamefont {Beurskens}, \citenamefont {Böttger}, \citenamefont
  {Bozhenkov}, \citenamefont {Brunner}, \citenamefont {Burhenn}, \citenamefont
  {Buttenschön}, \citenamefont {Damm}, \citenamefont {Edlund}, \citenamefont
  {Ford}, \citenamefont {Fuchert}, \citenamefont {Grulke}, \citenamefont
  {Huang}, \citenamefont {Knauer}, \citenamefont {Kunkel}, \citenamefont
  {Langenberg}, \citenamefont {Pablant}, \citenamefont {Pasch}, \citenamefont
  {Rahbarnia}, \citenamefont {Schilling}, \citenamefont {Thomsen},\ and\
  \citenamefont {Vano}}]{Wegner2020a}%
  \BibitemOpen
  \bibfield  {author} {\bibinfo {author} {\bibnamefont {{Th}omas Wegner}},
  \bibinfo {author} {\bibfnamefont {J.~A.}\ \bibnamefont {Alcuson}}, \bibinfo
  {author} {\bibfnamefont {B.}~\bibnamefont {Geiger}}, \bibinfo {author}
  {\bibfnamefont {A.}~\bibnamefont {von Stechow}}, \bibinfo {author}
  {\bibfnamefont {P.}~\bibnamefont {Xanthopoulos}}, \bibinfo {author}
  {\bibfnamefont {C.}~\bibnamefont {Angioni}}, \bibinfo {author} {\bibfnamefont
  {M.~N.~A.}\ \bibnamefont {Beurskens}}, \bibinfo {author} {\bibfnamefont
  {L.}~\bibnamefont {Böttger}}, \bibinfo {author} {\bibfnamefont {S.~A.}\
  \bibnamefont {Bozhenkov}}, \bibinfo {author} {\bibfnamefont {K.~J.}\
  \bibnamefont {Brunner}}, \bibinfo {author} {\bibfnamefont {R.}~\bibnamefont
  {Burhenn}}, \bibinfo {author} {\bibfnamefont {B.}~\bibnamefont
  {Buttenschön}}, \bibinfo {author} {\bibfnamefont {H.}~\bibnamefont {Damm}},
  \bibinfo {author} {\bibfnamefont {E.~M.}\ \bibnamefont {Edlund}}, \bibinfo
  {author} {\bibfnamefont {O.~P.}\ \bibnamefont {Ford}}, \bibinfo {author}
  {\bibfnamefont {G.}~\bibnamefont {Fuchert}}, \bibinfo {author} {\bibfnamefont
  {O.}~\bibnamefont {Grulke}}, \bibinfo {author} {\bibfnamefont
  {Z.}~\bibnamefont {Huang}}, \bibinfo {author} {\bibfnamefont {J.~P.}\
  \bibnamefont {Knauer}}, \bibinfo {author} {\bibfnamefont {F.}~\bibnamefont
  {Kunkel}}, \bibinfo {author} {\bibfnamefont {A.}~\bibnamefont {Langenberg}},
  \bibinfo {author} {\bibfnamefont {N.~A.}\ \bibnamefont {Pablant}}, \bibinfo
  {author} {\bibfnamefont {E.}~\bibnamefont {Pasch}}, \bibinfo {author}
  {\bibfnamefont {K.}~\bibnamefont {Rahbarnia}}, \bibinfo {author}
  {\bibfnamefont {J.}~\bibnamefont {Schilling}}, \bibinfo {author}
  {\bibfnamefont {H.}~\bibnamefont {Thomsen}}, \ and\ \bibinfo {author}
  {\bibfnamefont {L.}~\bibnamefont {Vano}},\ }\href {\doibase
  10.1088/1741-4326/abb869} {\bibfield  {journal} {\bibinfo  {journal} {Nuclear
  Fusion}\ } (\bibinfo {year} {2020}),\ 10.1088/1741-4326/abb869}\BibitemShut
  {NoStop}%
\bibitem [{\citenamefont {Wegner}\ \emph {et~al.}(2020)\citenamefont {Wegner},
  \citenamefont {Geiger}, \citenamefont {Foest}, \citenamefont {Jansen~van
  Vuuren}, \citenamefont {Winters}, \citenamefont {Biedermann}, \citenamefont
  {Burhenn}, \citenamefont {Buttenschön}, \citenamefont {Cseh}, \citenamefont
  {Joda}, \citenamefont {Kocsis}, \citenamefont {Kunkel}, \citenamefont
  {Quade}, \citenamefont {Schäfer}, \citenamefont {Schmitz},\ and\
  \citenamefont {Szepesi}}]{Wegner2020}%
  \BibitemOpen
  \bibfield  {author} {\bibinfo {author} {\bibfnamefont {T.}~\bibnamefont
  {Wegner}}, \bibinfo {author} {\bibfnamefont {B.}~\bibnamefont {Geiger}},
  \bibinfo {author} {\bibfnamefont {R.}~\bibnamefont {Foest}}, \bibinfo
  {author} {\bibfnamefont {A.}~\bibnamefont {Jansen~van Vuuren}}, \bibinfo
  {author} {\bibfnamefont {V.~R.}\ \bibnamefont {Winters}}, \bibinfo {author}
  {\bibfnamefont {C.}~\bibnamefont {Biedermann}}, \bibinfo {author}
  {\bibfnamefont {R.}~\bibnamefont {Burhenn}}, \bibinfo {author} {\bibfnamefont
  {B.}~\bibnamefont {Buttenschön}}, \bibinfo {author} {\bibfnamefont
  {G.}~\bibnamefont {Cseh}}, \bibinfo {author} {\bibfnamefont {I.}~\bibnamefont
  {Joda}}, \bibinfo {author} {\bibfnamefont {G.}~\bibnamefont {Kocsis}},
  \bibinfo {author} {\bibfnamefont {F.}~\bibnamefont {Kunkel}}, \bibinfo
  {author} {\bibfnamefont {A.}~\bibnamefont {Quade}}, \bibinfo {author}
  {\bibfnamefont {J.}~\bibnamefont {Schäfer}}, \bibinfo {author}
  {\bibfnamefont {O.}~\bibnamefont {Schmitz}}, \ and\ \bibinfo {author}
  {\bibfnamefont {T.}~\bibnamefont {Szepesi}},\ }\href {\doibase
  10.1063/1.5144943} {\bibfield  {journal} {\bibinfo  {journal} {Review of
  Scientific Instruments}\ }\textbf {\bibinfo {volume} {91}},\ \bibinfo {pages}
  {083503} (\bibinfo {year} {2020})},\ \Eprint
  {http://arxiv.org/abs/https://doi.org/10.1063/1.5144943}
  {https://doi.org/10.1063/1.5144943} \BibitemShut {NoStop}%
\bibitem [{\citenamefont {Ogawa}\ \emph {et~al.}(2019)\citenamefont {Ogawa},
  \citenamefont {Bozhenkov}, \citenamefont {Äkäslompolo}, \citenamefont
  {Killer}, \citenamefont {Grulke}, \citenamefont {Nicolai}, \citenamefont
  {Satheeswaran}, \citenamefont {Isobe}, \citenamefont {Osakabe}, \citenamefont
  {Yokoyama},\ and\ \citenamefont {Wolf}}]{Ogawa2019}%
  \BibitemOpen
  \bibfield  {author} {\bibinfo {author} {\bibfnamefont {K.}~\bibnamefont
  {Ogawa}}, \bibinfo {author} {\bibfnamefont {S.}~\bibnamefont {Bozhenkov}},
  \bibinfo {author} {\bibfnamefont {S.}~\bibnamefont {Äkäslompolo}}, \bibinfo
  {author} {\bibfnamefont {C.}~\bibnamefont {Killer}}, \bibinfo {author}
  {\bibfnamefont {O.}~\bibnamefont {Grulke}}, \bibinfo {author} {\bibfnamefont
  {D.}~\bibnamefont {Nicolai}}, \bibinfo {author} {\bibfnamefont
  {G.}~\bibnamefont {Satheeswaran}}, \bibinfo {author} {\bibfnamefont
  {M.}~\bibnamefont {Isobe}}, \bibinfo {author} {\bibfnamefont
  {M.}~\bibnamefont {Osakabe}}, \bibinfo {author} {\bibfnamefont
  {M.}~\bibnamefont {Yokoyama}}, \ and\ \bibinfo {author} {\bibfnamefont
  {R.}~\bibnamefont {Wolf}},\ }\href {\doibase 10.1088/1748-0221/14/09/c09021}
  {\bibfield  {journal} {\bibinfo  {journal} {Journal of Instrumentation}\
  }\textbf {\bibinfo {volume} {14}},\ \bibinfo {pages} {C09021} (\bibinfo
  {year} {2019})}\BibitemShut {NoStop}%
\bibitem [{\citenamefont {Äkäslompolo}\ \emph {et~al.}(2019)\citenamefont
  {Äkäslompolo}, \citenamefont {Drewelow}, \citenamefont {Gao}, \citenamefont
  {Ali}, \citenamefont {Biedermann}, \citenamefont {Bozhenkov}, \citenamefont
  {Dhard}, \citenamefont {Endler}, \citenamefont {Fellinger}, \citenamefont
  {Ford}, \citenamefont {Geiger}, \citenamefont {Geiger}, \citenamefont {den
  Harder}, \citenamefont {Hartmann}, \citenamefont {Hathiramani}, \citenamefont
  {Isobe}, \citenamefont {Jakubowski}, \citenamefont {Kazakov}, \citenamefont
  {Killer}, \citenamefont {Lazerson}, \citenamefont {Mayer}, \citenamefont
  {McNeely}, \citenamefont {Naujoks}, \citenamefont {Neelis}, \citenamefont
  {Kontula}, \citenamefont {Kurki-Suonio}, \citenamefont {Niemann},
  \citenamefont {Ogawa}, \citenamefont {Pisano}, \citenamefont {Poloskei},
  \citenamefont {Sitjes}, \citenamefont {Rahbarnia}, \citenamefont {Rust},
  \citenamefont {Schmitt}, \citenamefont {Sleczka}, \citenamefont {Vano},
  \citenamefont {van Vuuren}, \citenamefont {Wurden},\ and\ \citenamefont
  {Wolf}}]{Aekaeslompolo2019}%
  \BibitemOpen
  \bibfield  {author} {\bibinfo {author} {\bibfnamefont {S.}~\bibnamefont
  {Äkäslompolo}}, \bibinfo {author} {\bibfnamefont {P.}~\bibnamefont
  {Drewelow}}, \bibinfo {author} {\bibfnamefont {Y.}~\bibnamefont {Gao}},
  \bibinfo {author} {\bibfnamefont {A.}~\bibnamefont {Ali}}, \bibinfo {author}
  {\bibfnamefont {C.}~\bibnamefont {Biedermann}}, \bibinfo {author}
  {\bibfnamefont {S.}~\bibnamefont {Bozhenkov}}, \bibinfo {author}
  {\bibfnamefont {C.}~\bibnamefont {Dhard}}, \bibinfo {author} {\bibfnamefont
  {M.}~\bibnamefont {Endler}}, \bibinfo {author} {\bibfnamefont
  {J.}~\bibnamefont {Fellinger}}, \bibinfo {author} {\bibfnamefont
  {O.}~\bibnamefont {Ford}}, \bibinfo {author} {\bibfnamefont {B.}~\bibnamefont
  {Geiger}}, \bibinfo {author} {\bibfnamefont {J.}~\bibnamefont {Geiger}},
  \bibinfo {author} {\bibfnamefont {N.}~\bibnamefont {den Harder}}, \bibinfo
  {author} {\bibfnamefont {D.}~\bibnamefont {Hartmann}}, \bibinfo {author}
  {\bibfnamefont {D.}~\bibnamefont {Hathiramani}}, \bibinfo {author}
  {\bibfnamefont {M.}~\bibnamefont {Isobe}}, \bibinfo {author} {\bibfnamefont
  {M.}~\bibnamefont {Jakubowski}}, \bibinfo {author} {\bibfnamefont
  {Y.}~\bibnamefont {Kazakov}}, \bibinfo {author} {\bibfnamefont
  {C.}~\bibnamefont {Killer}}, \bibinfo {author} {\bibfnamefont
  {S.}~\bibnamefont {Lazerson}}, \bibinfo {author} {\bibfnamefont
  {M.}~\bibnamefont {Mayer}}, \bibinfo {author} {\bibfnamefont
  {P.}~\bibnamefont {McNeely}}, \bibinfo {author} {\bibfnamefont
  {D.}~\bibnamefont {Naujoks}}, \bibinfo {author} {\bibfnamefont
  {T.}~\bibnamefont {Neelis}}, \bibinfo {author} {\bibfnamefont
  {J.}~\bibnamefont {Kontula}}, \bibinfo {author} {\bibfnamefont
  {T.}~\bibnamefont {Kurki-Suonio}}, \bibinfo {author} {\bibfnamefont
  {H.}~\bibnamefont {Niemann}}, \bibinfo {author} {\bibfnamefont
  {K.}~\bibnamefont {Ogawa}}, \bibinfo {author} {\bibfnamefont
  {F.}~\bibnamefont {Pisano}}, \bibinfo {author} {\bibfnamefont
  {P.}~\bibnamefont {Poloskei}}, \bibinfo {author} {\bibfnamefont {A.~P.}\
  \bibnamefont {Sitjes}}, \bibinfo {author} {\bibfnamefont {K.}~\bibnamefont
  {Rahbarnia}}, \bibinfo {author} {\bibfnamefont {N.}~\bibnamefont {Rust}},
  \bibinfo {author} {\bibfnamefont {J.}~\bibnamefont {Schmitt}}, \bibinfo
  {author} {\bibfnamefont {M.}~\bibnamefont {Sleczka}}, \bibinfo {author}
  {\bibfnamefont {L.}~\bibnamefont {Vano}}, \bibinfo {author} {\bibfnamefont
  {A.}~\bibnamefont {van Vuuren}}, \bibinfo {author} {\bibfnamefont
  {G.}~\bibnamefont {Wurden}}, \ and\ \bibinfo {author} {\bibfnamefont
  {R.}~\bibnamefont {Wolf}},\ }\href {\doibase 10.1088/1748-0221/14/10/c10012}
  {\bibfield  {journal} {\bibinfo  {journal} {Journal of Instrumentation}\
  }\textbf {\bibinfo {volume} {14}},\ \bibinfo {pages} {C10012} (\bibinfo
  {year} {2019})}\BibitemShut {NoStop}%
\bibitem [{\citenamefont {Nagy}\ \emph {et~al.}(2019)\citenamefont {Nagy},
  \citenamefont {Bortolon}, \citenamefont {Gates}, \citenamefont {Gilson},
  \citenamefont {Killer}, \citenamefont {Klinger}, \citenamefont {Lunsford},
  \citenamefont {Maingi}, \citenamefont {Mansfield}, \citenamefont {Mauzey},
  \citenamefont {Nazikian}, \citenamefont {Roquemore},\ and\ \citenamefont
  {Wolfe}}]{Nagy2019}%
  \BibitemOpen
  \bibfield  {author} {\bibinfo {author} {\bibfnamefont {A.}~\bibnamefont
  {Nagy}}, \bibinfo {author} {\bibfnamefont {A.}~\bibnamefont {Bortolon}},
  \bibinfo {author} {\bibfnamefont {D.}~\bibnamefont {Gates}}, \bibinfo
  {author} {\bibfnamefont {E.}~\bibnamefont {Gilson}}, \bibinfo {author}
  {\bibfnamefont {C.}~\bibnamefont {Killer}}, \bibinfo {author} {\bibfnamefont
  {T.}~\bibnamefont {Klinger}}, \bibinfo {author} {\bibfnamefont
  {R.}~\bibnamefont {Lunsford}}, \bibinfo {author} {\bibfnamefont
  {R.}~\bibnamefont {Maingi}}, \bibinfo {author} {\bibfnamefont
  {D.}~\bibnamefont {Mansfield}}, \bibinfo {author} {\bibfnamefont
  {D.}~\bibnamefont {Mauzey}}, \bibinfo {author} {\bibfnamefont
  {R.}~\bibnamefont {Nazikian}}, \bibinfo {author} {\bibfnamefont
  {L.}~\bibnamefont {Roquemore}}, \ and\ \bibinfo {author} {\bibfnamefont
  {E.}~\bibnamefont {Wolfe}},\ }\href {\doibase
  https://doi.org/10.1016/j.fusengdes.2018.12.099} {\bibfield  {journal}
  {\bibinfo  {journal} {Fusion Engineering and Design}\ } (\bibinfo {year}
  {2019}),\ https://doi.org/10.1016/j.fusengdes.2018.12.099}\BibitemShut
  {NoStop}%
\bibitem [{\citenamefont {Lunsford}\ \emph {et~al.}(2021)\citenamefont
  {Lunsford}, \citenamefont {Killer}, \citenamefont {Nagy}, \citenamefont
  {Gates}, \citenamefont {Klinger}, \citenamefont {Dinklage}, \citenamefont
  {Satheeswaran}, \citenamefont {Kocsis}, \citenamefont {Lazerson},
  \citenamefont {Nespoli}, \citenamefont {Pablant}, \citenamefont {von
  Stechow}, \citenamefont {Alonso}, \citenamefont {Andreeva}, \citenamefont
  {Beurskens}, \citenamefont {Biedermann}, \citenamefont {Brezinsek},
  \citenamefont {Brunner}, \citenamefont {Buttenschön}, \citenamefont
  {Carralero}, \citenamefont {Cseh}, \citenamefont {Drewelow}, \citenamefont
  {Effenberg}, \citenamefont {Estrada}, \citenamefont {Ford}, \citenamefont
  {Grulke}, \citenamefont {Hergenhahn}, \citenamefont {Höefel}, \citenamefont
  {Knauer}, \citenamefont {Krause}, \citenamefont {Krychowiak}, \citenamefont
  {Kwak}, \citenamefont {Langenberg}, \citenamefont {Neuner}, \citenamefont
  {Nicolai}, \citenamefont {Pavone}, \citenamefont {Puig~Sitjes}, \citenamefont
  {Rahbarnia}, \citenamefont {Schilling}, \citenamefont {Svensson},
  \citenamefont {Szepesi}, \citenamefont {Thomsen}, \citenamefont {Wauters},
  \citenamefont {Windisch}, \citenamefont {Winters}, \citenamefont {Zhang},\
  and\ \citenamefont {Zsuga}}]{Lunsford2021}%
  \BibitemOpen
  \bibfield  {author} {\bibinfo {author} {\bibfnamefont {R.}~\bibnamefont
  {Lunsford}}, \bibinfo {author} {\bibfnamefont {C.}~\bibnamefont {Killer}},
  \bibinfo {author} {\bibfnamefont {A.}~\bibnamefont {Nagy}}, \bibinfo {author}
  {\bibfnamefont {D.~A.}\ \bibnamefont {Gates}}, \bibinfo {author}
  {\bibfnamefont {T.}~\bibnamefont {Klinger}}, \bibinfo {author} {\bibfnamefont
  {A.}~\bibnamefont {Dinklage}}, \bibinfo {author} {\bibfnamefont
  {G.}~\bibnamefont {Satheeswaran}}, \bibinfo {author} {\bibfnamefont
  {G.}~\bibnamefont {Kocsis}}, \bibinfo {author} {\bibfnamefont {S.~A.}\
  \bibnamefont {Lazerson}}, \bibinfo {author} {\bibfnamefont {F.}~\bibnamefont
  {Nespoli}}, \bibinfo {author} {\bibfnamefont {N.~A.}\ \bibnamefont
  {Pablant}}, \bibinfo {author} {\bibfnamefont {A.}~\bibnamefont {von
  Stechow}}, \bibinfo {author} {\bibfnamefont {A.}~\bibnamefont {Alonso}},
  \bibinfo {author} {\bibfnamefont {T.}~\bibnamefont {Andreeva}}, \bibinfo
  {author} {\bibfnamefont {M.}~\bibnamefont {Beurskens}}, \bibinfo {author}
  {\bibfnamefont {C.}~\bibnamefont {Biedermann}}, \bibinfo {author}
  {\bibfnamefont {S.}~\bibnamefont {Brezinsek}}, \bibinfo {author}
  {\bibfnamefont {K.~J.}\ \bibnamefont {Brunner}}, \bibinfo {author}
  {\bibfnamefont {B.}~\bibnamefont {Buttenschön}}, \bibinfo {author}
  {\bibfnamefont {D.}~\bibnamefont {Carralero}}, \bibinfo {author}
  {\bibfnamefont {G.}~\bibnamefont {Cseh}}, \bibinfo {author} {\bibfnamefont
  {P.}~\bibnamefont {Drewelow}}, \bibinfo {author} {\bibfnamefont
  {F.}~\bibnamefont {Effenberg}}, \bibinfo {author} {\bibfnamefont
  {T.}~\bibnamefont {Estrada}}, \bibinfo {author} {\bibfnamefont {O.~P.}\
  \bibnamefont {Ford}}, \bibinfo {author} {\bibfnamefont {O.}~\bibnamefont
  {Grulke}}, \bibinfo {author} {\bibfnamefont {U.}~\bibnamefont {Hergenhahn}},
  \bibinfo {author} {\bibfnamefont {U.}~\bibnamefont {Höefel}}, \bibinfo
  {author} {\bibfnamefont {J.}~\bibnamefont {Knauer}}, \bibinfo {author}
  {\bibfnamefont {M.}~\bibnamefont {Krause}}, \bibinfo {author} {\bibfnamefont
  {M.}~\bibnamefont {Krychowiak}}, \bibinfo {author} {\bibfnamefont
  {S.}~\bibnamefont {Kwak}}, \bibinfo {author} {\bibfnamefont {A.}~\bibnamefont
  {Langenberg}}, \bibinfo {author} {\bibfnamefont {U.}~\bibnamefont {Neuner}},
  \bibinfo {author} {\bibfnamefont {D.}~\bibnamefont {Nicolai}}, \bibinfo
  {author} {\bibfnamefont {A.}~\bibnamefont {Pavone}}, \bibinfo {author}
  {\bibfnamefont {A.}~\bibnamefont {Puig~Sitjes}}, \bibinfo {author}
  {\bibfnamefont {K.}~\bibnamefont {Rahbarnia}}, \bibinfo {author}
  {\bibfnamefont {J.}~\bibnamefont {Schilling}}, \bibinfo {author}
  {\bibfnamefont {J.}~\bibnamefont {Svensson}}, \bibinfo {author}
  {\bibfnamefont {T.}~\bibnamefont {Szepesi}}, \bibinfo {author} {\bibfnamefont
  {H.}~\bibnamefont {Thomsen}}, \bibinfo {author} {\bibfnamefont
  {T.}~\bibnamefont {Wauters}}, \bibinfo {author} {\bibfnamefont
  {T.}~\bibnamefont {Windisch}}, \bibinfo {author} {\bibfnamefont
  {V.}~\bibnamefont {Winters}}, \bibinfo {author} {\bibfnamefont
  {D.}~\bibnamefont {Zhang}}, \ and\ \bibinfo {author} {\bibfnamefont
  {L.}~\bibnamefont {Zsuga}},\ }\href {\doibase 10.1063/5.0047274} {\bibfield
  {journal} {\bibinfo  {journal} {Physics of Plasmas}\ }\textbf {\bibinfo
  {volume} {28}},\ \bibinfo {pages} {082506} (\bibinfo {year} {2021})},\
  \Eprint {http://arxiv.org/abs/https://doi.org/10.1063/5.0047274}
  {https://doi.org/10.1063/5.0047274} \BibitemShut {NoStop}%
\bibitem [{\citenamefont {Agostinetti}\ \emph {et~al.}(2018)\citenamefont
  {Agostinetti}, \citenamefont {Spolaore}, \citenamefont {Brombin},
  \citenamefont {Cervaro}, \citenamefont {Franchin}, \citenamefont {Grulke},
  \citenamefont {Killer}, \citenamefont {Martines}, \citenamefont {Moresco},
  \citenamefont {Peruzzo}, \citenamefont {Vianello},\ and\ \citenamefont
  {Visentin}}]{Agostinetti2018}%
  \BibitemOpen
  \bibfield  {author} {\bibinfo {author} {\bibfnamefont {P.}~\bibnamefont
  {Agostinetti}}, \bibinfo {author} {\bibfnamefont {M.}~\bibnamefont
  {Spolaore}}, \bibinfo {author} {\bibfnamefont {M.}~\bibnamefont {Brombin}},
  \bibinfo {author} {\bibfnamefont {V.}~\bibnamefont {Cervaro}}, \bibinfo
  {author} {\bibfnamefont {L.}~\bibnamefont {Franchin}}, \bibinfo {author}
  {\bibfnamefont {O.}~\bibnamefont {Grulke}}, \bibinfo {author} {\bibfnamefont
  {C.}~\bibnamefont {Killer}}, \bibinfo {author} {\bibfnamefont
  {E.}~\bibnamefont {Martines}}, \bibinfo {author} {\bibfnamefont
  {M.}~\bibnamefont {Moresco}}, \bibinfo {author} {\bibfnamefont
  {S.}~\bibnamefont {Peruzzo}}, \bibinfo {author} {\bibfnamefont
  {N.}~\bibnamefont {Vianello}}, \ and\ \bibinfo {author} {\bibfnamefont
  {M.}~\bibnamefont {Visentin}},\ }\href {\doibase 10.1109/TPS.2018.2799638}
  {\bibfield  {journal} {\bibinfo  {journal} {IEEE Transactions on Plasma
  Science}\ }\textbf {\bibinfo {volume} {46}},\ \bibinfo {pages} {1306}
  (\bibinfo {year} {2018})}\BibitemShut {NoStop}%
\bibitem [{\citenamefont {Spolaore}\ \emph {et~al.}(2019)\citenamefont
  {Spolaore}, \citenamefont {Agostinetti}, \citenamefont {Killer},
  \citenamefont {Moresco}, \citenamefont {Brombin}, \citenamefont {Cavazzana},
  \citenamefont {Ghirardelli}, \citenamefont {Grenfell}, \citenamefont
  {Grulke}, \citenamefont {Lazerson}, \citenamefont {Martines}, \citenamefont
  {Neubauer}, \citenamefont {Nicolai}, \citenamefont {Satheeswaran},
  \citenamefont {Schweer}, \citenamefont {Vianello},\ and\ \citenamefont
  {Visentin}}]{Spolaore2019}%
  \BibitemOpen
  \bibfield  {author} {\bibinfo {author} {\bibfnamefont {M.}~\bibnamefont
  {Spolaore}}, \bibinfo {author} {\bibfnamefont {P.}~\bibnamefont
  {Agostinetti}}, \bibinfo {author} {\bibfnamefont {C.}~\bibnamefont {Killer}},
  \bibinfo {author} {\bibfnamefont {M.}~\bibnamefont {Moresco}}, \bibinfo
  {author} {\bibfnamefont {M.}~\bibnamefont {Brombin}}, \bibinfo {author}
  {\bibfnamefont {R.}~\bibnamefont {Cavazzana}}, \bibinfo {author}
  {\bibfnamefont {R.}~\bibnamefont {Ghirardelli}}, \bibinfo {author}
  {\bibfnamefont {G.}~\bibnamefont {Grenfell}}, \bibinfo {author}
  {\bibfnamefont {O.}~\bibnamefont {Grulke}}, \bibinfo {author} {\bibfnamefont
  {S.}~\bibnamefont {Lazerson}}, \bibinfo {author} {\bibfnamefont
  {E.}~\bibnamefont {Martines}}, \bibinfo {author} {\bibfnamefont
  {O.}~\bibnamefont {Neubauer}}, \bibinfo {author} {\bibfnamefont
  {D.}~\bibnamefont {Nicolai}}, \bibinfo {author} {\bibfnamefont
  {G.}~\bibnamefont {Satheeswaran}}, \bibinfo {author} {\bibfnamefont
  {B.}~\bibnamefont {Schweer}}, \bibinfo {author} {\bibfnamefont
  {N.}~\bibnamefont {Vianello}}, \ and\ \bibinfo {author} {\bibfnamefont
  {M.}~\bibnamefont {Visentin}},\ }\href {\doibase
  10.1088/1748-0221/14/09/c09035} {\bibfield  {journal} {\bibinfo  {journal}
  {Journal of Instrumentation}\ }\textbf {\bibinfo {volume} {14}},\ \bibinfo
  {pages} {C09035} (\bibinfo {year} {2019})}\BibitemShut {NoStop}%
\bibitem [{\citenamefont {Lazerson}\ \emph {et~al.}(2019)\citenamefont
  {Lazerson}, \citenamefont {Gao}, \citenamefont {Hammond}, \citenamefont
  {Killer}, \citenamefont {Schlisio}, \citenamefont {Otte}, \citenamefont
  {Biedermann}, \citenamefont {Spolaore}, \citenamefont {Bozhenkov},
  \citenamefont {Geiger}, \citenamefont {Grulke}, \citenamefont {Nicolai},
  \citenamefont {Satheeswaran}, \citenamefont {Niemann}, \citenamefont
  {Jakubowski}, \citenamefont {Drewelow}, \citenamefont {Sitjes}, \citenamefont
  {Ali}, \citenamefont {Cannas}, \citenamefont {Pisano}, \citenamefont
  {König}, \citenamefont {Wurden}, \citenamefont {Kocsis}, \citenamefont
  {Szepesi}, \citenamefont {Wenzel}, \citenamefont {Mulsow}, \citenamefont
  {Rahbarnia}, \citenamefont {Schilling}, \citenamefont {Neuner}, \citenamefont
  {Andreeva}, \citenamefont {Thomsen}, \citenamefont {Knauer}, \citenamefont
  {Brunner}, \citenamefont {Blackwell}, \citenamefont {Endler}, \citenamefont
  {Klose},\ and\ \citenamefont {and}}]{Lazerson2019}%
  \BibitemOpen
  \bibfield  {author} {\bibinfo {author} {\bibfnamefont {S.~A.}\ \bibnamefont
  {Lazerson}}, \bibinfo {author} {\bibfnamefont {Y.}~\bibnamefont {Gao}},
  \bibinfo {author} {\bibfnamefont {K.}~\bibnamefont {Hammond}}, \bibinfo
  {author} {\bibfnamefont {C.}~\bibnamefont {Killer}}, \bibinfo {author}
  {\bibfnamefont {G.}~\bibnamefont {Schlisio}}, \bibinfo {author}
  {\bibfnamefont {M.}~\bibnamefont {Otte}}, \bibinfo {author} {\bibfnamefont
  {C.}~\bibnamefont {Biedermann}}, \bibinfo {author} {\bibfnamefont
  {M.}~\bibnamefont {Spolaore}}, \bibinfo {author} {\bibfnamefont
  {S.}~\bibnamefont {Bozhenkov}}, \bibinfo {author} {\bibfnamefont
  {J.}~\bibnamefont {Geiger}}, \bibinfo {author} {\bibfnamefont
  {O.}~\bibnamefont {Grulke}}, \bibinfo {author} {\bibfnamefont
  {D.}~\bibnamefont {Nicolai}}, \bibinfo {author} {\bibfnamefont
  {G.}~\bibnamefont {Satheeswaran}}, \bibinfo {author} {\bibfnamefont
  {H.}~\bibnamefont {Niemann}}, \bibinfo {author} {\bibfnamefont
  {M.}~\bibnamefont {Jakubowski}}, \bibinfo {author} {\bibfnamefont
  {P.}~\bibnamefont {Drewelow}}, \bibinfo {author} {\bibfnamefont {A.~P.}\
  \bibnamefont {Sitjes}}, \bibinfo {author} {\bibfnamefont {A.}~\bibnamefont
  {Ali}}, \bibinfo {author} {\bibfnamefont {B.}~\bibnamefont {Cannas}},
  \bibinfo {author} {\bibfnamefont {F.}~\bibnamefont {Pisano}}, \bibinfo
  {author} {\bibfnamefont {R.}~\bibnamefont {König}}, \bibinfo {author}
  {\bibfnamefont {G.}~\bibnamefont {Wurden}}, \bibinfo {author} {\bibfnamefont
  {G.}~\bibnamefont {Kocsis}}, \bibinfo {author} {\bibfnamefont
  {T.}~\bibnamefont {Szepesi}}, \bibinfo {author} {\bibfnamefont
  {U.}~\bibnamefont {Wenzel}}, \bibinfo {author} {\bibfnamefont
  {M.}~\bibnamefont {Mulsow}}, \bibinfo {author} {\bibfnamefont
  {K.}~\bibnamefont {Rahbarnia}}, \bibinfo {author} {\bibfnamefont
  {J.}~\bibnamefont {Schilling}}, \bibinfo {author} {\bibfnamefont
  {U.}~\bibnamefont {Neuner}}, \bibinfo {author} {\bibfnamefont
  {T.}~\bibnamefont {Andreeva}}, \bibinfo {author} {\bibfnamefont
  {H.}~\bibnamefont {Thomsen}}, \bibinfo {author} {\bibfnamefont
  {J.}~\bibnamefont {Knauer}}, \bibinfo {author} {\bibfnamefont {K.~J.}\
  \bibnamefont {Brunner}}, \bibinfo {author} {\bibfnamefont {B.}~\bibnamefont
  {Blackwell}}, \bibinfo {author} {\bibfnamefont {M.}~\bibnamefont {Endler}},
  \bibinfo {author} {\bibfnamefont {S.}~\bibnamefont {Klose}}, \ and\ \bibinfo
  {author} {\bibfnamefont {L.~R.}\ \bibnamefont {and}},\ }\href {\doibase
  10.1088/1741-4326/ab3df0} {\bibfield  {journal} {\bibinfo  {journal} {Nuclear
  Fusion}\ }\textbf {\bibinfo {volume} {59}},\ \bibinfo {pages} {126004}
  (\bibinfo {year} {2019})}\BibitemShut {NoStop}%
\bibitem [{\citenamefont {Nicolai}\ \emph {et~al.}(2017)\citenamefont
  {Nicolai}, \citenamefont {Borsuk}, \citenamefont {Drews}, \citenamefont
  {Grulke}, \citenamefont {Hollfeld}, \citenamefont {Krings}, \citenamefont
  {Liang}, \citenamefont {Linsmeier}, \citenamefont {Neubauer}, \citenamefont
  {Satheeswaran}, \citenamefont {Schweer},\ and\ \citenamefont
  {Offermanns}}]{Nicolai2017}%
  \BibitemOpen
  \bibfield  {author} {\bibinfo {author} {\bibfnamefont {D.}~\bibnamefont
  {Nicolai}}, \bibinfo {author} {\bibfnamefont {V.}~\bibnamefont {Borsuk}},
  \bibinfo {author} {\bibfnamefont {P.}~\bibnamefont {Drews}}, \bibinfo
  {author} {\bibfnamefont {O.}~\bibnamefont {Grulke}}, \bibinfo {author}
  {\bibfnamefont {K.}~\bibnamefont {Hollfeld}}, \bibinfo {author}
  {\bibfnamefont {T.}~\bibnamefont {Krings}}, \bibinfo {author} {\bibfnamefont
  {Y.}~\bibnamefont {Liang}}, \bibinfo {author} {\bibfnamefont
  {C.}~\bibnamefont {Linsmeier}}, \bibinfo {author} {\bibfnamefont
  {O.}~\bibnamefont {Neubauer}}, \bibinfo {author} {\bibfnamefont
  {G.}~\bibnamefont {Satheeswaran}}, \bibinfo {author} {\bibfnamefont
  {B.}~\bibnamefont {Schweer}}, \ and\ \bibinfo {author} {\bibfnamefont
  {G.}~\bibnamefont {Offermanns}},\ }\href {\doibase
  https://doi.org/10.1016/j.fusengdes.2017.03.013} {\bibfield  {journal}
  {\bibinfo  {journal} {Fusion Engineering and Design}\ }\textbf {\bibinfo
  {volume} {123}},\ \bibinfo {pages} {960 } (\bibinfo {year} {2017})},\
  \bibinfo {note} {proceedings of the 29th Symposium on Fusion Technology
  (SOFT-29) Prague, Czech Republic, September 5-9, 2016}\BibitemShut {NoStop}%
\bibitem [{\citenamefont {Satheeswaran}\ \emph {et~al.}(2017)\citenamefont
  {Satheeswaran}, \citenamefont {Hollfeld}, \citenamefont {Drews},
  \citenamefont {Nicolai}, \citenamefont {Neubauer}, \citenamefont {Schweer},\
  and\ \citenamefont {Grulke}}]{Satheeswaran2017}%
  \BibitemOpen
  \bibfield  {author} {\bibinfo {author} {\bibfnamefont {G.}~\bibnamefont
  {Satheeswaran}}, \bibinfo {author} {\bibfnamefont {K.}~\bibnamefont
  {Hollfeld}}, \bibinfo {author} {\bibfnamefont {P.}~\bibnamefont {Drews}},
  \bibinfo {author} {\bibfnamefont {D.}~\bibnamefont {Nicolai}}, \bibinfo
  {author} {\bibfnamefont {O.}~\bibnamefont {Neubauer}}, \bibinfo {author}
  {\bibfnamefont {B.}~\bibnamefont {Schweer}}, \ and\ \bibinfo {author}
  {\bibfnamefont {O.}~\bibnamefont {Grulke}},\ }\href {\doibase
  https://doi.org/10.1016/j.fusengdes.2017.05.125} {\bibfield  {journal}
  {\bibinfo  {journal} {Fusion Engineering and Design}\ }\textbf {\bibinfo
  {volume} {123}},\ \bibinfo {pages} {699 } (\bibinfo {year} {2017})},\
  \bibinfo {note} {proceedings of the 29th Symposium on Fusion Technology
  (SOFT-29) Prague, Czech Republic, September 5-9, 2016}\BibitemShut {NoStop}%
\bibitem [{\citenamefont {Schacht}\ \emph {et~al.}(2019)\citenamefont
  {Schacht}, \citenamefont {Laqua}, \citenamefont {Müller}, \citenamefont
  {Puttnies},\ and\ \citenamefont {Skodzik}}]{Schacht2019}%
  \BibitemOpen
  \bibfield  {author} {\bibinfo {author} {\bibfnamefont {J.}~\bibnamefont
  {Schacht}}, \bibinfo {author} {\bibfnamefont {H.}~\bibnamefont {Laqua}},
  \bibinfo {author} {\bibfnamefont {I.}~\bibnamefont {Müller}}, \bibinfo
  {author} {\bibfnamefont {H.}~\bibnamefont {Puttnies}}, \ and\ \bibinfo
  {author} {\bibfnamefont {J.}~\bibnamefont {Skodzik}},\ }\href {\doibase
  10.1109/TNS.2019.2913802} {\bibfield  {journal} {\bibinfo  {journal} {IEEE
  Transactions on Nuclear Science}\ }\textbf {\bibinfo {volume} {66}},\
  \bibinfo {pages} {969} (\bibinfo {year} {2019})}\BibitemShut {NoStop}%
\bibitem [{\citenamefont {Jakubowski}\ \emph {et~al.}(2021)\citenamefont
  {Jakubowski}, \citenamefont {Endler}, \citenamefont {Feng}, \citenamefont
  {Gao}, \citenamefont {Killer}, \citenamefont {König}, \citenamefont
  {Krychowiak}, \citenamefont {Perseo}, \citenamefont {Reimold}, \citenamefont
  {Schmitz}, \citenamefont {Pedersen}, \citenamefont {Brezinsek}, \citenamefont
  {Dinklage}, \citenamefont {Drewelow}, \citenamefont {Niemann}, \citenamefont
  {Otte}, \citenamefont {Gruca}, \citenamefont {Hammond}, \citenamefont
  {Kremeyer}, \citenamefont {Kubkowska}, \citenamefont {Jab{\l}o{\'{n}}ski},
  \citenamefont {Pandey}, \citenamefont {Wurden}, \citenamefont {Zhang},
  \citenamefont {Bozhenkov}, \citenamefont {Böckenhoff}, \citenamefont
  {Dhard}, \citenamefont {Baldzuhn}, \citenamefont {Gradic}, \citenamefont
  {Effenberg}, \citenamefont {Kornejew}, \citenamefont {Lazerson},
  \citenamefont {Lore}, \citenamefont {Naujoks}, \citenamefont {Sitjes},
  \citenamefont {Schlisio}, \citenamefont {Sleczka}, \citenamefont {Wenzel},
  \citenamefont {Winters},\ and\ \citenamefont {Team}}]{Jakubowski2021}%
  \BibitemOpen
  \bibfield  {author} {\bibinfo {author} {\bibfnamefont {M.}~\bibnamefont
  {Jakubowski}}, \bibinfo {author} {\bibfnamefont {M.}~\bibnamefont {Endler}},
  \bibinfo {author} {\bibfnamefont {Y.}~\bibnamefont {Feng}}, \bibinfo {author}
  {\bibfnamefont {Y.}~\bibnamefont {Gao}}, \bibinfo {author} {\bibfnamefont
  {C.}~\bibnamefont {Killer}}, \bibinfo {author} {\bibfnamefont
  {R.}~\bibnamefont {König}}, \bibinfo {author} {\bibfnamefont
  {M.}~\bibnamefont {Krychowiak}}, \bibinfo {author} {\bibfnamefont
  {V.}~\bibnamefont {Perseo}}, \bibinfo {author} {\bibfnamefont
  {F.}~\bibnamefont {Reimold}}, \bibinfo {author} {\bibfnamefont
  {O.}~\bibnamefont {Schmitz}}, \bibinfo {author} {\bibfnamefont
  {T.}~\bibnamefont {Pedersen}}, \bibinfo {author} {\bibfnamefont
  {S.}~\bibnamefont {Brezinsek}}, \bibinfo {author} {\bibfnamefont
  {A.}~\bibnamefont {Dinklage}}, \bibinfo {author} {\bibfnamefont
  {P.}~\bibnamefont {Drewelow}}, \bibinfo {author} {\bibfnamefont
  {H.}~\bibnamefont {Niemann}}, \bibinfo {author} {\bibfnamefont
  {M.}~\bibnamefont {Otte}}, \bibinfo {author} {\bibfnamefont {M.}~\bibnamefont
  {Gruca}}, \bibinfo {author} {\bibfnamefont {K.}~\bibnamefont {Hammond}},
  \bibinfo {author} {\bibfnamefont {T.}~\bibnamefont {Kremeyer}}, \bibinfo
  {author} {\bibfnamefont {M.}~\bibnamefont {Kubkowska}}, \bibinfo {author}
  {\bibfnamefont {S.}~\bibnamefont {Jab{\l}o{\'{n}}ski}}, \bibinfo {author}
  {\bibfnamefont {A.}~\bibnamefont {Pandey}}, \bibinfo {author} {\bibfnamefont
  {G.}~\bibnamefont {Wurden}}, \bibinfo {author} {\bibfnamefont
  {D.}~\bibnamefont {Zhang}}, \bibinfo {author} {\bibfnamefont
  {S.}~\bibnamefont {Bozhenkov}}, \bibinfo {author} {\bibfnamefont
  {D.}~\bibnamefont {Böckenhoff}}, \bibinfo {author} {\bibfnamefont
  {C.}~\bibnamefont {Dhard}}, \bibinfo {author} {\bibfnamefont
  {J.}~\bibnamefont {Baldzuhn}}, \bibinfo {author} {\bibfnamefont
  {D.}~\bibnamefont {Gradic}}, \bibinfo {author} {\bibfnamefont
  {F.}~\bibnamefont {Effenberg}}, \bibinfo {author} {\bibfnamefont
  {P.}~\bibnamefont {Kornejew}}, \bibinfo {author} {\bibfnamefont
  {S.}~\bibnamefont {Lazerson}}, \bibinfo {author} {\bibfnamefont
  {J.}~\bibnamefont {Lore}}, \bibinfo {author} {\bibfnamefont {D.}~\bibnamefont
  {Naujoks}}, \bibinfo {author} {\bibfnamefont {A.~P.}\ \bibnamefont {Sitjes}},
  \bibinfo {author} {\bibfnamefont {G.}~\bibnamefont {Schlisio}}, \bibinfo
  {author} {\bibfnamefont {M.}~\bibnamefont {Sleczka}}, \bibinfo {author}
  {\bibfnamefont {U.}~\bibnamefont {Wenzel}}, \bibinfo {author} {\bibfnamefont
  {V.}~\bibnamefont {Winters}}, \ and\ \bibinfo {author} {\bibfnamefont
  {W.-X.}\ \bibnamefont {Team}},\ }\href {\doibase 10.1088/1741-4326/ac1b68}
  {\bibfield  {journal} {\bibinfo  {journal} {Nuclear Fusion}\ }\textbf
  {\bibinfo {volume} {61}},\ \bibinfo {pages} {106003} (\bibinfo {year}
  {2021})}\BibitemShut {NoStop}%
\bibitem [{\citenamefont {Renner}\ \emph {et~al.}(2000)\citenamefont {Renner},
  \citenamefont {Boscary}, \citenamefont {Erckmann}, \citenamefont {Greuner},
  \citenamefont {Grote}, \citenamefont {Sapper}, \citenamefont {Speth},
  \citenamefont {Wesner}, \citenamefont {Wanner},\ and\ \citenamefont
  {Team}}]{Renner2000}%
  \BibitemOpen
  \bibfield  {author} {\bibinfo {author} {\bibfnamefont {H.}~\bibnamefont
  {Renner}}, \bibinfo {author} {\bibfnamefont {J.}~\bibnamefont {Boscary}},
  \bibinfo {author} {\bibfnamefont {V.}~\bibnamefont {Erckmann}}, \bibinfo
  {author} {\bibfnamefont {H.}~\bibnamefont {Greuner}}, \bibinfo {author}
  {\bibfnamefont {H.}~\bibnamefont {Grote}}, \bibinfo {author} {\bibfnamefont
  {J.}~\bibnamefont {Sapper}}, \bibinfo {author} {\bibfnamefont
  {E.}~\bibnamefont {Speth}}, \bibinfo {author} {\bibfnamefont
  {F.}~\bibnamefont {Wesner}}, \bibinfo {author} {\bibfnamefont
  {M.}~\bibnamefont {Wanner}}, \ and\ \bibinfo {author} {\bibfnamefont {W.-X.}\
  \bibnamefont {Team}},\ }\href {\doibase 10.1088/0029-5515/40/6/306}
  {\bibfield  {journal} {\bibinfo  {journal} {Nuclear Fusion}\ }\textbf
  {\bibinfo {volume} {40}},\ \bibinfo {pages} {1083} (\bibinfo {year}
  {2000})}\BibitemShut {NoStop}%
\bibitem [{\citenamefont {Gao}\ \emph {et~al.}(2019)\citenamefont {Gao},
  \citenamefont {Geiger}, \citenamefont {Jakubowski}, \citenamefont {Drewelow},
  \citenamefont {Endler}, \citenamefont {Rahbarnia}, \citenamefont {Bozhenkov},
  \citenamefont {Otte}, \citenamefont {Suzuki}, \citenamefont {Feng},
  \citenamefont {Niemann}, \citenamefont {Pisano}, \citenamefont {Ali},
  \citenamefont {Sitjes}, \citenamefont {Zanini}, \citenamefont {Laqua},
  \citenamefont {Stange}, \citenamefont {Marsen}, \citenamefont {Szepesi},
  \citenamefont {Zhang}, \citenamefont {Killer}, \citenamefont {Hammond},
  \citenamefont {Lazerson}, \citenamefont {Cannas}, \citenamefont {Thomsen},
  \citenamefont {Andreeva}, \citenamefont {Neuner}, \citenamefont {Schilling},
  \citenamefont {Knieps}, \citenamefont {Rack},\ and\ \citenamefont
  {and}}]{Gao2019a}%
  \BibitemOpen
  \bibfield  {author} {\bibinfo {author} {\bibfnamefont {Y.}~\bibnamefont
  {Gao}}, \bibinfo {author} {\bibfnamefont {J.}~\bibnamefont {Geiger}},
  \bibinfo {author} {\bibfnamefont {M.~W.}\ \bibnamefont {Jakubowski}},
  \bibinfo {author} {\bibfnamefont {P.}~\bibnamefont {Drewelow}}, \bibinfo
  {author} {\bibfnamefont {M.}~\bibnamefont {Endler}}, \bibinfo {author}
  {\bibfnamefont {K.}~\bibnamefont {Rahbarnia}}, \bibinfo {author}
  {\bibfnamefont {S.}~\bibnamefont {Bozhenkov}}, \bibinfo {author}
  {\bibfnamefont {M.}~\bibnamefont {Otte}}, \bibinfo {author} {\bibfnamefont
  {Y.}~\bibnamefont {Suzuki}}, \bibinfo {author} {\bibfnamefont
  {Y.}~\bibnamefont {Feng}}, \bibinfo {author} {\bibfnamefont {H.}~\bibnamefont
  {Niemann}}, \bibinfo {author} {\bibfnamefont {F.}~\bibnamefont {Pisano}},
  \bibinfo {author} {\bibfnamefont {A.}~\bibnamefont {Ali}}, \bibinfo {author}
  {\bibfnamefont {A.~P.}\ \bibnamefont {Sitjes}}, \bibinfo {author}
  {\bibfnamefont {M.}~\bibnamefont {Zanini}}, \bibinfo {author} {\bibfnamefont
  {H.}~\bibnamefont {Laqua}}, \bibinfo {author} {\bibfnamefont
  {T.}~\bibnamefont {Stange}}, \bibinfo {author} {\bibfnamefont
  {S.}~\bibnamefont {Marsen}}, \bibinfo {author} {\bibfnamefont
  {T.}~\bibnamefont {Szepesi}}, \bibinfo {author} {\bibfnamefont
  {D.}~\bibnamefont {Zhang}}, \bibinfo {author} {\bibfnamefont
  {C.}~\bibnamefont {Killer}}, \bibinfo {author} {\bibfnamefont
  {K.}~\bibnamefont {Hammond}}, \bibinfo {author} {\bibfnamefont
  {S.}~\bibnamefont {Lazerson}}, \bibinfo {author} {\bibfnamefont
  {B.}~\bibnamefont {Cannas}}, \bibinfo {author} {\bibfnamefont
  {H.}~\bibnamefont {Thomsen}}, \bibinfo {author} {\bibfnamefont
  {T.}~\bibnamefont {Andreeva}}, \bibinfo {author} {\bibfnamefont
  {U.}~\bibnamefont {Neuner}}, \bibinfo {author} {\bibfnamefont
  {J.}~\bibnamefont {Schilling}}, \bibinfo {author} {\bibfnamefont
  {A.}~\bibnamefont {Knieps}}, \bibinfo {author} {\bibfnamefont
  {M.}~\bibnamefont {Rack}}, \ and\ \bibinfo {author} {\bibfnamefont {Y.~L.}\
  \bibnamefont {and}},\ }\href {\doibase 10.1088/1741-4326/ab32c2} {\bibfield
  {journal} {\bibinfo  {journal} {Nuclear Fusion}\ }\textbf {\bibinfo {volume}
  {59}},\ \bibinfo {pages} {106015} (\bibinfo {year} {2019})}\BibitemShut
  {NoStop}%
\bibitem [{\citenamefont {Pedersen}\ \emph {et~al.}(2019)\citenamefont
  {Pedersen}, \citenamefont {König}, \citenamefont {Jakubowski}, \citenamefont
  {Krychowiak}, \citenamefont {Gradic}, \citenamefont {Killer}, \citenamefont
  {Niemann}, \citenamefont {Szepesi}, \citenamefont {Wenzel}, \citenamefont
  {Ali}, \citenamefont {Anda}, \citenamefont {Baldzuhn}, \citenamefont
  {Barbui}, \citenamefont {Biedermann}, \citenamefont {Blackwell},
  \citenamefont {Bosch}, \citenamefont {Bozhenkov}, \citenamefont {Brakel},
  \citenamefont {Brezinsek}, \citenamefont {Cai}, \citenamefont {Cannas},
  \citenamefont {Coenen}, \citenamefont {Cosfeld}, \citenamefont {Dinklage},
  \citenamefont {Dittmar}, \citenamefont {Drewelow}, \citenamefont {Drews},
  \citenamefont {Dunai}, \citenamefont {Effenberg}, \citenamefont {Endler},
  \citenamefont {Feng}, \citenamefont {Fellinger}, \citenamefont {Ford},
  \citenamefont {Frerichs}, \citenamefont {Fuchert}, \citenamefont {Gao},
  \citenamefont {Geiger}, \citenamefont {Goriaev}, \citenamefont {Hammond},
  \citenamefont {Harris}, \citenamefont {Hathiramani}, \citenamefont {Henkel},
  \citenamefont {Kazakov}, \citenamefont {Kirschner}, \citenamefont {Knieps},
  \citenamefont {Kobayashi}, \citenamefont {Kocsis}, \citenamefont {Kornejew},
  \citenamefont {Kremeyer}, \citenamefont {Lazerzon}, \citenamefont {LeViness},
  \citenamefont {Li}, \citenamefont {Li}, \citenamefont {Liang}, \citenamefont
  {Liu}, \citenamefont {Lore}, \citenamefont {Masuzaki}, \citenamefont
  {Moncada}, \citenamefont {Neubauer}, \citenamefont {Ngo}, \citenamefont
  {Oelmann}, \citenamefont {Otte}, \citenamefont {Perseo}, \citenamefont
  {Pisano}, \citenamefont {Sitjes}, \citenamefont {Rack}, \citenamefont
  {Rasinski}, \citenamefont {Romazanov}, \citenamefont {Rudischhauser},
  \citenamefont {Schlisio}, \citenamefont {Schmitt}, \citenamefont {Schmitz},
  \citenamefont {Schweer}, \citenamefont {Sereda}, \citenamefont {Sleczka},
  \citenamefont {Suzuki}, \citenamefont {Vecsei}, \citenamefont {Wang},
  \citenamefont {Wauters}, \citenamefont {Wiesen}, \citenamefont {Winters},
  \citenamefont {Wurden}, \citenamefont {Zhang},\ and\ \citenamefont
  {and}}]{Pedersen2019}%
  \BibitemOpen
  \bibfield  {author} {\bibinfo {author} {\bibfnamefont {T.~S.}\ \bibnamefont
  {Pedersen}}, \bibinfo {author} {\bibfnamefont {R.}~\bibnamefont {König}},
  \bibinfo {author} {\bibfnamefont {M.}~\bibnamefont {Jakubowski}}, \bibinfo
  {author} {\bibfnamefont {M.}~\bibnamefont {Krychowiak}}, \bibinfo {author}
  {\bibfnamefont {D.}~\bibnamefont {Gradic}}, \bibinfo {author} {\bibfnamefont
  {C.}~\bibnamefont {Killer}}, \bibinfo {author} {\bibfnamefont
  {H.}~\bibnamefont {Niemann}}, \bibinfo {author} {\bibfnamefont
  {T.}~\bibnamefont {Szepesi}}, \bibinfo {author} {\bibfnamefont
  {U.}~\bibnamefont {Wenzel}}, \bibinfo {author} {\bibfnamefont
  {A.}~\bibnamefont {Ali}}, \bibinfo {author} {\bibfnamefont {G.}~\bibnamefont
  {Anda}}, \bibinfo {author} {\bibfnamefont {J.}~\bibnamefont {Baldzuhn}},
  \bibinfo {author} {\bibfnamefont {T.}~\bibnamefont {Barbui}}, \bibinfo
  {author} {\bibfnamefont {C.}~\bibnamefont {Biedermann}}, \bibinfo {author}
  {\bibfnamefont {B.}~\bibnamefont {Blackwell}}, \bibinfo {author}
  {\bibfnamefont {H.-S.}\ \bibnamefont {Bosch}}, \bibinfo {author}
  {\bibfnamefont {S.}~\bibnamefont {Bozhenkov}}, \bibinfo {author}
  {\bibfnamefont {R.}~\bibnamefont {Brakel}}, \bibinfo {author} {\bibfnamefont
  {S.}~\bibnamefont {Brezinsek}}, \bibinfo {author} {\bibfnamefont
  {J.}~\bibnamefont {Cai}}, \bibinfo {author} {\bibfnamefont {B.}~\bibnamefont
  {Cannas}}, \bibinfo {author} {\bibfnamefont {J.}~\bibnamefont {Coenen}},
  \bibinfo {author} {\bibfnamefont {J.}~\bibnamefont {Cosfeld}}, \bibinfo
  {author} {\bibfnamefont {A.}~\bibnamefont {Dinklage}}, \bibinfo {author}
  {\bibfnamefont {T.}~\bibnamefont {Dittmar}}, \bibinfo {author} {\bibfnamefont
  {P.}~\bibnamefont {Drewelow}}, \bibinfo {author} {\bibfnamefont
  {P.}~\bibnamefont {Drews}}, \bibinfo {author} {\bibfnamefont
  {D.}~\bibnamefont {Dunai}}, \bibinfo {author} {\bibfnamefont
  {F.}~\bibnamefont {Effenberg}}, \bibinfo {author} {\bibfnamefont
  {M.}~\bibnamefont {Endler}}, \bibinfo {author} {\bibfnamefont
  {Y.}~\bibnamefont {Feng}}, \bibinfo {author} {\bibfnamefont {J.}~\bibnamefont
  {Fellinger}}, \bibinfo {author} {\bibfnamefont {O.}~\bibnamefont {Ford}},
  \bibinfo {author} {\bibfnamefont {H.}~\bibnamefont {Frerichs}}, \bibinfo
  {author} {\bibfnamefont {G.}~\bibnamefont {Fuchert}}, \bibinfo {author}
  {\bibfnamefont {Y.}~\bibnamefont {Gao}}, \bibinfo {author} {\bibfnamefont
  {J.}~\bibnamefont {Geiger}}, \bibinfo {author} {\bibfnamefont
  {A.}~\bibnamefont {Goriaev}}, \bibinfo {author} {\bibfnamefont
  {K.}~\bibnamefont {Hammond}}, \bibinfo {author} {\bibfnamefont
  {J.}~\bibnamefont {Harris}}, \bibinfo {author} {\bibfnamefont
  {D.}~\bibnamefont {Hathiramani}}, \bibinfo {author} {\bibfnamefont
  {M.}~\bibnamefont {Henkel}}, \bibinfo {author} {\bibfnamefont
  {Y.}~\bibnamefont {Kazakov}}, \bibinfo {author} {\bibfnamefont
  {A.}~\bibnamefont {Kirschner}}, \bibinfo {author} {\bibfnamefont
  {A.}~\bibnamefont {Knieps}}, \bibinfo {author} {\bibfnamefont
  {M.}~\bibnamefont {Kobayashi}}, \bibinfo {author} {\bibfnamefont
  {G.}~\bibnamefont {Kocsis}}, \bibinfo {author} {\bibfnamefont
  {P.}~\bibnamefont {Kornejew}}, \bibinfo {author} {\bibfnamefont
  {T.}~\bibnamefont {Kremeyer}}, \bibinfo {author} {\bibfnamefont
  {S.}~\bibnamefont {Lazerzon}}, \bibinfo {author} {\bibfnamefont
  {A.}~\bibnamefont {LeViness}}, \bibinfo {author} {\bibfnamefont
  {C.}~\bibnamefont {Li}}, \bibinfo {author} {\bibfnamefont {Y.}~\bibnamefont
  {Li}}, \bibinfo {author} {\bibfnamefont {Y.}~\bibnamefont {Liang}}, \bibinfo
  {author} {\bibfnamefont {S.}~\bibnamefont {Liu}}, \bibinfo {author}
  {\bibfnamefont {J.}~\bibnamefont {Lore}}, \bibinfo {author} {\bibfnamefont
  {S.}~\bibnamefont {Masuzaki}}, \bibinfo {author} {\bibfnamefont
  {V.}~\bibnamefont {Moncada}}, \bibinfo {author} {\bibfnamefont
  {O.}~\bibnamefont {Neubauer}}, \bibinfo {author} {\bibfnamefont
  {T.}~\bibnamefont {Ngo}}, \bibinfo {author} {\bibfnamefont {J.}~\bibnamefont
  {Oelmann}}, \bibinfo {author} {\bibfnamefont {M.}~\bibnamefont {Otte}},
  \bibinfo {author} {\bibfnamefont {V.}~\bibnamefont {Perseo}}, \bibinfo
  {author} {\bibfnamefont {F.}~\bibnamefont {Pisano}}, \bibinfo {author}
  {\bibfnamefont {A.~P.}\ \bibnamefont {Sitjes}}, \bibinfo {author}
  {\bibfnamefont {M.}~\bibnamefont {Rack}}, \bibinfo {author} {\bibfnamefont
  {M.}~\bibnamefont {Rasinski}}, \bibinfo {author} {\bibfnamefont
  {J.}~\bibnamefont {Romazanov}}, \bibinfo {author} {\bibfnamefont
  {L.}~\bibnamefont {Rudischhauser}}, \bibinfo {author} {\bibfnamefont
  {G.}~\bibnamefont {Schlisio}}, \bibinfo {author} {\bibfnamefont
  {J.}~\bibnamefont {Schmitt}}, \bibinfo {author} {\bibfnamefont
  {O.}~\bibnamefont {Schmitz}}, \bibinfo {author} {\bibfnamefont
  {B.}~\bibnamefont {Schweer}}, \bibinfo {author} {\bibfnamefont
  {S.}~\bibnamefont {Sereda}}, \bibinfo {author} {\bibfnamefont
  {M.}~\bibnamefont {Sleczka}}, \bibinfo {author} {\bibfnamefont
  {Y.}~\bibnamefont {Suzuki}}, \bibinfo {author} {\bibfnamefont
  {M.}~\bibnamefont {Vecsei}}, \bibinfo {author} {\bibfnamefont
  {E.}~\bibnamefont {Wang}}, \bibinfo {author} {\bibfnamefont {T.}~\bibnamefont
  {Wauters}}, \bibinfo {author} {\bibfnamefont {S.}~\bibnamefont {Wiesen}},
  \bibinfo {author} {\bibfnamefont {V.}~\bibnamefont {Winters}}, \bibinfo
  {author} {\bibfnamefont {G.}~\bibnamefont {Wurden}}, \bibinfo {author}
  {\bibfnamefont {D.}~\bibnamefont {Zhang}}, \ and\ \bibinfo {author}
  {\bibfnamefont {S.~Z.}\ \bibnamefont {and}},\ }\href {\doibase
  10.1088/1741-4326/ab280f} {\bibfield  {journal} {\bibinfo  {journal} {Nuclear
  Fusion}\ }\textbf {\bibinfo {volume} {59}},\ \bibinfo {pages} {096014}
  (\bibinfo {year} {2019})}\BibitemShut {NoStop}%
\bibitem [{\citenamefont {Feng}\ \emph {et~al.}(2021)\citenamefont {Feng},
  \citenamefont {Jakubowski}, \citenamefont {König}, \citenamefont
  {Krychowiak}, \citenamefont {Otte}, \citenamefont {Reimold}, \citenamefont
  {Reiter}, \citenamefont {Schmitz}, \citenamefont {Zhang}, \citenamefont
  {Beidler}, \citenamefont {Biedermann}, \citenamefont {Bozhenkov},
  \citenamefont {Brunner}, \citenamefont {Dinklage}, \citenamefont {Drewelow},
  \citenamefont {Effenberg}, \citenamefont {Endler}, \citenamefont {Fuchert},
  \citenamefont {Gao}, \citenamefont {Geiger}, \citenamefont {Hammond},
  \citenamefont {Helander}, \citenamefont {Killer}, \citenamefont {Knauer},
  \citenamefont {Kremeyer}, \citenamefont {Pasch}, \citenamefont
  {Rudischhauser}, \citenamefont {Schlisio}, \citenamefont {Pedersen},
  \citenamefont {Wenzel}, \citenamefont {Winters},\ and\ \citenamefont
  {team}}]{Feng2021}%
  \BibitemOpen
  \bibfield  {author} {\bibinfo {author} {\bibfnamefont {Y.}~\bibnamefont
  {Feng}}, \bibinfo {author} {\bibfnamefont {M.}~\bibnamefont {Jakubowski}},
  \bibinfo {author} {\bibfnamefont {R.}~\bibnamefont {König}}, \bibinfo
  {author} {\bibfnamefont {M.}~\bibnamefont {Krychowiak}}, \bibinfo {author}
  {\bibfnamefont {M.}~\bibnamefont {Otte}}, \bibinfo {author} {\bibfnamefont
  {F.}~\bibnamefont {Reimold}}, \bibinfo {author} {\bibfnamefont
  {D.}~\bibnamefont {Reiter}}, \bibinfo {author} {\bibfnamefont
  {O.}~\bibnamefont {Schmitz}}, \bibinfo {author} {\bibfnamefont
  {D.}~\bibnamefont {Zhang}}, \bibinfo {author} {\bibfnamefont
  {C.}~\bibnamefont {Beidler}}, \bibinfo {author} {\bibfnamefont
  {C.}~\bibnamefont {Biedermann}}, \bibinfo {author} {\bibfnamefont
  {S.}~\bibnamefont {Bozhenkov}}, \bibinfo {author} {\bibfnamefont
  {K.}~\bibnamefont {Brunner}}, \bibinfo {author} {\bibfnamefont
  {A.}~\bibnamefont {Dinklage}}, \bibinfo {author} {\bibfnamefont
  {P.}~\bibnamefont {Drewelow}}, \bibinfo {author} {\bibfnamefont
  {F.}~\bibnamefont {Effenberg}}, \bibinfo {author} {\bibfnamefont
  {M.}~\bibnamefont {Endler}}, \bibinfo {author} {\bibfnamefont
  {G.}~\bibnamefont {Fuchert}}, \bibinfo {author} {\bibfnamefont
  {Y.}~\bibnamefont {Gao}}, \bibinfo {author} {\bibfnamefont {J.}~\bibnamefont
  {Geiger}}, \bibinfo {author} {\bibfnamefont {K.}~\bibnamefont {Hammond}},
  \bibinfo {author} {\bibfnamefont {P.}~\bibnamefont {Helander}}, \bibinfo
  {author} {\bibfnamefont {C.}~\bibnamefont {Killer}}, \bibinfo {author}
  {\bibfnamefont {J.}~\bibnamefont {Knauer}}, \bibinfo {author} {\bibfnamefont
  {T.}~\bibnamefont {Kremeyer}}, \bibinfo {author} {\bibfnamefont
  {E.}~\bibnamefont {Pasch}}, \bibinfo {author} {\bibfnamefont
  {L.}~\bibnamefont {Rudischhauser}}, \bibinfo {author} {\bibfnamefont
  {G.}~\bibnamefont {Schlisio}}, \bibinfo {author} {\bibfnamefont {T.~S.}\
  \bibnamefont {Pedersen}}, \bibinfo {author} {\bibfnamefont {U.}~\bibnamefont
  {Wenzel}}, \bibinfo {author} {\bibfnamefont {V.}~\bibnamefont {Winters}}, \
  and\ \bibinfo {author} {\bibfnamefont {W.-X.}\ \bibnamefont {team}},\ }\href
  {\doibase 10.1088/1741-4326/ac0772} {\bibfield  {journal} {\bibinfo
  {journal} {Nuclear Fusion}\ }\textbf {\bibinfo {volume} {61}},\ \bibinfo
  {pages} {086012} (\bibinfo {year} {2021})}\BibitemShut {NoStop}%
\bibitem [{\citenamefont {Baldzuhn}\ \emph {et~al.}(2019)\citenamefont
  {Baldzuhn}, \citenamefont {Damm}, \citenamefont {Beidler}, \citenamefont
  {McCarthy}, \citenamefont {Panadero}, \citenamefont {Biedermann},
  \citenamefont {Bozhenkov}, \citenamefont {Brunner}, \citenamefont {Fuchert},
  \citenamefont {Kazakov}, \citenamefont {Beurskens}, \citenamefont {Dibon},
  \citenamefont {Geiger}, \citenamefont {Grulke}, \citenamefont {Höfel},
  \citenamefont {Klinger}, \citenamefont {Köchl}, \citenamefont {Knauer},
  \citenamefont {Kocsis}, \citenamefont {Kornejew}, \citenamefont {Lang},
  \citenamefont {Langenberg}, \citenamefont {Laqua}, \citenamefont {Pablant},
  \citenamefont {Pasch}, \citenamefont {Pedersen}, \citenamefont {Ploeckl},
  \citenamefont {Rahbarnia}, \citenamefont {Schlisio}, \citenamefont {Scott},
  \citenamefont {Stange}, \citenamefont {von Stechow}, \citenamefont {Szepesi},
  \citenamefont {Turkin}, \citenamefont {Wagner}, \citenamefont {Winters},
  \citenamefont {Wurden},\ and\ \citenamefont {and}}]{Baldzuhn2019}%
  \BibitemOpen
  \bibfield  {author} {\bibinfo {author} {\bibfnamefont {J.}~\bibnamefont
  {Baldzuhn}}, \bibinfo {author} {\bibfnamefont {H.}~\bibnamefont {Damm}},
  \bibinfo {author} {\bibfnamefont {C.~D.}\ \bibnamefont {Beidler}}, \bibinfo
  {author} {\bibfnamefont {K.}~\bibnamefont {McCarthy}}, \bibinfo {author}
  {\bibfnamefont {N.}~\bibnamefont {Panadero}}, \bibinfo {author}
  {\bibfnamefont {C.}~\bibnamefont {Biedermann}}, \bibinfo {author}
  {\bibfnamefont {S.~A.}\ \bibnamefont {Bozhenkov}}, \bibinfo {author}
  {\bibfnamefont {K.~J.}\ \bibnamefont {Brunner}}, \bibinfo {author}
  {\bibfnamefont {G.}~\bibnamefont {Fuchert}}, \bibinfo {author} {\bibfnamefont
  {Y.}~\bibnamefont {Kazakov}}, \bibinfo {author} {\bibfnamefont
  {M.}~\bibnamefont {Beurskens}}, \bibinfo {author} {\bibfnamefont
  {M.}~\bibnamefont {Dibon}}, \bibinfo {author} {\bibfnamefont
  {J.}~\bibnamefont {Geiger}}, \bibinfo {author} {\bibfnamefont
  {O.}~\bibnamefont {Grulke}}, \bibinfo {author} {\bibfnamefont
  {U.}~\bibnamefont {Höfel}}, \bibinfo {author} {\bibfnamefont
  {T.}~\bibnamefont {Klinger}}, \bibinfo {author} {\bibfnamefont
  {F.}~\bibnamefont {Köchl}}, \bibinfo {author} {\bibfnamefont
  {J.}~\bibnamefont {Knauer}}, \bibinfo {author} {\bibfnamefont
  {G.}~\bibnamefont {Kocsis}}, \bibinfo {author} {\bibfnamefont
  {P.}~\bibnamefont {Kornejew}}, \bibinfo {author} {\bibfnamefont {P.~T.}\
  \bibnamefont {Lang}}, \bibinfo {author} {\bibfnamefont {A.}~\bibnamefont
  {Langenberg}}, \bibinfo {author} {\bibfnamefont {H.}~\bibnamefont {Laqua}},
  \bibinfo {author} {\bibfnamefont {N.~A.}\ \bibnamefont {Pablant}}, \bibinfo
  {author} {\bibfnamefont {E.}~\bibnamefont {Pasch}}, \bibinfo {author}
  {\bibfnamefont {T.~S.}\ \bibnamefont {Pedersen}}, \bibinfo {author}
  {\bibfnamefont {B.}~\bibnamefont {Ploeckl}}, \bibinfo {author} {\bibfnamefont
  {K.}~\bibnamefont {Rahbarnia}}, \bibinfo {author} {\bibfnamefont
  {G.}~\bibnamefont {Schlisio}}, \bibinfo {author} {\bibfnamefont {E.~R.}\
  \bibnamefont {Scott}}, \bibinfo {author} {\bibfnamefont {T.}~\bibnamefont
  {Stange}}, \bibinfo {author} {\bibfnamefont {A.}~\bibnamefont {von Stechow}},
  \bibinfo {author} {\bibfnamefont {T.}~\bibnamefont {Szepesi}}, \bibinfo
  {author} {\bibfnamefont {Y.}~\bibnamefont {Turkin}}, \bibinfo {author}
  {\bibfnamefont {F.}~\bibnamefont {Wagner}}, \bibinfo {author} {\bibfnamefont
  {V.}~\bibnamefont {Winters}}, \bibinfo {author} {\bibfnamefont
  {G.}~\bibnamefont {Wurden}}, \ and\ \bibinfo {author} {\bibfnamefont {D.~Z.}\
  \bibnamefont {and}},\ }\href {\doibase 10.1088/1361-6587/ab3567} {\bibfield
  {journal} {\bibinfo  {journal} {Plasma Physics and Controlled Fusion}\
  }\textbf {\bibinfo {volume} {61}},\ \bibinfo {pages} {095012} (\bibinfo
  {year} {2019})}\BibitemShut {NoStop}%
\bibitem [{\citenamefont {Bozhenkov}\ \emph {et~al.}(2020)\citenamefont
  {Bozhenkov}, \citenamefont {Kazakov}, \citenamefont {Ford}, \citenamefont
  {Beurskens}, \citenamefont {Alcus{\'{o}}n}, \citenamefont {Alonso},
  \citenamefont {Baldzuhn}, \citenamefont {Brandt}, \citenamefont {Brunner},
  \citenamefont {Damm}, \citenamefont {Fuchert}, \citenamefont {Geiger},
  \citenamefont {Grulke}, \citenamefont {Hirsch}, \citenamefont {Höfel},
  \citenamefont {Huang}, \citenamefont {Knauer}, \citenamefont {Krychowiak},
  \citenamefont {Langenberg}, \citenamefont {Laqua}, \citenamefont {Lazerson},
  \citenamefont {Marushchenko}, \citenamefont {Moseev}, \citenamefont {Otte},
  \citenamefont {Pablant}, \citenamefont {Pasch}, \citenamefont {Pavone},
  \citenamefont {Proll}, \citenamefont {Rahbarnia}, \citenamefont {Scott},
  \citenamefont {Smith}, \citenamefont {Stange}, \citenamefont {von Stechow},
  \citenamefont {Thomsen}, \citenamefont {Turkin}, \citenamefont {Wurden},
  \citenamefont {Xanthopoulos}, \citenamefont {Zhang},\ and\ \citenamefont
  {and}}]{Bozhenkov2020}%
  \BibitemOpen
  \bibfield  {author} {\bibinfo {author} {\bibfnamefont {S.}~\bibnamefont
  {Bozhenkov}}, \bibinfo {author} {\bibfnamefont {Y.}~\bibnamefont {Kazakov}},
  \bibinfo {author} {\bibfnamefont {O.}~\bibnamefont {Ford}}, \bibinfo {author}
  {\bibfnamefont {M.}~\bibnamefont {Beurskens}}, \bibinfo {author}
  {\bibfnamefont {J.}~\bibnamefont {Alcus{\'{o}}n}}, \bibinfo {author}
  {\bibfnamefont {J.}~\bibnamefont {Alonso}}, \bibinfo {author} {\bibfnamefont
  {J.}~\bibnamefont {Baldzuhn}}, \bibinfo {author} {\bibfnamefont
  {C.}~\bibnamefont {Brandt}}, \bibinfo {author} {\bibfnamefont
  {K.}~\bibnamefont {Brunner}}, \bibinfo {author} {\bibfnamefont
  {H.}~\bibnamefont {Damm}}, \bibinfo {author} {\bibfnamefont {G.}~\bibnamefont
  {Fuchert}}, \bibinfo {author} {\bibfnamefont {J.}~\bibnamefont {Geiger}},
  \bibinfo {author} {\bibfnamefont {O.}~\bibnamefont {Grulke}}, \bibinfo
  {author} {\bibfnamefont {M.}~\bibnamefont {Hirsch}}, \bibinfo {author}
  {\bibfnamefont {U.}~\bibnamefont {Höfel}}, \bibinfo {author} {\bibfnamefont
  {Z.}~\bibnamefont {Huang}}, \bibinfo {author} {\bibfnamefont
  {J.}~\bibnamefont {Knauer}}, \bibinfo {author} {\bibfnamefont
  {M.}~\bibnamefont {Krychowiak}}, \bibinfo {author} {\bibfnamefont
  {A.}~\bibnamefont {Langenberg}}, \bibinfo {author} {\bibfnamefont
  {H.}~\bibnamefont {Laqua}}, \bibinfo {author} {\bibfnamefont
  {S.}~\bibnamefont {Lazerson}}, \bibinfo {author} {\bibfnamefont {N.~B.}\
  \bibnamefont {Marushchenko}}, \bibinfo {author} {\bibfnamefont
  {D.}~\bibnamefont {Moseev}}, \bibinfo {author} {\bibfnamefont
  {M.}~\bibnamefont {Otte}}, \bibinfo {author} {\bibfnamefont {N.}~\bibnamefont
  {Pablant}}, \bibinfo {author} {\bibfnamefont {E.}~\bibnamefont {Pasch}},
  \bibinfo {author} {\bibfnamefont {A.}~\bibnamefont {Pavone}}, \bibinfo
  {author} {\bibfnamefont {J.}~\bibnamefont {Proll}}, \bibinfo {author}
  {\bibfnamefont {K.}~\bibnamefont {Rahbarnia}}, \bibinfo {author}
  {\bibfnamefont {E.}~\bibnamefont {Scott}}, \bibinfo {author} {\bibfnamefont
  {H.}~\bibnamefont {Smith}}, \bibinfo {author} {\bibfnamefont
  {T.}~\bibnamefont {Stange}}, \bibinfo {author} {\bibfnamefont
  {A.}~\bibnamefont {von Stechow}}, \bibinfo {author} {\bibfnamefont
  {H.}~\bibnamefont {Thomsen}}, \bibinfo {author} {\bibfnamefont
  {Y.}~\bibnamefont {Turkin}}, \bibinfo {author} {\bibfnamefont
  {G.}~\bibnamefont {Wurden}}, \bibinfo {author} {\bibfnamefont
  {P.}~\bibnamefont {Xanthopoulos}}, \bibinfo {author} {\bibfnamefont
  {D.}~\bibnamefont {Zhang}}, \ and\ \bibinfo {author} {\bibfnamefont {R.~W.}\
  \bibnamefont {and}},\ }\href {\doibase 10.1088/1741-4326/ab7867} {\bibfield
  {journal} {\bibinfo  {journal} {Nuclear Fusion}\ }\textbf {\bibinfo {volume}
  {60}},\ \bibinfo {pages} {066011} (\bibinfo {year} {2020})}\BibitemShut
  {NoStop}%
\bibitem [{\citenamefont {Rahbarnia}\ \emph {et~al.}(2020)\citenamefont
  {Rahbarnia}, \citenamefont {Thomsen}, \citenamefont {Schilling},
  \citenamefont {vaz Mendes}, \citenamefont {Endler}, \citenamefont {Kleiber},
  \citenamefont {Könies}, \citenamefont {Borchardt}, \citenamefont {Slaby},
  \citenamefont {Bluhm}, \citenamefont {Zilker}, \citenamefont {Carvalho},\
  and\ \citenamefont {Team}}]{Rahbarnia2020}%
  \BibitemOpen
  \bibfield  {author} {\bibinfo {author} {\bibfnamefont {K.}~\bibnamefont
  {Rahbarnia}}, \bibinfo {author} {\bibfnamefont {H.}~\bibnamefont {Thomsen}},
  \bibinfo {author} {\bibfnamefont {J.}~\bibnamefont {Schilling}}, \bibinfo
  {author} {\bibfnamefont {S.}~\bibnamefont {vaz Mendes}}, \bibinfo {author}
  {\bibfnamefont {M.}~\bibnamefont {Endler}}, \bibinfo {author} {\bibfnamefont
  {R.}~\bibnamefont {Kleiber}}, \bibinfo {author} {\bibfnamefont
  {A.}~\bibnamefont {Könies}}, \bibinfo {author} {\bibfnamefont
  {M.}~\bibnamefont {Borchardt}}, \bibinfo {author} {\bibfnamefont
  {C.}~\bibnamefont {Slaby}}, \bibinfo {author} {\bibfnamefont
  {T.}~\bibnamefont {Bluhm}}, \bibinfo {author} {\bibfnamefont
  {M.}~\bibnamefont {Zilker}}, \bibinfo {author} {\bibfnamefont {B.~B.}\
  \bibnamefont {Carvalho}}, \ and\ \bibinfo {author} {\bibfnamefont {W.-X.}\
  \bibnamefont {Team}},\ }\href {\doibase 10.1088/1361-6587/abc395} {\bibfield
  {journal} {\bibinfo  {journal} {Plasma Physics and Controlled Fusion}\
  }\textbf {\bibinfo {volume} {63}},\ \bibinfo {pages} {015005} (\bibinfo
  {year} {2020})}\BibitemShut {NoStop}%
\bibitem [{\citenamefont {Sheehan}\ and\ \citenamefont
  {Hershkowitz}(2011)}]{Sheehan2011}%
  \BibitemOpen
  \bibfield  {author} {\bibinfo {author} {\bibfnamefont {J.~P.}\ \bibnamefont
  {Sheehan}}\ and\ \bibinfo {author} {\bibfnamefont {N.}~\bibnamefont
  {Hershkowitz}},\ }\href {\doibase 10.1088/0963-0252/20/6/063001} {\bibfield
  {journal} {\bibinfo  {journal} {Plasma Sources Science and Technology}\
  }\textbf {\bibinfo {volume} {20}},\ \bibinfo {pages} {063001} (\bibinfo
  {year} {2011})}\BibitemShut {NoStop}%
\bibitem [{\citenamefont {Winslow}\ \emph {et~al.}(1998)\citenamefont
  {Winslow}, \citenamefont {Bengtson}, \citenamefont {Richards},\ and\
  \citenamefont {Wootton}}]{Winslow1998}%
  \BibitemOpen
  \bibfield  {author} {\bibinfo {author} {\bibfnamefont {D.~L.}\ \bibnamefont
  {Winslow}}, \bibinfo {author} {\bibfnamefont {R.~D.}\ \bibnamefont
  {Bengtson}}, \bibinfo {author} {\bibfnamefont {B.}~\bibnamefont {Richards}},
  \ and\ \bibinfo {author} {\bibfnamefont {A.~J.}\ \bibnamefont {Wootton}},\
  }\href {\doibase 10.1063/1.872779} {\bibfield  {journal} {\bibinfo  {journal}
  {Physics of Plasmas}\ }\textbf {\bibinfo {volume} {5}},\ \bibinfo {pages}
  {752} (\bibinfo {year} {1998})},\ \Eprint
  {http://arxiv.org/abs/https://doi.org/10.1063/1.872779}
  {https://doi.org/10.1063/1.872779} \BibitemShut {NoStop}%
\bibitem [{\citenamefont {Thomsen}\ \emph {et~al.}(2005)\citenamefont
  {Thomsen}, \citenamefont {Endler}, \citenamefont {Klinger},\ and\
  \citenamefont {the W7-AS~Team}}]{Thomsen2005}%
  \BibitemOpen
  \bibfield  {author} {\bibinfo {author} {\bibfnamefont {H.}~\bibnamefont
  {Thomsen}}, \bibinfo {author} {\bibfnamefont {M.}~\bibnamefont {Endler}},
  \bibinfo {author} {\bibfnamefont {T.}~\bibnamefont {Klinger}}, \ and\
  \bibinfo {author} {\bibnamefont {the W7-AS~Team}},\ }\href {\doibase
  10.1088/0741-3335/47/9/003} {\bibfield  {journal} {\bibinfo  {journal}
  {Plasma Physics and Controlled Fusion}\ }\textbf {\bibinfo {volume} {47}},\
  \bibinfo {pages} {1401} (\bibinfo {year} {2005})}\BibitemShut {NoStop}%
\bibitem [{\citenamefont {Rapson}\ \emph {et~al.}(2014)\citenamefont {Rapson},
  \citenamefont {Grulke}, \citenamefont {Matyash},\ and\ \citenamefont
  {Klinger}}]{Rapson2014}%
  \BibitemOpen
  \bibfield  {author} {\bibinfo {author} {\bibfnamefont {C.}~\bibnamefont
  {Rapson}}, \bibinfo {author} {\bibfnamefont {O.}~\bibnamefont {Grulke}},
  \bibinfo {author} {\bibfnamefont {K.}~\bibnamefont {Matyash}}, \ and\
  \bibinfo {author} {\bibfnamefont {T.}~\bibnamefont {Klinger}},\ }\href
  {\doibase 10.1063/1.4875577} {\bibfield  {journal} {\bibinfo  {journal}
  {Physics of Plasmas}\ }\textbf {\bibinfo {volume} {21}},\ \bibinfo {pages}
  {052103} (\bibinfo {year} {2014})},\ \Eprint
  {http://arxiv.org/abs/https://doi.org/10.1063/1.4875577}
  {https://doi.org/10.1063/1.4875577} \BibitemShut {NoStop}%
\bibitem [{\citenamefont {Medina}\ \emph {et~al.}(2001)\citenamefont {Medina},
  \citenamefont {Pedrosa}, \citenamefont {Ochando}, \citenamefont {Rodriguez},
  \citenamefont {Hidalgo}, \citenamefont {Fraguas},\ and\ \citenamefont
  {Carreras}}]{Medina2001}%
  \BibitemOpen
  \bibfield  {author} {\bibinfo {author} {\bibfnamefont {F.}~\bibnamefont
  {Medina}}, \bibinfo {author} {\bibfnamefont {M.~A.}\ \bibnamefont {Pedrosa}},
  \bibinfo {author} {\bibfnamefont {M.~A.}\ \bibnamefont {Ochando}}, \bibinfo
  {author} {\bibfnamefont {L.}~\bibnamefont {Rodriguez}}, \bibinfo {author}
  {\bibfnamefont {C.}~\bibnamefont {Hidalgo}}, \bibinfo {author} {\bibfnamefont
  {A.~L.}\ \bibnamefont {Fraguas}}, \ and\ \bibinfo {author} {\bibfnamefont
  {B.~A.}\ \bibnamefont {Carreras}},\ }\href {\doibase 10.1063/1.1310579}
  {\bibfield  {journal} {\bibinfo  {journal} {Review of Scientific
  Instruments}\ }\textbf {\bibinfo {volume} {72}},\ \bibinfo {pages} {471}
  (\bibinfo {year} {2001})},\ \Eprint
  {http://arxiv.org/abs/https://doi.org/10.1063/1.1310579}
  {https://doi.org/10.1063/1.1310579} \BibitemShut {NoStop}%
\bibitem [{\citenamefont {Hirsch}\ \emph {et~al.}(2008)\citenamefont {Hirsch},
  \citenamefont {Baldzuhn}, \citenamefont {Beidler}, \citenamefont {Brakel},
  \citenamefont {Burhenn}, \citenamefont {Dinklage}, \citenamefont {Ehmler},
  \citenamefont {Endler}, \citenamefont {Erckmann}, \citenamefont {Feng},
  \citenamefont {Geiger}, \citenamefont {Giannone}, \citenamefont {Grieger},
  \citenamefont {Grigull}, \citenamefont {Hartfuß}, \citenamefont {Hartmann},
  \citenamefont {Jaenicke}, \citenamefont {König}, \citenamefont {Laqua},
  \citenamefont {Maaßberg}, \citenamefont {McCormick}, \citenamefont {Sardei},
  \citenamefont {Speth}, \citenamefont {Stroth}, \citenamefont {Wagner},
  \citenamefont {Weller}, \citenamefont {Werner}, \citenamefont {Wobig},
  \citenamefont {Zoletnik},\ and\ \citenamefont {for~the
  W7-AS~Team}}]{Hirsch2008}%
  \BibitemOpen
  \bibfield  {author} {\bibinfo {author} {\bibfnamefont {M.}~\bibnamefont
  {Hirsch}}, \bibinfo {author} {\bibfnamefont {J.}~\bibnamefont {Baldzuhn}},
  \bibinfo {author} {\bibfnamefont {C.}~\bibnamefont {Beidler}}, \bibinfo
  {author} {\bibfnamefont {R.}~\bibnamefont {Brakel}}, \bibinfo {author}
  {\bibfnamefont {R.}~\bibnamefont {Burhenn}}, \bibinfo {author} {\bibfnamefont
  {A.}~\bibnamefont {Dinklage}}, \bibinfo {author} {\bibfnamefont
  {H.}~\bibnamefont {Ehmler}}, \bibinfo {author} {\bibfnamefont
  {M.}~\bibnamefont {Endler}}, \bibinfo {author} {\bibfnamefont
  {V.}~\bibnamefont {Erckmann}}, \bibinfo {author} {\bibfnamefont
  {Y.}~\bibnamefont {Feng}}, \bibinfo {author} {\bibfnamefont {J.}~\bibnamefont
  {Geiger}}, \bibinfo {author} {\bibfnamefont {L.}~\bibnamefont {Giannone}},
  \bibinfo {author} {\bibfnamefont {G.}~\bibnamefont {Grieger}}, \bibinfo
  {author} {\bibfnamefont {P.}~\bibnamefont {Grigull}}, \bibinfo {author}
  {\bibfnamefont {H.-J.}\ \bibnamefont {Hartfuß}}, \bibinfo {author}
  {\bibfnamefont {D.}~\bibnamefont {Hartmann}}, \bibinfo {author}
  {\bibfnamefont {R.}~\bibnamefont {Jaenicke}}, \bibinfo {author}
  {\bibfnamefont {R.}~\bibnamefont {König}}, \bibinfo {author} {\bibfnamefont
  {H.~P.}\ \bibnamefont {Laqua}}, \bibinfo {author} {\bibfnamefont
  {H.}~\bibnamefont {Maaßberg}}, \bibinfo {author} {\bibfnamefont
  {K.}~\bibnamefont {McCormick}}, \bibinfo {author} {\bibfnamefont
  {F.}~\bibnamefont {Sardei}}, \bibinfo {author} {\bibfnamefont
  {E.}~\bibnamefont {Speth}}, \bibinfo {author} {\bibfnamefont
  {U.}~\bibnamefont {Stroth}}, \bibinfo {author} {\bibfnamefont
  {F.}~\bibnamefont {Wagner}}, \bibinfo {author} {\bibfnamefont
  {A.}~\bibnamefont {Weller}}, \bibinfo {author} {\bibfnamefont
  {A.}~\bibnamefont {Werner}}, \bibinfo {author} {\bibfnamefont
  {H.}~\bibnamefont {Wobig}}, \bibinfo {author} {\bibfnamefont
  {S.}~\bibnamefont {Zoletnik}}, \ and\ \bibinfo {author} {\bibnamefont
  {for~the W7-AS~Team}},\ }\href
  {http://stacks.iop.org/0741-3335/50/i=5/a=053001} {\bibfield  {journal}
  {\bibinfo  {journal} {Plasma Physics and Controlled Fusion}\ }\textbf
  {\bibinfo {volume} {50}},\ \bibinfo {pages} {053001} (\bibinfo {year}
  {2008})}\BibitemShut {NoStop}%
\bibitem [{\citenamefont {Laqua}\ \emph {et~al.}(2014)\citenamefont {Laqua},
  \citenamefont {Chlechowitz}, \citenamefont {Otte},\ and\ \citenamefont
  {Stange}}]{Laqua2014}%
  \BibitemOpen
  \bibfield  {author} {\bibinfo {author} {\bibfnamefont {H.~P.}\ \bibnamefont
  {Laqua}}, \bibinfo {author} {\bibfnamefont {E.}~\bibnamefont {Chlechowitz}},
  \bibinfo {author} {\bibfnamefont {M.}~\bibnamefont {Otte}}, \ and\ \bibinfo
  {author} {\bibfnamefont {T.}~\bibnamefont {Stange}},\ }\href {\doibase
  10.1088/0741-3335/56/7/075022} {\bibfield  {journal} {\bibinfo  {journal}
  {Plasma Physics and Controlled Fusion}\ }\textbf {\bibinfo {volume} {56}},\
  \bibinfo {pages} {075022} (\bibinfo {year} {2014})}\BibitemShut {NoStop}%
\end{thebibliography}

%

\end{document}